\UseRawInputEncoding
\documentclass[preprint,12pt,authoryear]{elsarticle}

\usepackage[utf8]{inputenc}
\usepackage[T1]{fontenc}
\usepackage{amsmath,amssymb}
\usepackage{graphicx}
\usepackage[usenames,dvipsnames]{color}
\usepackage{enumitem}
\usepackage{hyperref}
\usepackage{lineno}
\DeclareUnicodeCharacter{03BA}{\ensuremath{\kappa}}
\DeclareUnicodeCharacter{22C6}{\ensuremath{\star}}

\journal{Advances in Space Research}

\begin{document}


\begin{frontmatter}

\title{Collisionless and collisional kinetics of a plasma atmosphere with spatially and temporally intermittent heating at its base}

\author[inst1]{Luca Barbieri\corref{cor1}}
\ead{luca.barbieri@obspm.fr}
\author[inst1]{Pascal D\'emoulin}
\author[inst1]{Filippo Pantellini}

\cortext[cor1]{Corresponding author}
\address[inst1]{LIRA, Observatoire de Paris, Universit\'e PSL, CNRS, Sorbonne Universit\'e, Universit\'e Paris-Cit\'e, 5 place Jules Janssen, 92195 Meudon, France}

\begin{abstract}
The solar corona exhibits a pronounced temperature inversion, with plasma temperatures increasing by nearly two orders of magnitude from the chromosphere to the corona. The solar atmosphere hosts spatially sparse and temporally intermittent heating events whose role in establishing this thermal structure remains an open question. In this work, we investigate how the combined effects of spatial intermittency and temporal variability of stochastic heating localized at the base of the transition region shape the temperature and density structure of coronal loops within a kinetic framework. Stochastic thermal boundary conditions are introduced together with a surface coarse-graining procedure. In the collisionless limit, analytical solutions are obtained for two stationary regimes corresponding to heating-event time scales shorter or longer than the particle crossing time along the loop. Coulomb collisions are then incorporated through a reduced kinetic model describing the gradual thermalization of suprathermal particles with height.

In the collisionless short-time-scale regime, spatial filling factor and temporal intermittency enter symmetrically through a single effective parameter controlling the abundance of suprathermal particles, leading to the formation of a transition region and a hot corona both within individual heated loops and at the coarse-grained level. When Coulomb collisions are included, this thermal structure is preserved, although the coronal density is significantly reduced by the progressive thermalization of the suprathermal population. In the long-time-scale regime, where individual loops become nearly isothermal, the temperature inversion emerges only after surface coarse graining and is controlled solely by the spatial filling factor. In this case, Coulomb collisions and optically thin radiative losses produce only minor modifications of the loop structure, while the plasma remains in a low-beta state with density and temperature profiles broadly consistent with coronal observations.

These results show that spatially sparse and temporally intermittent heating naturally generates suprathermal particle distributions and reproduces the observed thermal structure of the solar corona within a kinetic framework, while highlighting the markedly different sensitivity of the short- and long-time-scale regimes to collisional effects.
\end{abstract}

\end{frontmatter}


\section{Introduction}

From the chromosphere to the solar corona, the outermost layer of the solar atmosphere, the temperature rises from a few thousand to several million kelvin with increasing altitude, while the density simultaneously decreases by more than two orders of magnitude. This opposite behaviour of the temperature and density profiles is referred to as a temperature inversion. Understanding how the coronal plasma attains such a stationary configuration remains one of the key unresolved problems in solar physics, commonly known as the coronal heating problem \citep{Aschwandensolarcorona}.

Several mechanisms have been proposed to explain how the corona is heated to such extreme temperatures. Among them are 
\setlist{nolistsep}
\begin{itemize}
     \item impulsive magnetic reconnection such as nanoflares \citep{Parker:1972wu,2005ApJ...618.1020G,Mason2023,Bose2024,Johnston_2025,Mondal_2025}, 
    \item wave dissipation \citep{Ionson_1978,Heyvaerts_Priest_1983,Ayaz_2024,Ayaz2025b,Ayaz2025a,Ayaz2026a,Ayaz2026b},
    \item magnetohydrodynamic turbulence \citep{Dmitruk:1997uf,Rappazzo:2008vl,Dahlburg2012,2013ApJ...773L...2R,Usmanov_2025}, 
    \item wave-particle interactions and kinetic heating processes \citep{Vocks2001,Vocks_2002a,Vocks_2002b,Marsch2003,Chandran_2010,Bian2010,Bian2011,Vocks2021,Ayaz_2024,Ayaz2025b,Ayaz2025a,Ayaz2026a,Ayaz2026b}.
\end{itemize}
Global magnetic models incorporating several of the above effects have been developed by \citet{Yalim_2023,Usmanov_2025}.  
Finally, models based on gravitational filtering of suprathermal particles provide an alternative to heating the corona directly (\cite{Scudder1992a,Scudder1992b,Meyer-Vernet_2007,Pierrard2014,barbieri2023temperature,Barbieri2024b,Hau_2025,Barbieri2025c,Banik2026}; see also \cite{Mullan_2025} and references therein). 

For detailed reviews of the coronal heating problem, we refer the reader to \citet{Klimchuk_2006,Marsch2006,Reale2010,2015RSPTA.37340265W,Klimchuk2015,Viall2021,Arregui2024}.
Coronal heating models have also been successfully applied to study the hot coronae of low-mass (i.e. $M<1.5  \,M_{\odot}$) main-sequence stars \citep{Shoda2021,Airapetian_2021,Shoda2024,barbieri2024temperaturedensityprofilescorona}.

The upper chromosphere and the base of the transition region are highly dynamic layers, characterized by intense, short-lived, and small-scale heating events \citep[for extended reviews, see][]{Young2018-sp,Harra2025,parenti2025}. 
Spectroscopic and imaging studies of transition-region brightenings have been carried out using several instruments over the past decades: 
\setlist{nolistsep}
\begin{itemize}
  \item Observations with \textit{SOHO} revealed a variety of transient brightenings \citep{Dere:1989ux,Teriaca:2004wy}. 
  \item With \textit{IRIS}, numerous classes of small-scale events, including UV bursts, nanojets, and transition-region brightenings, have been identified and analyzed \citep{Peter:2014uz,Tiwari:2019us,Lee:ApJ2020,Bhatnagar2025,Milanovic_2025,Dolliou2026}. 
  \item EUV and UV brightenings, as well as nanojets occurring in both active regions and the quiet Sun, have been detected with \textit{SDO}/AIA \citep{Rauoafi:ApJ2023,Huang2023,Dolliou_2024,Wang2025,Dolliou2026}. 
  \item \textit{Hinode}/EIS has provided high-resolution spectroscopic and imaging diagnostics of active-region dynamics \citep{Lee:ApJ2020,Huang2023,Dolliou_2024}. 
  \item More recently, \textit{Solar Orbiter}/EUI has revealed a wealth of small-scale phenomena such as “campfires” and EUV brightenings \citep{Berghmans:2021wl,Zhukov2021,Dolliou2023,narang2025,Lim2025a,Lim2025}. 
  \item Additional spectroscopic diagnostics from \textit{Solar Orbiter}/SPICE have further contributed to the characterization of these events \citep{Chitta2022,Dolliou_2024}.
\end{itemize} ~\\

Motivated by these observational findings, two new kinetic models have recently been proposed to investigate the coronal heating problem. The first one \citep{barbieri2023temperature,Barbieri2024b} employs a one-dimensional kinetic approach and shows that intense, intermittent, and short-lived heating events can drive the system toward a stationary state in which temperature and density are anti-correlated. The resulting temperature profiles closely reproduce those observed in the corona. In this model, the temporal characteristics of the heating events are crucial in producing and shaping the temperature inversion.  

The second model \citep{Barbieri2025c} combines a three-dimensional kinetic framework with an analytical coarse-graining method. It demonstrates that small-scale, spatially distributed heating events can also produce a temperature inversion  provided that the events occupy only a small fraction of the total surface area. 
In this scenario, the spatial distribution, that is, the surface fraction covered by heating events, plays a key role in determining the temperature inversion.  

In this paper, we explore the connection between these two approaches. We analyse two distinct time-scale regimes of the heating process. The first is a short time-scale regime, in which the characteristic time scales of the heating events are shorter than the electron crossing time. This regime is compatible with recent observations of “campfire” brightenings \citep{narang2025,Lim2025}. 

The second is a long time-scale regime, in which the characteristic time scales of the heating events are comparable to or longer than the proton crossing time. This regime is consistent with recent spectroscopic studies \citep{Dolliou_2024}. Heating events of this duration have also been extensively observed and analysed over the past decades  \citep[see][ and references therein]{parenti2025,Harra2025} .

We show that, in the fast time-scale regime, the temperature inversion can be governed by the combined effects of the temporal intermittency of the heating and its spatial distribution. By contrast, in the long time-scale regime, it is controlled solely by the spatial filling factor. We derive the role of both effects in shaping the resulting temperature and density profiles.

We also examine the mechanical equilibrium of the plasma structure. We then include the effect of Coulomb collisions, discussing in which regime they can modify the results and analysing their impact on the density and temperature profiles.  Finally we estimate the effect of radiative cooling in the regime robust to Coulomb collisions.

The paper is structured as follows. 
Section~\ref{sec1} introduces the collisionless model and we summarise the two regimes that lead to temperature inversion, based on previous theoretical studies \citep{Barbieri2025b,Barbieri2025c} and relevant observations \citep{Dolliou_2024,narang2025,Lim2025}. We define the boundary condition at the loop base, then derive from it the resulting velocity distribution in the volume.
In Section~\ref{sec3}, we introduce a local spatial averaging (coarse graining), then derive the corresponding  velocity distribution functions in the two timescale regimes, as well as the corresponding density and temperature profiles. We also verify mechanical equilibrium between neighbouring loops.  In Section~\ref{sec4}, we model the effect of Coulomb collisions and quantify their impact in the two timescale regimes.
In Section~\ref{sec5} we estimate the effect of radiative cooling in the long-time-scale regime. In Section~\ref{secLimitations} we outline the limitations and future perspectives of present model.  Finally, Section~\ref{secSummary} summarises and discusses the main results of the present work.

\section{The collisionless model}  
\label{sec1}

\subsection{Basic equations} \label{subsec:basic}

We consider a two-species plasma atmosphere composed of electrons and protons confined within a closed magnetic loop. The plasma is assumed to be globally at rest. In the present treatment, the two loop footpoints are symmetric with respect to the top, with the same boundary conditions and equal plasma pressure, so that no net field-aligned mass flow develops.

For simplicity, we restrict the analysis to the nearly vertical portions of the loop and approximate the magnetic field as vertical. This approximation is appropriate for sufficiently large coronal loops whose height exceeds the gravitational scale height of the plasma. It allows the particle dynamics to be described using the vertical Cartesian coordinate $z$, rather than a curvilinear coordinate along the loop, while retaining the gravitational
stratification that is essential to the present model.

The plasma of a heated loop is treated as a collisionless, two-species system composed of electrons and protons, which interact through self-consistent electrostatic fields. Throughout the present work, electrons and protons are assumed to share the same temperature profile, $T_e=T_p$. Under this assumption, quasi-neutrality is maintained by the classical Pannekoek--Rosseland polarization electric field
\citep{Pannekoek_1922,Rosseland_1924,Neslusan2001-rp,
belmont2013collisionless,Barbieri_2025c},
\begin{equation}\label{PRfield}
    E_{\rm PR} = \frac{(m_p-m_e)g}{2\,e},
\end{equation}
where $g$ is the magnitude of gravity and $m_e$ and $m_p$ are the electron and proton masses, respectively. The combined gravitational and electrostatic forces therefore result in the same effective gravitational force for both species, with magnitude
\begin{equation}
    F_g = m\, g,
    \qquad
    m = \frac{m_p+m_e}{2}.
\end{equation}
If the electron and proton temperature profiles differ, the classical Pannekoek--Rosseland field is no longer sufficient to ensure quasi-neutrality and the ambipolar electric field must instead be determined self-consistently. This more general case is discussed in Subsect.~\ref{secSpecies-dependent}.

The distribution function $f_{\alpha}$ for each species $\alpha$ evolves according to the collisionless Vlasov equation
\begin{equation}\label{Vlasovdynamics}
    \frac{\partial f_{\alpha}}{\partial t}
    + \textbf{v} \cdot \nabla f_{\alpha}
    + \frac{\textbf{F}_{\alpha}}{m_{\alpha}}
      \cdot \nabla_{\textbf{v}} f_{\alpha} = 0 \,,
\end{equation}
where $\textbf{F}_{\alpha}$ is the total force acting on particles of species $\alpha$, defined as
\begin{equation}\label{totalforce}
    \textbf{F}_{\alpha} = m \,\textbf{g}
      + \frac{e_{\alpha}}{c} \textbf{v} \times \textbf{B} \,,
      \qquad m = \frac{m_p + m_e}{2} \,.
\end{equation}

We assume that the loop cross-section is constant. Magnetic-flux conservation implies a constant magnetic-field
strength along the loop. The particle motion conserves the first adiabatic invariant,
\begin{equation}\label{eq:magneticmoment}
    \mu_\alpha = \frac{m_\alpha\, v_\perp^2}{2\,B}.
\end{equation}
Since $B$ is constant along the loop, conservation of $\mu_\alpha$ does not produce any exchange between perpendicular and parallel particle energy, and
the associated magnetic-mirror force, vanishes. The magnetic field therefore does not modify the velocity distribution function within the assumptions of present model, although it confines particle motion along the magnetic field lines.

For typical coronal conditions ($B = 10~\mathrm{G}$,
$T \sim 10^6~\mathrm{K}$), the Larmor radii
$r_{L,e} \approx 0.03~\mathrm{m}$ and $r_{L,p} \approx 1.3~\mathrm{m}$ are much smaller than the loop length by a
factor of less than $10^{-6}$. If the loop cross-section varies, however, magnetic-flux conservation implies a spatially varying $B$, so that conservation of $\mu_\alpha$ affects the particle dynamics through the magnetic-mirror force and can modify the velocity distribution function.
This more general case is discussed in Subsect.~\ref{secBexpansion} and will be the subject of future work.

\begin{figure*}
    \centering
    \includegraphics[width=0.99\textwidth]{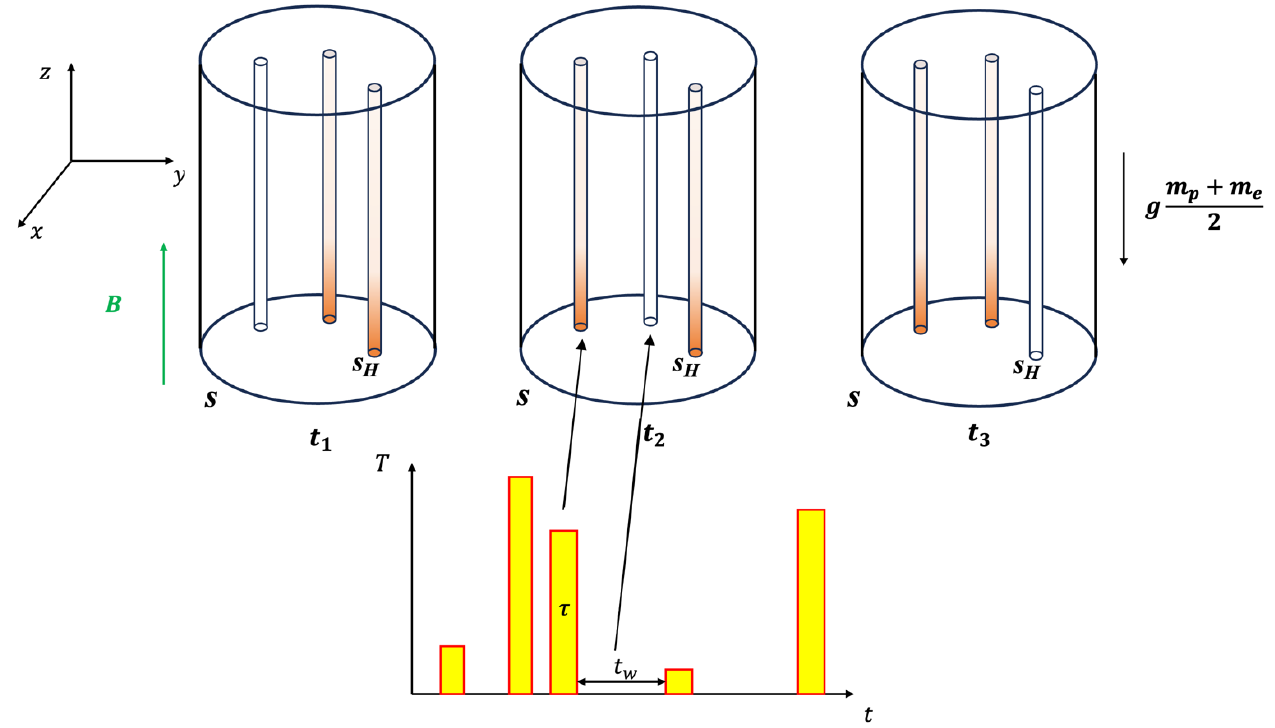}
    \caption{Schematic representation of the coronal plasma model. The plasma is confined within a symmetric closed magnetic loop with respect to the top and is assumed to be globally at rest; the model describes the nearly vertical portions of the loop, approximated by a vertical magnetic field. \textit{Top:} the coronal plasma above the coarse-graining surface $S$, located at the base of the transition region where stochastic heating events occur. 
    Individual heating events are represented by circles of area $s_H$: filled orange circles indicate active heating events, whereas white circles with black contours mark regions without heating at that time. 
   The column above heating events have a colour gradient representing schematically the plasma density or temperature gradients with the larger intensity for the largest or lowest values respectively.
    Three snapshots at successive times $t_1$, $t_2$, and $t_3$ illustrate the temporal evolution of the spatial distribution of heating events at the base.  
    \textit{Bottom:} time series of a single heating event. 
    The heating events have a heating phase of mean duration $\tau$ and a mean waiting time $t_w$ between two successive events. 
    The coronal plasma is embedded in a uniform magnetic field $\mathbf{B}$ (green). 
    Particles experience a net external force $\mathbf{g}(m_p + m_e)/2$ (black), which combines gravity and the Pannekoek–Rosseland electric field. 
    The Cartesian coordinate system $(x, y, z)$ used throughout the paper is shown in the top-left corner.
    }
    \label{fig:model}
\end{figure*}

\subsection{Boundary conditions} \label{subsec:boundary_conditions}

At the base of the loop ($z = 0$), the plasma is in thermal contact with a fully collisional boundary, which is modeled as a Maxwellian thermal reservoir
\begin{equation}\label{maxwellianboundary}
    f_{T,\alpha}= \hat{n}_{T}(x,y)\left(\frac{m_{\alpha}}{2\pi\, k_B \, T(x,y)}\right)^{3/2} 
    e^{-\frac{m_{\alpha} \, v^2}{2\,k_B \, T(x,y)}}   \,,
\end{equation}
where $\hat{n}_{T}(x,y)$ and $T(x,y)$ are the density and temperature within the upper chromosphere in the position $(x,y)$ of the horizontal plane. Outside heating events, this temperature remains fixed at the chromospheric background value $T_0 = 10^4\,\mathrm{K}$.

A wide range of observations consistently demonstrates that the base of the transition region is highly dynamic, hosting frequent, localized, and short-duration heating events (see the observational references compiled in the Introduction). Consequently, the effective boundary temperature at $z = 0$ is expected to fluctuate in time.  
We model this stochastic heating as follows. On the short time scale $\tau(x,y)$, a localized heating event raises the temperature to $T(x,y) = T_0 + \Delta T(x,y)$, where $T(x,y)$ is drawn from a probability distribution $\hat\gamma(x,y,T)$. After this interval, the temperature returns to the chromospheric background value $T_0$ for a waiting time $t_w(x,y)$. A new event then occurs, and the process repeats. A schematic illustration of the model is shown in Fig.~\ref{fig:model}.

For simplicity, we assume that each heating event induces a Maxwellian distribution function at temperature $T(x,y)$, while between events the distribution remains Maxwellian at $T_0$. This defines the boundary condition at $z = 0$ of our model. Stochastic heating induces different dynamical regimes in the overlying coronal plasma, depending on the characteristic time scales $\tau(x,y)$ and $t_w(x,y)$ of the temperature fluctuations.

The assumption that each heating event is described by a Maxwellian distribution should be regarded as a conservative approximation rather than as a realistic description of the microscopic heating process. Indeed, the short-time-scale regime considered later in this work corresponds to heating events whose characteristic duration is comparable to, or shorter than, the local collisional relaxation time at the base of transition region. Under these conditions, the heated particle populations at the base of the transition region are unlikely to reach local thermodynamic equilibrium. More realistically, the resulting velocity distribution functions are expected to develop suprathermal tails or extended wings \citep{Che_2021}, as commonly found in kinetic models of the solar atmosphere \citep[e.g.][]{Hau_2025,Banik2026} and stellar winds \citep{Chamberlain1960,Jockers1970,Lemaire1971,Maksimovic1997,Lamy2003,Zouganelis2004,Pierrard2011,Pierrard2023,PtersdeBonhome2025,Vinogradov2026}.

By representing each heating event with a Maxwellian distribution, the present model deliberately minimizes the non-thermal content of the heated particle population and therefore provides a conservative lower bound on the effects of stochastic heating. Nevertheless, the results presented below show that even under this conservative assumption the combined action of stochastic heating and gravitational filtering naturally produces a temperature inversion and the formation of a hot corona. The inclusion of more realistic non-Maxwellian heating events would therefore be expected to reinforce, rather than suppress, the kinetic effects described here.

Finally, in the present treatment, we assume that the two footpoints of the loop are statistically symmetric and both undergo stochastic heating, i.e. the same heating statistics and boundary conditions are imposed at the two loop footpoints. However, a more realistic situation may arise when only one footpoint experiences stochastic heating, while the other remains at the background chromospheric temperature $T_0$. In this case, the system is no longer symmetric with respect to the loop apex. As a consequence, a net flow is expected to develop along the loop. To preserve charge neutrality, the associated electric field must then deviate from the classical Pannekoek--Rosseland field \citep{belmont2013collisionless}. These additional effects lie beyond the scope of the present work. Nevertheless, in Section \ref{secAsymmetric} 
we briefly discuss the consequences of breaking the symmetry of the system.

\subsection{Short time-scales regime}

Previous analytical and numerical studies have shown that when the characteristic time scales $\tau(x,y)$ and $t_w(x,y)$ are shorter than the electron loop crossing time, $t_{c,e}$, the injected energy cannot be redistributed across the entire system \citep{Barbieri2024b,Barbieri2025b}. 
As a consequence, neither electrons nor protons are able to relax toward thermal equilibrium during the heating phases, which last a time interval $\tau(x,y)$. 
The same holds during the cooling phases, which last a time interval $t_w(x,y)$. 
Rather, the plasma evolves toward a non-thermal configuration arising from the coexistence and mixing of particle populations at different temperatures.
Formally, the condition required to reach such a stationary state can be expressed as
\begin{equation}\label{shorttimescalesregime}
    \tau(x,y), \, t_w(x,y) < t_{c,e}  \,.
\end{equation}

For loops shorter than the coronal gravitational scale height ($= k_B\, T / mg \approx 50$ Mm) and of length $L$, the thermal electron crossing time is
\begin{equation}\label{thermalcrossing}
   t_{c,e} = \frac{L}{v_{T,e}} \,, \qquad v_{T,e} = \sqrt{\frac{k_B \, T}{m_e}}   \,,
\end{equation}
In order to reach a stationary state (with a loop top at coronal temperature $T = 10^6\,\mathrm{K}$), the heating events must occur on time scales shorter than approximately $t_{c,e} \sim 15~\mathrm{s}$ for $L = 0.1 R_{\odot}$. Recent observations of small-scale brightenings have shown the existence of heating events with such short time scales \citep[commonly referred to as \textit{campfires}]{narang2025,Lim2025}. 
These studies reveal that the probability of detecting brightenings increases as the event duration decreases, reaching values as short as $\sim 1~\mathrm{s}$ (the temporal resolution limit of these observations). 

In this non-thermal stationary regime without collisions, the mixing of particle populations at different temperatures self-consistently generates suprathermal velocity distributions.  
In the resulting stationary state, the particle distribution function at the base ($z=0$) is given by \citet{Barbieri2024b,Barbieri2025b}:
\begin{equation}\label{multitempboundary}
\begin{aligned}
f_{\alpha}(x,y,z=0,v)
=
\mathcal{N}_{\alpha}
\Bigg[
&
\hat{A}_{t}(x,y)
\int_{T_0}^{\infty}
\frac{\hat{\gamma}(x,y,T)}{T^{3/2}}
e^{-\frac{m_{\alpha}\,v^2}{2k_B\, T}}
\,dT
\\
&
+\,
\frac{1-\hat{A}_t(x,y)}{T_0^{3/2}}
e^{-\frac{m_{\alpha}\,v^2}{2k_B\, T_0}}
\Bigg].
\end{aligned}
\end{equation}
where $\hat{A}_{t}(x,y)$ denotes the fraction of time during which a flux tube located in $(x,y)$ experiences a heating event,
\begin{equation}
    \hat{A}_t(x,y) = \frac{\tau(x,y)}{\tau(x,y) + t_w(x,y)}   \,,
\end{equation}
and the normalization constant $\mathcal{N}_{\alpha}$ is given by
\begin{equation} \nonumber
    \mathcal{N}_{\alpha} = n_0 \left(\frac{m_{\alpha}}{2\pi\, k_B}\right)^{3/2}  \,,
\end{equation}
with $n_0$ the particle density at $z=0$.

The interpretation of Eq.~\eqref{multitempboundary} is as follows. Since the time scales $\tau(x,y)$ and $t_w(x,y)$ are shorter than the electron crossing time, the particles generated by temperature increments coexist and mix with those produced at the background temperature $T_0$. Consequently, Eq.~\eqref{multitempboundary} consists of a thermal population at $T = T_0$ plus a non-thermal component arising from the superposition of Maxwellian populations at higher temperatures, each weighted by the probability $\hat{\gamma}(x,y,T)$ of a heating event occurring at temperature $T > T_0$. The relative importance of the non-thermal component is set by $A_{t,(x,y)}$, which represents the fraction of time during which the thermostat is not at the background temperature $T_0$. Since $f_{\alpha}(x,y,z,v)$ follows the collisionless Vlasov equation (Eq.~\eqref{Vlasovdynamics}), Liouville mapping can be applied to extend Eq.~\eqref{multitempboundary} in height, yielding
\begin{equation}\label{VDFshortTau}
\begin{aligned}
f_{\alpha}(x,y,z,v)
=
\mathcal{N}_{\alpha}
\Bigg(
&
\hat{A}_{t}(x,y)
\int_{T_0}^{\infty}
\frac{\hat{\gamma}(x,y,T)}{T^{3/2}}
e^{-\frac{H_{\alpha}}{k_B\, T}}
\,dT
\\
&
+\,
\frac{1-\hat{A}_{t}(x,y)}{T_0^{3/2}}
e^{-\frac{H_{\alpha}}{k_B\, T_0}}
\Bigg).
\end{aligned}
\end{equation}
where $H_{\alpha}$ is the single-particle energy
\begin{equation} \label{Halpha}
    H_{\alpha} = \frac{1}{2} m_{\alpha}\, v^2 + m \, g \, z  \,.
\end{equation}

At low altitudes $z$, the thermal population dominates, but it progressively decreases with height as the external potential in $H_{\alpha}$ filters out the slower particles.  
In contrast, suprathermal particles populate higher altitudes, producing a temperature inversion through gravitational filtering.

The detailed shape of the distribution function given by Eq.~\eqref{VDFshortTau} has been previously discussed (see Fig.~3 in \citealt{barbieri2023temperature}, Figs.~11--12 in \citealt{Barbieri2024b}, and Fig.~4 in \citet{Barbieri2025c}. Its form is qualitatively similar to a superposition of two Maxwellians at different temperatures, as in \citet{Meyer-Vernet_2007}, where the density of the cold population is proportional to $(1 - \hat{A}_{t}(x,y))$ and that of the hot population is proportional to $\hat{A}_{t}(x,y)$. A more detailed comparison with the bi-Maxwellian model is included in Section \ref{subsec:Coarse-grained_n_T}. 

\subsection{Long time-scales regime}\label{subsec:longtimescales}

If the characteristic time scales $\tau(x,y)$ and $t_w(x,y)$ are increased beyond the regime defined by Eq.~\eqref{shorttimescalesregime}, the loops no longer reach a stationary state and cease to display temperature inversion \citep{Barbieri2025b}. Instead, they evolve dynamically in time unless the time scales $\tau(x,y)$ and $t_w(x,y)$ become sufficiently long to satisfy 
\begin{equation}   \label{longtimescalesregime}
\tau(x,y), \, t_w(x,y) \gtrsim t_{c,p},
\end{equation}
where $t_{c,p}$ is the proton crossing time. This time scale is related to the electron crossing time by
\begin{equation}
t_{c,p} = \sqrt{\frac{m_p}{m_e}} \, t_{c,e} \sim 400 \,\mathrm{s} \sim 6\text{--}7 \,\mathrm{min}.
\end{equation}

In this limit, during a heating event, the boundary condition at $z=0$ is assumed to be a single Maxwellian distribution at temperature $T$. 
This regime of time scales lies well above the temporal resolution limit of $1\,\mathrm{s}$. Brightening events on such time scales have recently been observed and analyzed \citep{Dolliou_2024,narang2025,Lim2025} \citep[see for extended reviews][and references therein]{parenti2025,Harra2025}. Using again the Liouville mapping as in the previous subsection, the extension of the boundary condition with height is simply a Maxwellian distribution at temperature $T(x,y)$ filtered by gravity:
\begin{equation} \label{VDFlongTau}
    f_{\alpha}(x,y,z,v)= \hat{n}_{T}(x,y)\left(\frac{m_{\alpha}}{2\pi\, k_B\, T(x,y)}\right)^{3/2} e^{-\frac{H_{\alpha}}{k_B \, T(x,y)}} \,,
\end{equation}
where $H_{\alpha}$ is still given by Eq.~\eqref{Halpha}.

In summary, during a long-duration event, we assume that the heating input is sufficient to maintain a constant temperature at the loop base and along the loop. In this regime, nearly isothermal loops characterized by different temperatures are embedded in a cool medium (at $\approx T_0$). The heating is also assumed to be sufficient to set the loop base density to $\hat{n}_{T}(x,y)$, which is not directly related to $n_0$ but instead depends on the available heating. Observations of coronal brightening events, such as campfires, indicate typical densities at the base of heated plasma columns of
\(\sim 10^{9}\,\mathrm{cm}^{-3}\)
\citep{Berghmans:2021wl,Zhukov2021}.

\section{Coarse-grained model} 
\label{sec3}
The analysis of previous section considers an elementary loop with its footpoints undergoing stochastic heating.  Still, within the instrumental spatial resolution several loops, with different heating series, can be present. Then, a spatial averaging (coarse graining) of previous results is introduced below.

\subsection{The regime of short-time scales}\label{subsec:shorttimescale}

Under the conditions defined by Eq.~\eqref{shorttimescalesregime}, the duration of individual stochastic heating events is short compared with the particle crossing time. Consequently, at any given time a magnetic loop contains the superposition of particle populations generated by heating events occurring at different times and characterized by different temperatures. Averaging over an observational surface \(S\) yields the coarse-grained boundary distribution

\begin{equation}
\tilde f_\alpha(z=0,v)
=
\frac1S
\int_S
f_\alpha(x,y,z=0,v)\,dS
+
(1-A_S)\,
f_{T_0,\alpha}(v),
\label{coarsegrainedboundaryfasttimescales}
\end{equation}
where \(A_S\) denotes the fractional surface occupied by heating events,
\begin{equation}
A_S
=
\frac{\sum_{i=1}^{N_h}s_{H,i}}{S},
\end{equation}
with \(s_{H,i}\) the area of the \(i\)-th heating event while $N_h$ is the total number of heating events. The present work is therefore not concerned with the microscopic heating statistics and duty cycles of individual plasma columns, described by the quantities \(\hat{\gamma}(x,y,T)\) and \(\hat{A}_{t}(x,y)\), but rather with their statistical properties averaged over an observational window of surface area \(S\). We therefore introduce the surface-averaged heating-temperature distribution,

\begin{equation}
\gamma(T)
=
\frac{1}{S}
\int_S
\hat{\gamma}(x,y,T)\,dS,
\end{equation}
and the corresponding surface-averaged duty cycle,
\begin{equation}
A_t
=
\frac{1}{S}
\int_S
\hat{A}_{t}(x,y)\,dS.
\end{equation}
Throughout the present work, we therefore replace the microscopic heating distributions
\(\hat{\gamma}(x,y,T)\)
by their surface-averaged counterpart
\(\gamma(T)\), and the microscopic duty cycles
\(\hat{A}_{t}(x,y)\)
by their surface-averaged value
\(A_t\). 

This approximation should not be interpreted as assuming that all plasma columns undergo identical heating histories. Rather, it assumes that the unresolved plasma columns can be regarded as independent realizations of the same stationary stochastic process and are therefore statistically characterized by a common heating-temperature distribution and a common average duty cycle. In this sense, the approximation follows the same philosophy as the ergodic hypothesis of statistical mechanics \citep{huang1987statistical}: when the microscopic realization of the system is inaccessible, macroscopic quantities are described by the statistical properties of the ensemble rather than by individual realizations.

In practice, however, reconstructing the complete heating history of every individual plasma column represents an exceedingly challenging observational task \citep{Dolliou2023,Dolliou_2024}. Such a task would require identifying and tracking the temporal evolution of every elementary heating event occurring within each magnetic flux tube, together with the corresponding heating temperatures. Current observational studies instead focus on the statistical properties of ensembles of heating events contained within an observational window of surface area \(S\), such as their occurrence rates, characteristic lifetimes, spatial extents, and the distribution of their inferred heating temperatures \citep{Berghmans:2021wl,narang2025,Lim2025,Lim2025a}.

Before proceeding, we emphasize that the exact functional form of the heating-temperature distribution \(\gamma(T)\) has not yet been established observationally. In the following, we adopt a distribution motivated by the available observations. More importantly, we will show that, under the conditions considered in the present work, the resulting temperature and density profiles depend only weakly on the particular choice of \(\gamma(T)\), provided that its mean temperature is kept fixed. The average duty cycle \(A_t\) is instead constrained by requiring that the average temperature at the top of the chromosphere, corresponding to the lower boundary of the present model, remains consistent with the values inferred from observations.

Because of this assumption, we have

\begin{equation}
\frac1S
\int_S
f_\alpha(x,y,z=0,v)\,dS
=
A_t
\int_{T_0}^{\infty}
\frac{\gamma(T)}
{T^{3/2}}
e^{-\frac{m_{\alpha}\, v^2}{2\,k_B \, T}}   
\,dT.
\end{equation}
Liouville mapping the boundary distribution throughout the phase space, we get
\begin{equation}
\tilde f_\alpha(z,v)
=
\mathcal N_\alpha
\left[
A
\int_{T_0}^{\infty}
\frac{\gamma(T)}
{T^{3/2}}
e^{-\frac{H_\alpha}{k_B\, T}}
\,dT
+
\frac{1-A}
{T_0^{3/2}}
e^{-\frac{H_\alpha}{k_B\, T_0}}
\right],
\label{VDFphasespace}
\end{equation}
where
\begin{equation}
A=A_SA_t
\end{equation}
combines the spatial filling factor and the temporal intermittency of the heating events. Equation~\eqref{VDFphasespace} is formally identical to Eq.~\eqref{VDFshortTau}, except that the spatial filling factor \(A_S\) is replaced by the effective parameter \(A=A_SA_t\). This demonstrates that spatial intermittency and temporal intermittency enter the collisionless kinetic description in exactly the same way, contributing only through their product for the regime of short-time scales.

\subsection{The regime of long-time scales}

In the long time-scale regime defined by Eq.~\eqref{longtimescalesregime}, the system effectively consists of a mixture of isothermal loops characterized by different temperatures. For simplicity, we suppose that the distribution function of these events is still described by $\gamma(T)$.  In fact, it was shown in  \citet{Barbieri2025c} that the results depend weakly on the shape of $\gamma(T)$ provide it contains an extended enough tail at coronal temperature. Finally, this problem was treated analytically, leading to an explicit expression for the particle distribution function at the surface coarse-graining level. The resulting expression is mathematically equivalent to Eq.~\eqref{VDFphasespace}, with the replacement $A \rightarrow A_S$ and $\gamma(T) \rightarrow \gamma(T)\,n_T/n_0$.

In principle, each individual heating event is expected to produce a different base density \(\hat{n}_T(x,y)\), depending on local plasma conditions. Consistent with the statistical approach adopted throughout this work, we replace these microscopic values by their ensemble-averaged counterpart,
\begin{equation}
    n_T=\frac{1}{S}
\int_S \hat{n}_{T}(x,y)\,dS,
\end{equation}
whose value is inferred from the available observations $n_T \sim 10^{9} \mathrm{cm}^{-3}$ \citep{Berghmans:2021wl,Zhukov2021}. As for the heating-temperature distribution and the duty cycle, the present model therefore describes the average properties of an ensemble of unresolved plasma columns rather than the detailed evolution of each individual heating event.

In other words, in the long time-scale regime the relative abundance of suprathermal tails generated by the heating events is determined solely by the filling factor. In this regime, the assumption that individual heating events can be modeled as Maxwellian distributions is fully justified, since the collisional relaxation time is much shorter than the minimum time-scale of the heating events given by Eq.~\eqref{longtimescalesregime}.

\subsection{Coarse-grained density and temperature}
\label{subsec:Coarse-grained_n_T}

Since, under coarse graining, the two time scales are described by an equivalent distribution function (with only the interpretation of $A$ differing), the results below apply to both temporal regimes in the collisionless limit. From the coarse-grained distribution function given by Eq.~\eqref{VDFphasespace}, the standard kinetic definitions yield the number density
\begin{equation}\label{densityfasttime}
\tilde{n}(z) = n_{0} \left[ 
     A \int_{T_0}^{\infty} \gamma(T)  \, 
     e^{-\frac{m \, g \, z}{k_B \, T}} dT 
   + (1 - A)  \, 
     e^{-\frac{m \, g \, z}{k_B \, T_0}} 
   \right]   \,,
\end{equation}
and the kinetic temperature
\begin{equation}\label{temperaturefasttime}
\tilde{T}(z) = \frac{ 
      A \int_{T_0}^{+\infty} T \, \gamma(T)  \, 
      e^{-\frac{m \, g \, z}{k_B \, T}} dT 
    + (1-A) \, T_0 \, 
      e^{-\frac{m \, g \, z}{k_B \, T_0}}
             }{
     A \int_{T_0}^{+\infty}\gamma(T)  \, 
     e^{-\frac{m \, g \, z}{k_B \, T}} dT 
     + (1-A) \, 
     e^{-\frac{m \, g \, z}{k_B \, T_0}}
            }  \,.
\end{equation}

The latter expression is obtained from the standard definition of plasma temperature, based on the variance of the velocity distribution function (VDF). It describes how the different contributions to the VDF combine to define a global temperature. To illustrate its meaning, we consider the special case of a bi-Gaussian distribution by setting
\begin{equation}
    \gamma(T) = \delta(T - T_h)  \,.
\end{equation}
In this case, the total distribution function reduces to a bi-Gaussian with temperatures $T_h$ and $T_0$:
\begin{equation}\label{VDFphasespace2T}
    \tilde{f}_{\alpha}(z,v)=\mathcal{N}_{\alpha}\left[
       \frac{A}{T_h^{3/2}} 
      e^{-\frac{H_{\alpha}}{k_B \, T_h}} 
      + \frac{1 - A}{T_0^{3/2}} 
      e^{-\frac{H_{\alpha}}{k_B \, T_0}} \right]   \,.
\end{equation}
This is classically interpreted as a core distribution at temperature $T_{Co} = T_0$ and a halo distribution at temperature $T_h$. 

The number density is
\begin{equation}\label{densitycorehalo}
\tilde{n}(z) = n_{0} \left[ 
     A \, e^{-\frac{m \, g \, z}{k_B \, T_h}} 
   + (1 - A) \, e^{-\frac{m \, g \, z}{k_B \, T_{Co}}} 
   \right] 
   = n_h + n_{Co}   \,,
\end{equation}
where $n_h$ and $n_{Co}$ are the halo and core densities (both functions of $z$).
The kinetic temperature is then given by
\begin{equation}\label{temperaturecorehalo}
\tilde{T}(z) = \frac{ 
       A \, e^{-\frac{m \, g \, z}{k_B \, T_h}} \, T_h
 + (1-A) \, e^{-\frac{m \, g \, z}{k_B \, T_{Co}}} \, T_{Co}
             }{
       A \, e^{-\frac{m \, g \, z}{k_B \, T_h}} 
 + (1-A) \, e^{-\frac{m \, g \, z}{k_B \, T_{Co}}}
            } 
  = \frac{n_h\, T_h + n_{Co}\, T_{Co}}{n_h + n_{Co}}   \,,          
\end{equation}

Equation~\eqref{temperaturecorehalo} is precisely the same expression used to compute in-situ temperatures from spacecraft observations, such as those from \textit{Helios} or the \textit{Parker Solar Probe} \citep{Stverak2008,Maksimovic_al_2020,lazar2021kappa,Jeong_2022}. 
Therefore, Eq.~\eqref{temperaturefasttime} naturally generalizes the standard method of deriving kinetic temperatures for bi-Gaussian (core--halo) distributions from in-situ measurements to the more general case of multi-temperature (i.e., multi-Gaussian) distributions characterizing suprathermal populations \citep{lazar2021kappa}.

\begin{figure}
    \centering
    \includegraphics[width=0.99\columnwidth]{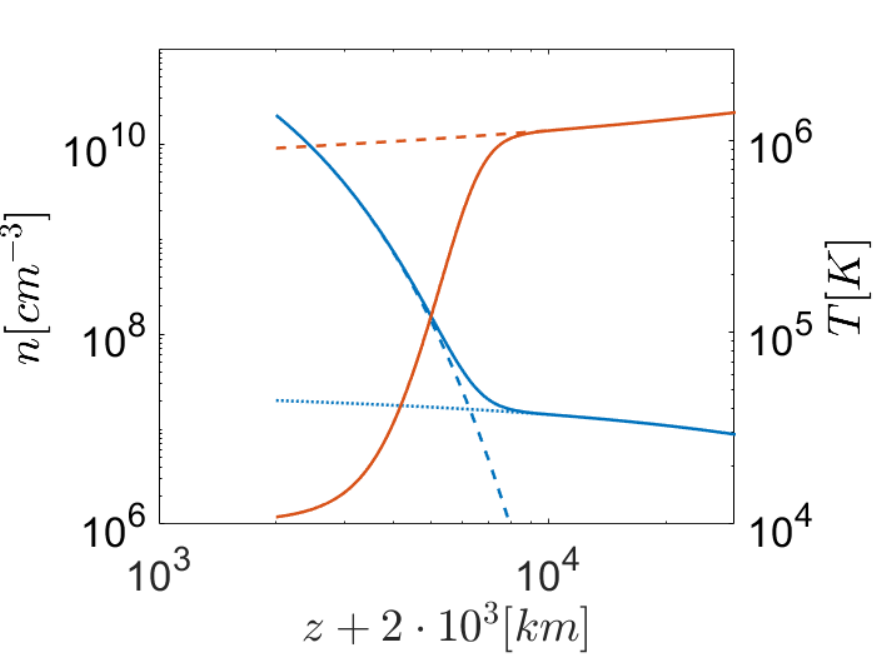}
    \caption{
    Number density (blue solid line, in $\mathrm{cm}^{-3}$) and temperature (red solid line, in $\mathrm{K}$) as functions of height above the chromosphere (in km). The number density is obtained from Eq.~\eqref{densityfasttime} and the temperature from Eq.~\eqref{temperaturefasttime}. The temperature contribution from heated particles alone is shown by the red dashed line (computed from Eq.~\eqref{temperaturefasttime} with $A = 1$). The density of cold particles only is shown by the blue dashed line (computed from Eq.~\eqref{densityfasttime} with $A=0$), while the density of hot particles only is shown by the blue dotted line (corresponding to the first term in Eq.~\eqref{densityfasttime}). The profiles are computed assuming a distribution of heating-event intensities $\gamma(T)$ described by Eq.~\eqref{exponentialincrements}, with parameters $n_0 = 2 \times 10^{10}\,\mathrm{cm}^{-3}$, $T_0 = 10^4\,\mathrm{K}$, $\Delta T = 90\,T_0$, and $A = 0.001$, chosen to satisfy observational constraints.
    }
    \label{fig:fig5}
\end{figure}

Equations~\eqref{densityfasttime} and \eqref{temperaturefasttime}, which describe the density and temperature profiles in the collisionless case, are formally valid both in the fast time-scale regime defined by Eq.~\eqref{shorttimescalesregime} and in the long time-scale regime given by Eq.~\eqref{longtimescalesregime}. The difference lies in their physical interpretation.

In the fast time-scale regime, the temperature inversion develops within individual magnetic loops undergoing stochastic heating and is described by Eqs.~\eqref{densityfasttime} and \eqref{temperaturefasttime} after replacing \(A\rightarrow A_t\), where \(A_t\) denotes the temporal filling factor of the heating events. At the level of surface coarse graining, the same expressions remain valid with the replacement $A_t \rightarrow A$. Therefore, a coronal temperature of the order of \(10^6\,\mathrm{K}\) can be obtained either when heating events are spatially sparse (\(A_S\ll1\)), temporally intermittent (\(A_t\ll1\)), or through any combination of the two leading to an effective filling factor \(A\ll1\), provided that the average temperature increase associated with the heating events,

\begin{equation}
\Delta T
=
\int_{T_0}^{+\infty}
(T-T_0)\,
\gamma(T)\,dT,
\end{equation}
is of the order of \(10^6\,\mathrm{K}\).

In the long time-scale regime, by contrast, the same coarse-grained expressions are recovered after replacing
\(A\rightarrow A_S\) and
\(\gamma(T)\rightarrow\gamma(T)\,n_T/n_0\).
In this limit, each heating event lasts much longer than the particle crossing time, allowing an individual magnetic loop to relax toward an approximately isothermal state. Consequently, the temperature inversion no longer develops within a single loop but emerges only after performing the surface coarse graining over many unresolved magnetic flux tubes.

As shown in \citet{Barbieri2025c}, the resulting temperature and density profiles are remarkably insensitive to the detailed shape of the probability distribution \(\gamma(T)\). Throughout this work, we adopt the illustrative choice

\begin{equation}\label{exponentialincrements}
\gamma(T)
=
\frac{1}{\Delta T}
\exp\!\left(
-\frac{T-T_0}{\Delta T}
\right),
\qquad
T>T_0,
\end{equation}
which corresponds to a situation in which weak heating events are considerably more frequent than those reaching coronal temperatures (\(\sim10^6\,\mathrm{K}\)). This choice is motivated by recent observations indicating that the vast majority of brightenings remain confined to transition-region temperatures and only a small fraction reaches coronal temperatures \citep{Huang2023,Dolliou2023,Dolliou_2024,Dolliou2025}. An explicit example of the resulting temperature inversion is shown in Fig.~\ref{fig:fig5}.

The requirement that \(A_t\ll1\), \(A_S\ll1\), or equivalently \(A=A_SA_t\ll1\), is also supported by the observed thermodynamic conditions at the base of the transition region, which constitutes the lower boundary of the present model \citep{Cauzzi:2009ta}. Indeed, the boundary temperature is given by

\begin{equation}
\tilde{T}(z=0) =
T_0+A\,(\Delta T +T_0).
\end{equation}

Since \(\Delta T\sim10^6\,\mathrm{K}\), while the observed temperature at the base of the transition region remains of the order of \(10^4\,\mathrm{K}\), consistency with observations requires the contribution of the heated plasma to remain a small perturbation of the chromospheric background. This condition implies \(A \ll T_0/\Delta T\) then \(A \ll 1\), which may be realized through a small spatial filling factor, a small temporal filling factor, or a combination of both. Here we emphasize that $\tilde{T}(z)$ denotes the kinetic temperature of the coarse-grained plasma distribution, obtained from the second velocity moment of $\tilde{f}_{\alpha}$. It should not be confused with $T$, which labels the temperature of an individual Maxwellian component associated with a heating event and is distributed according to $\gamma(T)$. Thus, $\tilde{T}(z)$ is a macroscopic quantity resulting from the superposition of populations characterized by different values of $T$.

The rapid initial increase in temperature, accompanied by a corresponding decrease in density, marks the transition region. This behaviour arises from gravitational filtering. Specifically, the thermal component at $T = T_0$ of the distribution function in Eq.~\eqref{VDFphasespace} quickly vanishes with increasing height due to energy conservation, while the suprathermal component at $T > T_0$ becomes dominant above the transition region.

This behaviour is evident in Fig.~\ref{fig:fig5}, where the temperature increases from $T_0$, set by the cold population, up to $T \sim T_0 + \Delta T \sim 10^6\,\mathrm{K}$. At coronal heights, the distribution is dominated by particles generated by heating events, as indicated by the superposition of the red dashed and solid curves.

The decreasing importance of the cold population with height is also apparent in the density profiles. Near the base, the density is dominated by cold particles at $T_0 = 10^4\,\mathrm{K}$, while above the transition region their contribution becomes negligible. In this region, the density is entirely determined by the particles produced by heating events, as indicated by the overlap of the blue solid and dotted curves.

\subsection{Mechanical equilibrium}\label{subsec:mechanical equilibrium}

We now examine the mechanical stability of the plasma configuration in both the short and long time-scale regimes defined by Eqs.~\eqref{shorttimescalesregime} and \eqref{longtimescalesregime}. Under the assumptions adopted throughout this work, namely isotropic velocity distribution functions, equal electron and proton densities and temperatures, and a magnetic field with only a vertical component, the momentum balance equation reduces to

\begin{equation}
\nabla
\left(
p+\frac{B^2}{8\pi}
\right)
=
-
m\,n\,\mathbf g,
\label{momentum_balance}
\end{equation}
where $p=nk_B\,T$ denotes the plasma pressure, while the right-hand side represents the force per unit volume resulting from the combined effects of gravity and the Pannekoek--Rosseland electric field. Along the magnetic field, Eq.~(\ref{momentum_balance}) simply expresses hydrostatic equilibrium, with gravity balanced by the plasma pressure gradient together with the Pannekoek--Rosseland electric field.

\begin{figure}
    \centering
    \includegraphics[width=0.99\columnwidth]{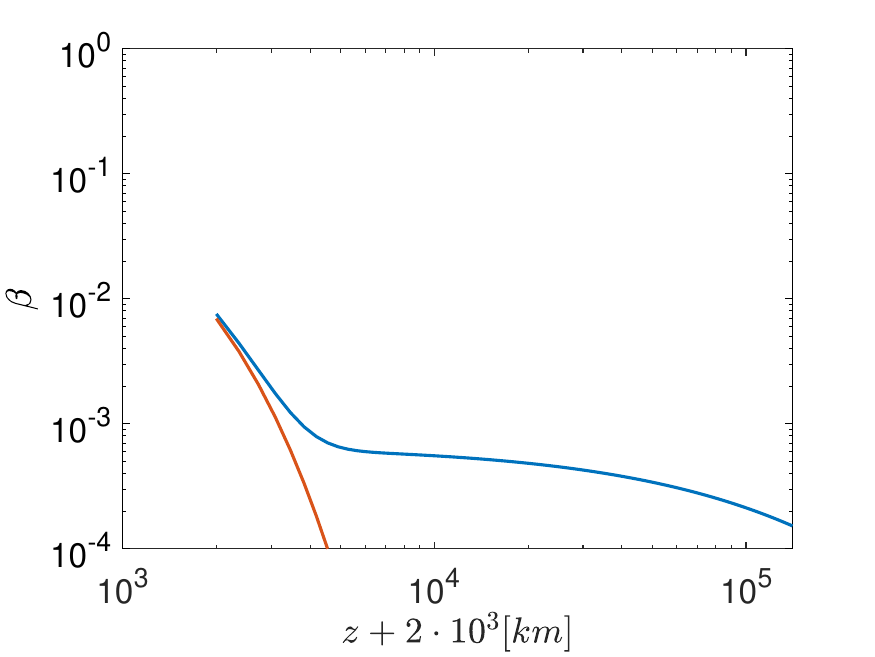}
    \caption{
    Plasma beta as a function of height above the chromosphere. The blue curve corresponds to heated loops, computed from Eq.~(\ref{betaheatedmain}), while the red curve represents non-heated loops, computed from Eq.~(\ref{betacoldmain}). The parameters are the same as those adopted in Fig.~\ref{fig:fig5}, with a magnetic field strength of $B=10\,\mathrm{G}$.
    }
    \label{fig:fig9}
\end{figure}

The more critical issue concerns the mechanical equilibrium between neighbouring magnetic loops that experience different heating histories. Since adjacent loops may possess different temperatures and densities, pressure differences naturally arise. Mechanical stability therefore requires the magnetic pressure to dominate over the plasma pressure, namely
\begin{equation}
\beta
=
\frac{8\pi p}{B^2}
\ll1.
\end{equation}

In the short time-scale regime, the plasma beta of heated loops is
\begin{equation}
\beta_H(z)
=
\frac{8\pi\, k_B\, n(z)\,T(z)}{B^2},
\label{betaheatedmain}
\end{equation}
where $n(z)$ and $T(z)$ are given by
Eqs.~\eqref{densityfasttime} and
\eqref{temperaturefasttime}. For non-heated loops,

\begin{equation}
\beta_0(z)
=
\frac{
8\pi\, k_B\, n_0\,T_0
e^{-\frac{m\,g\,z}{k_B\, T_0}}
}{B^2}.
\label{betacoldmain}
\end{equation}

The corresponding profiles are shown in Fig.~\ref{fig:fig9}. In both cases the plasma beta remains much smaller than unity throughout the transition region and the corona. The rapid decrease of $\beta_H$ across the transition region is a direct consequence of the strong density decrease, which dominates over the simultaneous temperature increase.

In the long time-scale regime, each loop becomes isothermal. The expression for non-heated loops remains Eq.~(\ref{betacoldmain}), while for heated loops one simply replaces $n_0\rightarrow n_T$ and $T_0\rightarrow T$. Since $\beta$ decreases monotonically with height, its maximum value is attained at the loop footpoint. For representative coronal parameters,
$n_T=10^9\,\mathrm{cm^{-3}}$,
$T=10^6\,\mathrm{K}$ and
$B=10\,\mathrm{G}$,
we obtain

\begin{equation}
\beta(z=0)\simeq3\times10^{-2},
\end{equation}
which remains well below unity over the range of temperatures considered in the present work.

These results demonstrate that magnetic pressure overwhelmingly dominates over plasma pressure in both dynamical regimes. Consequently, even if neighbouring loops possess different thermal properties, the magnetic field is sufficiently strong to maintain lateral mechanical equilibrium and prevent significant transverse expansion.

The magnetic field strength adopted here,
$B\simeq10\,\mathrm{G}$,
is representative of typical coronal conditions
\citep[e.g.][]{GolubPasachoff:book}. Moving downward toward the chromosphere, magnetic flux tubes progressively converge and their cross-sectional area decreases
\citep[e.g.][]{Solanki_1991,Mandrini_2000}, leading to stronger magnetic fields. At the same time, however, the plasma density and gas pressure increase much more rapidly, causing the plasma beta to approach and eventually exceed unity
\citep{Gary2001}. Consequently, the lower chromosphere becomes progressively gas-pressure dominated and the assumptions underlying the present magnetically guided kinetic model cease to be valid. This transition occurs below the lower boundary of the present model, which is located at the base of the transition region.

\section{The effect of Coulomb collisions}\label{sec4}

Up to this point, Coulomb collisions have been neglected in our analysis. As will be shown below, their inclusion leads to different behaviour in the short- and long-time-scale regimes. Moreover, we note that, so far, the results are independent of the fact that we are modelling a loop with a curvilinear geometry. In the collisionless limit, the distribution functions depend on velocity and height only through the mean-field Hamiltonian. As a consequence, the specific geometrical shape of the loop does not affect the solution.

In the presence of collisions, however, this is no longer true. When passing from a straight vertical geometry to a curvilinear one, the effective path length travelled by a particle along the loop increases. This, in turn, increases the number of Coulomb collisions experienced by the particle at coronal heights, modifying the coronal density and temperature.
To avoid introducing additional geometrical complications, in the following we suppose an approximate vertical orientation for the loop part analysed, so this minimize the effect of collisions.  

\subsection{Short time-scale regime}

In the short time-scale regime defined by Eq.~\eqref{shorttimescalesregime}, loops undergoing stochastic heating reach a non-equilibrium stationary state. In this regime, Coulomb collisions can modify the velocity distribution functions and, consequently, alter the resulting density and temperature profiles in each elementary loop.

Coulomb collisions act through the cumulative effect of many small-angle deflections, producing dynamical friction and velocity-space diffusion of the suprathermal tails generated by stochastic heating. Rather than solving the full Fokker--Planck collision operator, we adopt a reduced two-component description in which collisions redistribute particles between a suprathermal halo and a Maxwellian core. Under these assumptions, the distribution function in Eq.~\eqref{VDFshortTau}
is written as (we add the subscript c to notify the collisions)
\begin{equation}\label{VDFcollisions}
\begin{aligned}
f_{\alpha,c}(z,v)
= \mathcal{N}_\alpha \Bigg[
& A_t
\int_{T_0}^{\infty}
\gamma(T)\, w(T,z)\,
\frac{1}{T^{3/2}}
e^{-\frac{H_\alpha}{k_B\, T}}\, dT
\\
& + C(z)\,
\frac{1}{T_M(z)^{3/2}}
e^{-\frac{H_\alpha}{k_B\, T_M(z)}}
\Bigg] \, ,
\end{aligned}
\end{equation}
where $w(T,z)$ is the collisional survival function of a Maxwellian population generated by a heating event at temperature $T$, $C(z)$ is the amplitude of the Maxwellian core, and $T_M(z)$ is the core temperature.

The corresponding density and pressure moments are 
\begin{equation}\label{densitycollisions}
\begin{aligned}
n_c(z) = n_0 \Bigg[
& A_t \int_{T_0}^{\infty}
\gamma(T)\, w(T,z)\,
e^{-\frac{m\, g\, z}{k_B\, T}}\, dT+ C(z)
e^{-\frac{m\, g\, z}{k_B\, T_M(z)}}
\Bigg] \, ,
\end{aligned}
\end{equation}
and
\begin{equation}\label{pressurecollisions}
\begin{aligned}
p_c(z)
= n_0 k_B \Bigg[
& A_t \int_{T_0}^{\infty}
T\,\gamma(T)\, w(T,z)\,
e^{-\frac{m\, g\, z}{k_B\, T}}\, dT \\ & +C(z)\,T_M(z)
e^{-\frac{m\, g\, z}{k_B\, T_M(z)}}
\Bigg] \, .
\end{aligned}
\end{equation}
The local temperature then follows from
\begin{equation}\label{temperaturecollisions}
T_c(z)=\frac{p_c(z)}{k_B\,n_c(z)} \, .
\end{equation}

Gravitational stratification enters the density and pressure moments through the Boltzmann factors in Eqs.~\eqref{densitycollisions} and \eqref{pressurecollisions}. By contrast, collisional redistribution between halo and core is entirely described by the coupled evolution of \(C(z)\), \(w(T,z)\), and \(T_M(z)\).

\subsubsection{Determination of \texorpdfstring{$C(z)$}{C(z)}}

We first determine the evolution of the Maxwellian core amplitude \(C(z)\). Coulomb collisions redistribute particles locally between the suprathermal halo and the Maxwellian core without creating or destroying particles. This condition is imposed over an infinitesimal displacement \(\delta z\), assumed to be much smaller than the characteristic gravitational scale height, so that gravitational stratification can be neglected at that local scale. Under this assumption, the collisional redistribution must conserve the local particle number,
\begin{equation}
\delta n_c\big|_{\delta z}=0 \, .
\end{equation}

Using Eq.~\eqref{densitycollisions} and retaining only the collisional variations of \(w(T,z)\) and \(C(z)\), while treating the gravitational factors as fixed in the local collisional balance, we obtain
\begin{equation}\label{C_equation}
A_t
\int_{T_0}^{\infty}
\gamma(T)\,
\frac{\partial w(T,z)}{\partial z}
e^{-\frac{m\,g\,z}{k_B\, T}}\, dT
+
\frac{dC}{dz}
e^{-\frac{m\,g\,z}{k_B\, T_M(z)}}
=
0 \, .
\end{equation}

The determination of \(C(z)\) follows a logic conceptually analogous to multiscale perturbative methods commonly used in celestial mechanics, where slow secular variations are extracted by separating the fast orbital motion from the long-time evolution associated with precession \citep{pollard1966celestial}. In the present case, the same idea is applied by separating the local collisional redistribution, encoded in \(w(T,z)\) and \(C(z)\), from the slower large-scale gravitational stratification. Within this analogy, the gravitational terms play the role of the slow contribution, whereas the collisional redistribution corresponds to the fast dynamics.

\subsubsection{Determination of $w(T,z)$}
We model \(w(T,z)\) as an energy-dependent survival probability in space. For particles generated by a heating event at temperature \(T\), the quantity \(w(T,z)\) is defined as the fraction that remains in the suprathermal halo up to the height \(z\) without being thermalized. Following standard collisional transport theory \citep{LandiDeglInnocenti2019}, we assume that the probability of thermalization over an infinitesimal displacement \(dz\) is proportional to \(dz\) and depends only on the particle state at the local position. This corresponds to a Markovian survival process in space.

To complete the model, it is therefore necessary to determine the local probability
\(P_{\rm loss}(T,z,dz)\) that a particle generated by a heating event of temperature \(T\) undergoes Coulomb thermalization while propagating over the infinitesimal distance \(dz\). This probability can be derived from a standard kinetic-theory argument (see \ref{appendix_lossprobability} for the detailed derivation) and we get
\begin{equation}\label{lossprobability}
P_{\mathrm{loss}}(T,z,dz)=\frac{dz}{\lambda(T,z)} \, ,
\end{equation}
where $\lambda(T,z)$ is the effective Coulomb mean free path of a Maxwellian population generated by a heating event at temperature $T$, propagating in a background plasma of density $n_c(z)$, and is given by
\begin{equation}\label{lambdaH}
\lambda(T,z)=\frac{(3\,k_B\, T)^2}{4\pi\, e^4\, n_c(z)\,\ln\Lambda} \, .
\end{equation}

To derive the evolution equation for the survival probability \(w(T,z)\), we consider a particle generated by a heating event of temperature \(T\) that has already propagated up to the height \(z\). We now follow its evolution over an additional infinitesimal displacement \(dz\). Since the thermalization process is assumed to be Markovian, the evolution of the survival probability depends only on the particle state at the position \(z\), independently of its previous history. During this infinitesimal displacement \(dz\), two mutually exclusive events may occur:

\begin{itemize}
\item the particle survives without being thermalized, with probability

\begin{equation}\label{survprob}
    P_{\rm surv}(T,z,dz)=1-P_{\rm loss}(T,z,dz) = 1-dz/\lambda(T,z);
\end{equation}

\item the particle undergoes a Coulomb collision leading to thermalization, with probability given by Eq. \eqref{lossprobability}

\end{itemize}

The probability of surviving up to the height \(z+dz\) is therefore given by the product of the survival probability up to the height \(z\), Eq.~\eqref{survprob}, and the conditional probability of not being thermalized over the additional displacement \(dz\), Eq.~\eqref{lossprobability}. Hence,

\begin{equation}
\begin{aligned}
w(T,z+dz)
&=
\underbrace{w(T,z)}_{\substack{\text{survival up}\\ \text{to }z}}
\,
\underbrace{\left(
1-\frac{dz}{\lambda(T,z)}
\right)}_{\substack{\text{survival over}\\ \text{the interval }dz}} .
\end{aligned}
\end{equation}

Finally, subtracting \(w(T,z)\) from both sides, dividing by \(dz\)  and taking the limit \(dz\rightarrow0\), one obtains
\begin{equation}\label{A_diffusive}
\frac{\partial w(T,z)}{\partial z}
=
-\frac{w(T,z)}{\lambda(T,z)}.
\end{equation}

Equation~\eqref{A_diffusive} does not admit a closed analytical solution because the mean free path \(\lambda\) depends on \(z\) through \(n_c(z)\), which itself depends on both \(w\) and \(C\). The problem must therefore be solved numerically. Nevertheless, a formal solution for \(w(T,z)\) can be written as
\begin{equation}\label{A_diffusiveexplicit}
w(T,z)=w(T,0)\,
\exp\!\left(
-\int_0^z \frac{d\zeta}{\lambda(T,\zeta)}
\right) \, .
\end{equation}

The exponential damping factor \(\exp(-\tau)\) is controlled by
\begin{equation}
\tau(T,z)=\int_0^z \frac{d\zeta}{\lambda(T,\zeta)} \, .
\end{equation}

Let us introduce the local Knudsen number
\begin{equation}\label{KnusdenLn}
    K_n = \frac{\lambda}{L_n} , \qquad L_n =\left|
\frac{d\ln n_c}{dz}
\right|^{-1}.
\end{equation}

Substituting into the definition of $\tau$ yields the continuous representation
\begin{equation}
\tau(T,z)
=
\int_{0}^{z}
\frac{d\zeta}{L_n(T,\zeta)}\,
\frac{1}{K_n(T,\zeta)} .
\end{equation}

It is convenient to introduce the dimensionless coordinate
\begin{equation}
s(z)=\int_0^{z}\frac{d\zeta}{L_n(T,\zeta)},
\end{equation}
which measures the distance along the column in units of the local density-gradient scale length.  
In terms of this variable the cumulative damping factor becomes
\begin{equation}\label{tauknusden}
\tau(T,z)=
\int_{0}^{s(z)}
\frac{ds'}{K_n(T,s')}.
\end{equation}
This expression shows that the exponential attenuation factor $\exp[-\tau(T,z)]$
represents the cumulative effect of the inverse local Knudsen number integrated along the column when distances are measured in units of the macroscopic gradient scale.

\subsubsection{Determination of \texorpdfstring{$T_M(z)$}{TM(z)}}

The core temperature \(T_M(z)\) is determined by imposing hydrostatic equilibrium,
\begin{equation}\label{pressurebalance}
\frac{dp_c}{dz}=-m\, g\, n_c \, .
\end{equation}
Using Eqs.~\eqref{densitycollisions}, \eqref{pressurecollisions}, \eqref{A_diffusive}, and \eqref{C_equation}, we obtain after some algebra
\begin{equation}\label{TM_temperaturecore}
\frac{dT_M}{dz}
=
\frac{A\left[I_1(z)-T_M(z)\,I_0(z)\right]}
{C(z)\,e^{-\frac{m\,g\,z}{k_B\, T_M(z)}}
\left(1+\frac{m\,g\,z}{k_B\, T_M(z)}\right)} \, ,
\end{equation}
where
\begin{equation}
I_1(z)=
\int_{T_0}^{\infty}
T\,\gamma(T)\,
\frac{w(T,z)}{\lambda(T,z)}
e^{-\frac{m\,g\,z}{k_B\, T}}\, dT
\end{equation}
and
\begin{equation}
I_0(z)=
\int_{T_0}^{\infty}
\gamma(T)\,
\frac{w(T,z)}{\lambda(T,z)}
e^{-\frac{m\,g\,z}{k_B\, T}}\, dT \, .
\end{equation}
Then, Eq. \eqref{TM_temperaturecore} provides $T_M(z)$ when integrated in $z$. 

It is instructive to rewrite Eq.~\eqref{TM_temperaturecore} in a more transparent form. To this end, we first discuss the physical meaning of the quantities \(I_0(z)\) and \(I_1(z)\). The integral \(I_0(z)\) represents the local rate at which suprathermal particles are transferred from the halo to the Maxwellian core through Coulomb collisions. Indeed, the factor \(\gamma(T)\) gives the probability of generating a suprathermal population at temperature \(T\), \(w(T,z)\) is the fraction of these particles that survives up to the height \(z\), \(1/\lambda(T,z)\) is the probability per unit length that a surviving particle thermalizes through Coulomb collisions, and the Boltzmann factor accounts for the gravitational filtering of particles reaching the altitude \(z\). Therefore, \(I_0(z)\) measures the total number of particles that leave the suprathermal halo and are incorporated into the Maxwellian core per unit height.

The quantity \(I_1(z)\) is the first temperature moment of the same transfer process. It therefore represents the total thermal energy carried by the particles injected into the Maxwellian core per unit height. Consequently, the ratio

\begin{equation}
T_{\rm inj}(z)
=
\frac{I_1(z)}{I_0(z)},
\label{Tinj}
\end{equation}
defines the mean temperature of the particles that are thermalized at the altitude \(z\). In other words, \(T_{\rm inj}(z)\) is the effective temperature of the suprathermal particles continuously injected into the Maxwellian component.

Using this definition, Eq.~\eqref{TM_temperaturecore} can be rewritten as

\begin{equation}
\frac{dT_M}{dz}
=
\frac{
A\,I_0(z)
\left[
T_{\rm inj}(z)-T_M(z)
\right]
}{
C(z)
e^{-\frac{m\,g\,z}{k_B\,T_M(z)}}
\left(
1+\dfrac{m\,g\,z}{k_B\,T_M(z)}
\right)
}.
\label{TM_equation_rewritten}
\end{equation}

Equation~\eqref{TM_equation_rewritten} admits a straightforward physical interpretation. The factor \(A\,I_0(z)\) measures the local rate at which suprathermal particles are incorporated into the Maxwellian core, while the difference
\(
T_{\rm inj}(z)-T_M(z)
\)
determines whether these newly thermalized particles heat or cool the core. Whenever \(T_{\rm inj}>T_M\), the injected particles are, on average, hotter than the Maxwellian population and therefore increase its temperature. Conversely, when \(T_{\rm inj}<T_M\), they cool the Maxwellian component.

The denominator is proportional to the local density of the Maxwellian core,

\begin{equation}
n_M(z)
=
n_0\,C(z)
e^{-\frac{m\,g\,z}{k_B\, T_M(z)}},
\end{equation}
so that, for a fixed energy input, a larger Maxwellian population undergoes a smaller temperature variation because of its larger thermal inertia. The additional factor
\begin{equation}
1+\frac{m\,g\,z}{k_B\,T_M(z)}
\end{equation}
originates from differentiating the Boltzmann stratification factor with respect to the height while allowing the Maxwellian temperature to vary with altitude. Consequently, Eq.~\eqref{TM_equation_rewritten} has the form of a relaxation equation: the Maxwellian core continuously adjusts its temperature toward the effective temperature of the suprathermal particles transferred into it through Coulomb collisions, while the increasing size of the Maxwellian population progressively reduces the temperature response to the incoming particles.

\begin{figure}
    \centering
    \includegraphics[width=0.99\columnwidth]{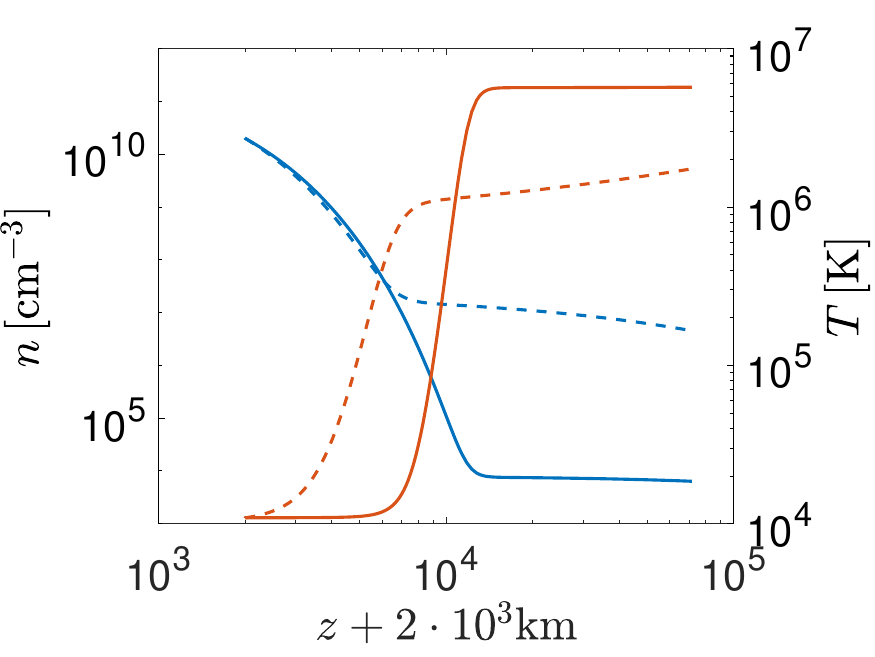}
    \caption{
    Number density (blue, in $\mathrm{cm}^{-3}$) and temperature (red, in $\mathrm{K}$) as functions of height above the chromosphere (in km).
    The profiles including Coulomb collisions (solid lines) are obtained by computing the number density from Eq.~\eqref{densitycollisions} and the temperature from Eq.~\eqref{temperaturecollisions}. 
    The corresponding collisionless profiles (dashed lines) are the same as those shown in Fig.~\ref{fig:fig5}.
    All profiles are computed using the same parameters and function $\gamma(T)$ listed in Fig. \ref{fig:fig5}.
    }
    \label{fig:Collisionsprofiles}
\end{figure}

\begin{figure}
    \centering
    \includegraphics[width=0.99\columnwidth]{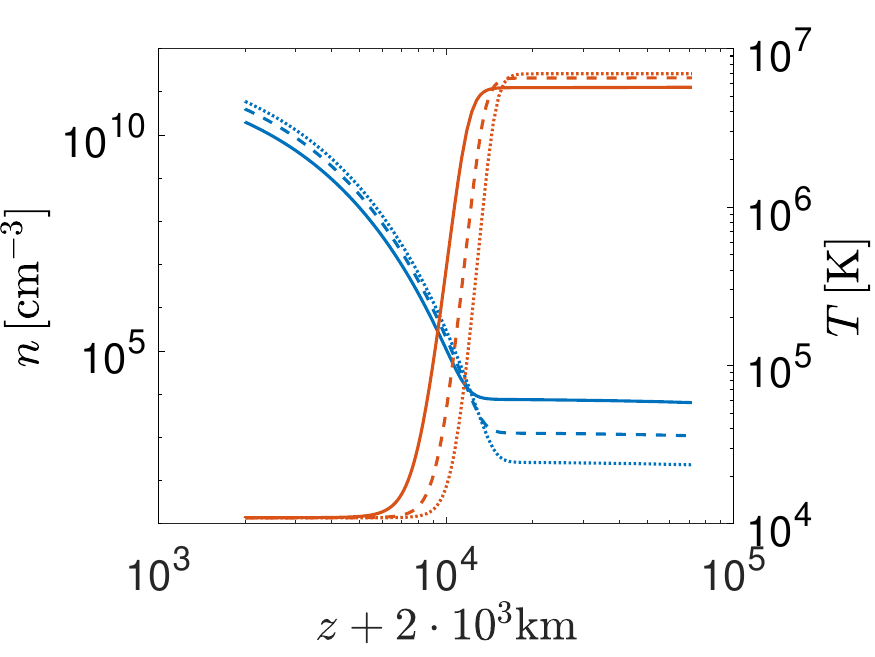}
    \caption{
    Number density (blue, in $\mathrm{cm}^{-3}$) and temperature (red, in $\mathrm{K}$) as functions of height above the chromosphere (in km).
    The profiles are obtained by computing the number density from Eq.~\eqref{densitycollisions} and the temperature from Eq.~\eqref{temperaturecollisions}. 
    Solid lines correspond to a base density $n_0 = 2 \times 10^{10}  \,\mathrm{cm}^{-3}$, dashed lines to $n_0 = 4 \times 10^{10}  \,\mathrm{cm}^{-3}$, and dotted lines to $n_0 = 6 \times 10^{10}  \,\mathrm{cm}^{-3}$. All the remaining parameters are the same as in Fig. \ref{fig:fig5}.
    }
    \label{fig:Collisionsprofilesvaryingn0}
\end{figure}

\subsection{Modification of temperature and density profiles and the intensity of the filtering effect}

Equations~\eqref{C_equation}, \eqref{A_diffusive}, and \eqref{TM_temperaturecore} form a closed system of coupled integro-differential equations for \(C(z)\), \(w(T,z)\), and \(T_M(z)\). The system is solved numerically \footnote{For details on the numerical scheme see ~\ref{AppendixNumerics}.} with the boundary conditions
\begin{equation}\label{Initialconditions}
w(T,0)=1 \quad \forall T \, ,
\qquad
C(0)=1-A_t \, ,
\qquad
T_M(0)=T_0 \, .
\end{equation}
Once \(w(T,z)\), \(C(z)\), and \(T_M(z)\) are obtained, the density and temperature profiles follow from Eqs.~\eqref{densitycollisions} and \eqref{temperaturecollisions}.

Figure~\ref{fig:Collisionsprofiles} shows the resulting density and temperature profiles in the presence of Coulomb collisions (solid lines), compared with the collisionless case (dashed lines). Near the base of the atmosphere, where the density is largest, collisional effects are stronger and efficiently reduce the suprathermal population at low heights. As a consequence, the transition from chromospheric to coronal temperatures is shifted toward higher altitudes.  Only the hottest particles, with the largest mean free path $\lambda$, remain mostly collisionless and they can reach the corona. As a result, the coronal density is much lower than in the collisionless case, while the coronal temperature is higher. 
This behaviour is consistent with the action of dynamical friction in more general Fokker--Planck treatments, where only particles with sufficiently high energies are able to reach large heights. In the present model, this selective filtering is encoded in the temperature-dependent attenuation factor $w(T,z)$.

Figure~\ref{fig:Collisionsprofilesvaryingn0} illustrates the effect of varying the base density on the resulting density and temperature profiles. As the base density increases, Coulomb collisions become stronger in the lower atmosphere. Consequently, the transition from chromospheric to coronal temperatures is shifted to progressively higher altitudes. At the same time, the enhanced collisionality reduces the number of particles that can reach coronal heights and further selecting the hottest components of the distribution. This leads to a decrease in the coronal density and an increase in the coronal temperature.

\begin{figure}
    \centering
    \includegraphics[width=0.99\columnwidth]{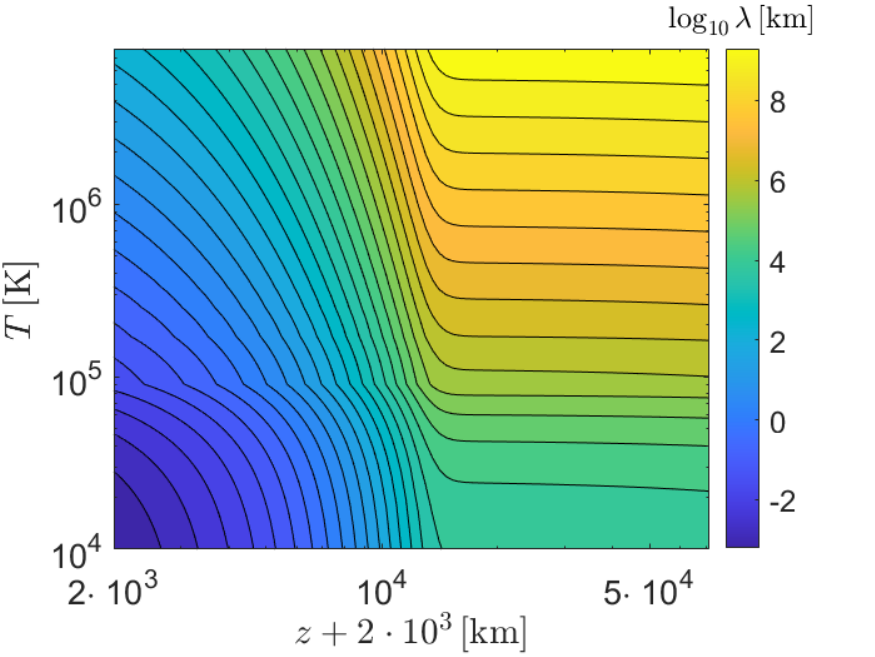}
    \caption{Contour plot of the mean free path $\lambda$ (computed from Eq.~\eqref{lambdaH}) as a function of the temperature $T$ (in $\mathrm{K}$) and the height $z$ (in $\mathrm{km}$). Unlike in the previous figures, here $T$ denotes the temperature appearing in the numerator of Eq.~\eqref{lambdaH}, i.e. the temperature associated with heating events drawn from the distribution $\gamma(T)$. The same parameters as in Fig.~\ref{fig:fig5} are used.}
    \label{fig:meanfreepath}
\end{figure}

Within the present model, this behaviour can be traced back to the behavior of the Knusden number as shown in Eq. \eqref{tauknusden}. Since the Knusden number is proportional to the mean free path, Eq.~\eqref{lambdaH}, we first analyze it with Fig.~\ref{fig:meanfreepath}.  
For a fixed value of $T$, the mean free path increases with height while passing through the transition region and saturates once coronal heights are reached. 
Indeed, from Eq.~\eqref{lambdaH}, the mean free path scales approximately as $n_c(z)^{-1}$, since the Coulomb logarithm, $\ln \Lambda$, provides only a subdominant contribution. Therefore, as the density decreases through the transition region, Fig.~\ref{fig:Collisionsprofiles}, the mean free path increases. In the corona the density approaches a plateau, and consequently the mean free path also tends to a constant value. Next, at fixed height,
increasing $T$ (i.e. considering progressively hotter Maxwellians) also increases the mean free path as it scales approximately as $T^2$.

\begin{figure}
    \centering
    \includegraphics[width=0.99\columnwidth]{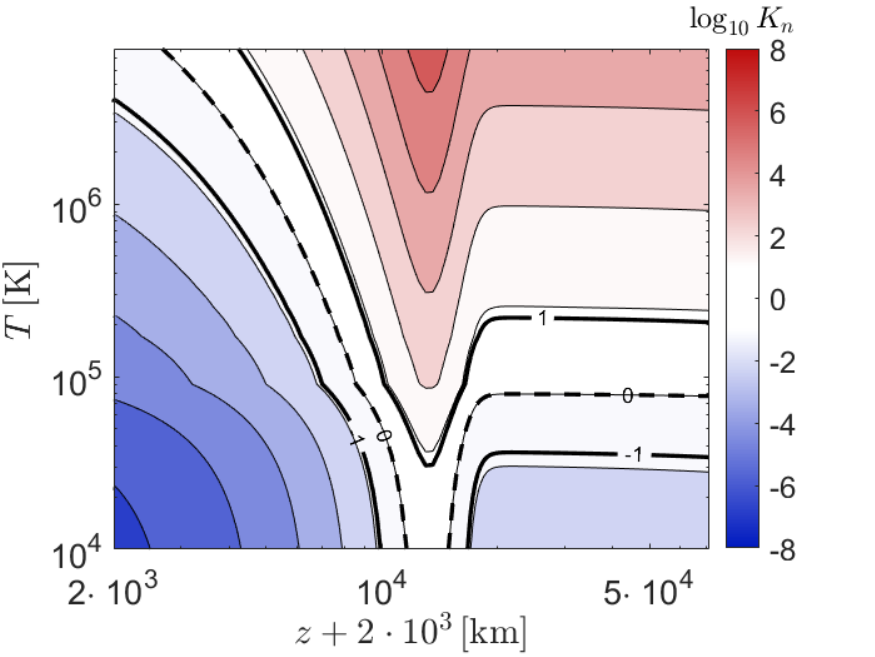}
    \caption{Contour plot of the Knudsen number $K_n$ (computed from Eq.~\eqref{KnusdenLn}) as a function of the temperature $T$ (in $\mathrm{K}$) and the height $z$ (in $\mathrm{km}$). As in Fig.~\ref{fig:meanfreepath}, $T$ denotes the temperature associated with heating events drawn from the distribution $\gamma(T)$. The colour scale represents $\log_{10} K_n$. The white region marks the transition regime $0.1 < K_n < 10$, separating the collisional regime ($K_n < 0.1$, blue shades) from the collisionless regime ($K_n > 10$, red shades). The black contours indicate the boundaries $K_n = 0.1$ and $K_n = 10$, while the dashed contour shows the surface $K_n = 1$, corresponding to the classical transition between fluid and kinetic behaviour. The same parameters as in Fig.~\ref{fig:fig5} are used.}
    \label{fig:Knusden}
\end{figure}

Having clarified the behaviour of the mean free path, we can now understand the behaviour of the Knudsen number, Eq.~\eqref{KnusdenLn}, illustrated in Fig.~\ref{fig:Knusden}. 
Below $z \approx 10^4$ km and above $z \approx 2 \times 10^4$ km
the Knudsen number behaves as the mean free path since the density scale height is weakly changing in both domains.
In contrast with the mean free path, the Knudsen number exhibits a non-monotonic behaviour in a narrow range below $z \approx 2 \times 10^4$ km since it decreases rapidly there. This behaviour is determined by the variation of the density scale height $L_n$ which increase strongly when entering in the corona.

The exponential damping is controlled by the dimensionless exponent $\tau(T,z)$ given by Eq.~\eqref{tauknusden}. Since it involves the integration of the inverse Knudsen number, which is larger at low heights, the strongest collisional effects occur in this region. Particles at temperature $T$, injected at $z=0$, are therefore strongly affected by collisions already within the first gravitational scale height. Only particles with the largest values of $T$, for which $\mathrm{Kn}(T,z=0)$ is of order unity or larger, can cross the dense region. At greater heights, particles become progressively less collisional and can therefore reach the corona with only limited density attenuation. 

In summary, the collisional effects strongly alter the coronal properties in this regime. In particular, the resulting coronal densities become substantially smaller than the lowest observed values in coronal holes, which are of the order of $10^{7}  \,\mathrm{cm}^{-3}$. We therefore conclude that very fast heating events can produce large coronal temperatures, but the associated density remains much too low to account for the observed coronal structures.

The mechanical stability discussed in Sect.~\ref{subsec:mechanical equilibrium} for the collisionless limit remains valid when Coulomb collisions are included. Indeed, the plasma beta of the columns undergoing stochastic heating remains much smaller than unity. Close to the base of the transition region, collisions have only a minor effect on the plasma pressure, so that the plasma beta is essentially unchanged with respect to the collisionless case. At larger heights, however, Coulomb collisions progressively deplete the suprathermal particle population, leading to a significant reduction of the coronal density. Although the collisional filtering of suprathermal particles produces an increase of the coronal temperature, this effect is not sufficient to compensate for the decrease in density. Consequently, the gas pressure decreases with height, while the magnetic field strength is assumed to remain unchanged, causing the plasma beta to become even smaller than in the collisionless case. Columns that do not undergo stochastic heating are unaffected by collisions, and therefore their plasma beta remains identical to that obtained in the collisionless limit. Consequently, magnetic pressure continues to dominate over gas pressure in all plasma columns, ensuring the mechanical stability of the configuration.

\subsection{Long time-scale regime}

In the long-timescale regime defined by Eq.~\eqref{longtimescalesregime}, the system can be described as a superposition of many parallel loops, 
each in thermal equilibrium at a different temperature $T$. In this regime, the results are not significantly affected by Coulomb collisions, and the collisionless solution remains essentially unchanged. This implicitly supposes that enough heating is available to maintain such temperature profiles.

In the collisionless case, the fact that individual loops are isothermal depends on the assumption that heating events at the base are modeled by thermal (Maxwellian) distributions. One may therefore ask whether this property persists 
if a single heating event is instead described by a suprathermal distribution.

In the short-timescale regime, the representation of suprathermal tails as a superposition of Maxwellians arises naturally from the mixing of particles heated at different temperatures at the base of the system. 
More generally, any suprathermal distribution (e.g., a $\kappa$ distribution) can be decomposed into a superposition of Maxwellians. This procedure is known as superstatistics \citep{BECKsuper,Chavanis2006,Davis2019,Davis2020,lazar2021kappa,Tamburrini2025}. We therefore impose at the base a suprathermal distribution decomposed into a sum of Maxwellians and apply the collisional formalism of Eq.~\eqref{VDFcollisions}. We set $A=1$ in order to study only an heated loop (which has no cold component in the short-timescale regime).
Moreover, as already stated in Sect.~\ref{subsec:longtimescales}, we assume a low density at the base of $n_T \sim 10^{9}\,\mathrm{cm}^{-3}$ due to the dilution of the surrounding plasma produced by brightening events \citep{Berghmans:2021wl,Zhukov2021}. 

In this configuration, even though a suprathermal distribution with a kinetic temperature of $10^6\,\mathrm{K}$ is imposed at the base, the characteristic density scale height at the base of the model is much larger than in the previous case. As a consequence, the Knudsen number remains extremely small, with 
$K_n \sim 10^{-3}$ for particles with temperature $T \sim 10^6\,\mathrm{K}$ ($K_n \propto T^2$ for other $T$). The system is therefore extremely collisional, so that the suprathermal population rapidly thermalizes and relaxes to a nearly isothermal atmosphere at $T \sim 10^6\,\mathrm{K}$. As a result, even in the case of events described by a suprathermal distribution, the global configuration of the system is still a superposition of nearly isothermal loops.

We therefore conclude that Coulomb collisions do not affect the temperature inversion produced by surface coarse-graining in the long-time-scale regime.

\section{Estimate of radiative losses in the long-time-scale regime}\label{sec5}

Since the long-time-scale regime is the most robust in the presence of Coulomb collisions, we estimate here the possible impact of radiative losses. We consider a heated loop whose boundary distribution function is the Maxwellian reservoir introduced in Eq.~\eqref{maxwellianboundary}. In this regime, the hot particles injected at the base are responsible for producing the coronal temperatures that appear in the increasing temperature profile at the surface coarse-graining level.

We first estimate the upward kinetic-energy flux carried by particles with positive vertical velocity ($v_z>0$), associated with the hot population injected at the base. For a Maxwellian distribution, the total upward energy flux carried by electrons and protons is
\begin{equation}\label{upwardflux}
\begin{aligned}
F_{\rm hot}^+
&=
\frac{1}{2}\sum_{\alpha=\{e,p\}}m_\alpha
\int_{v_z>0} v^2\, v_z\, f_{T,\alpha}(\mathbf{v})\, d^3\textbf{v} \\
&=
2\,n_T\, k_B\, T \sqrt{\frac{k_B\, T}{2\pi}}
\left(
\frac{1}{\sqrt{m_e}}+\frac{1}{\sqrt{m_p}}
\right).
\end{aligned}
\end{equation}

For $T=10^6\,{\rm K}$ and $n_T=10^9\,{\rm cm^{-3}}$, consistent with the dilution of the surrounding plasma produced by brightening events \citep{Berghmans:2021wl,Zhukov2021}, this yields
\begin{equation}
F_{\rm hot}^+ \simeq
4.4\times10^7\,{\rm erg\,cm^{-2}\,s^{-1}},
\end{equation}
with the dominant contribution coming from the electrons.

The volumetric radiative loss rate can then be written as
\begin{equation}
Q_{\rm rad}=n_T^2\phi(T).
\end{equation}

The radiative loss function $\phi(T)$ is evaluated using the CHIANTI atomic database \citep{DelZanna2021}. For $T=10^6\,{\rm K}$ we adopt
\begin{equation}
\phi(10^6\,{\rm K})\simeq10^{-22}\,{\rm erg\,cm^{3}\,s^{-1}} .
\end{equation}
With $n_T=10^9\,{\rm cm^{-3}}$, this gives
\begin{equation}
Q_{\rm rad}\simeq10^{-4}\,{\rm erg\,cm^{-3}\,s^{-1}}.
\end{equation}

Since $F_{\rm hot}^+$ is a flux while $Q_{\rm rad}$ is a volumetric loss rate, their ratio defines a characteristic radiative length scale,
\begin{equation}
\ell_{\rm rad}=\frac{F_{\rm hot}^+}{Q_{\rm rad}} .
\end{equation}

Using the values above we obtain
\begin{equation}
\ell_{\rm rad}\simeq4.4\times10^{11}\,{\rm cm}
\simeq6.3\,R_\odot .
\end{equation}

This scale is much larger than the typical loop length considered in this work ($L\sim0.1R_\odot\simeq7\times10^9\,{\rm cm}$). In other words, the characteristic radiative length satisfies $\ell_{\rm rad}\gg L$, indicating that the energy carried upward by the hot particle population largely exceeds the energy lost through radiation over the loop length. 

The above estimate is evaluated at the base of the columns, where the density is maximal and radiative losses are therefore strongest. At larger heights, both the upward kinetic-energy flux and the density decrease approximately exponentially because of gravitational stratification. In the long-time-scale regime, the density in the hot columns behaves as
\begin{equation}
n(z)\propto e^{-\frac{m\,g\,z}{k_B\, T}},
\end{equation}
so that the local radiative loss rate scales as
\begin{equation}
Q_{\rm rad}(z)\propto n^2(z)\propto e^{-\frac{2\,m\,g\,z}{k_B\, T}},
\end{equation}
while the upward kinetic-energy flux scales as
\begin{equation}
F_{\rm hot}^+(z)\propto e^{-\frac{m\,g\,z}{k_B\, T}}.
\end{equation}
The corresponding radiative length therefore becomes
\begin{equation}
\ell_{\rm rad}(z)=\frac{F_{\rm hot}^+(z)}{Q_{\rm rad}(z)}\propto e^{\frac{m\,g\,z}{k_B\, T}}.
\end{equation}

An analogous scaling holds for the cold columns upon replacing $T \rightarrow T_0$. Hence, the characteristic radiative length increases with altitude for the hot populations, implying that radiative cooling becomes progressively less efficient above the base. Hence, within this order-of-magnitude estimate, radiative losses do not significantly affect the heated columns.

Since the hot columns are responsible for the formation of the corona at a coarse-grained level, this estimate suggests, within the framework of our model, that radiative effects do not significantly affect the coronal density and temperature.

We have not explicitly addressed the effect of radiation on the cold component, which dominates at lower densities and is therefore more sensitive to radiative losses. However, due to the strong variability of the loss function $\phi(T)$ at temperatures close to $T_0$, an estimate based on a fixed value of $T_0$, such as the one presented above, may be misleading.

Radiative losses may reduce the scale height of the cold population. Since this scale height controls the spatial extent of the transition from chromospheric to coronal densities, such an effect could lead to a reduction in the thickness of the transition region at a coarse-grained level.

\section{Limitations and future perspectives} \label{secLimitations}

The model presented in this work should be regarded as a first step toward a kinetic study of stochastic heating in plasma atmospheres evolving in a regime that is neither fully collisional nor fully collisionless. Several simplifying assumptions have been introduced to isolate the combined effects of key physical processes on the thermal structure of coronal loops. The included processes are the spatially sparse and temporally intermittent stochastic heating localized at the base of the transition region, gravitational filtering, and Coulomb collisions. Although these ingredients already reproduce properties of the solar atmosphere, 
a number of limitations remain. The following developments outline the most important physical processes that should be incorporated in future works to extend the present framework toward a self-consistent kinetic model of the solar atmosphere.

\subsection{Magnetic expansion and kinetic instabilities} \label{secBexpansion}

An important extension concerns the inclusion of magnetic flux-tube expansion  with height \citep[e.g.][]{Gabriel1976,Dowdy1986,Solanki_1991,Mandrini_2000,Peter2001,Wiegelmann2014,Judge_2021}. 
In this case, particle dynamics is governed not only by energy conservation but also by the conservation of the first adiabatic invariant (magnetic moment) given by Eq. \eqref{eq:magneticmoment}.

As particles propagate toward regions of weaker magnetic field, conservation of the magnetic moment implies a progressive reduction of the perpendicular velocity. Combined with total energy conservation, this reduction is accompanied by a corresponding increase of the parallel velocity. Consequently, the particle distribution function naturally develops a pressure anisotropy with
  \begin{equation}
  T_\parallel>T_\perp,
  \end{equation}
an effect that is absent from the present isotropic treatment.

The inclusion of magnetic-field expansion also introduces a non-trivial topology in phase space. In collisionless plasmas, the combined effects of energy conservation and conservation of the magnetic moment naturally give rise to a region of velocity space that cannot be populated by particles. This inaccessible region is commonly referred to as the trapped-particle region in velocity space \citep{Verscharen2026}. As a consequence, the velocity distribution function develops a characteristic anisotropic structure that departs significantly from a Maxwellian distribution. Coulomb collisions partially smooth these phase-space structures by continuously scattering particles in pitch angle. Since the collisional scattering rate decreases rapidly with increasing particle velocity, this isotropization is expected to be efficient mainly for the low-energy core population, whereas the suprathermal particles produced by the strongest stochastic heating events are expected to preserve a significant fraction of their collisionless anisotropy.

Such anisotropic velocity distribution functions are susceptible to a variety of kinetic instabilities \citep{Hellinger2006,Verscharen2019,Verscharen2022,Opie_2022}. Because the expansion-induced anisotropy is characterized by an excess of parallel pressure, the firehose instability is expected to be the most likely candidate, although the instability threshold depends on both the local plasma beta and the detailed shape of the velocity distribution function. Extending the present model to include non-trivial magnetic geometries and investigating the interplay between magnetic expansion, Coulomb collisions, and stochastic heating therefore represents an important direction for future work. Such an extension would also make it possible to assess quantitatively whether the predicted anisotropies are sufficient to trigger the firehose instability or other kinetic instabilities in the solar atmosphere.

\subsection{Species-dependent temperature profiles} \label{secSpecies-dependent}

In the present work, the stochastic heating process is assumed to possess the same statistical properties for electrons and protons, leading to identical boundary distributions apart from the particle masses. Under these assumptions, the classical Pannekoek--Rosseland electric potential provides a consistent description of quasi-neutrality.

A more general formulation would allow the stochastic heating statistics to differ between the two species, for example through different probability distributions of heating events, $\gamma(T)$. In this case, the ambipolar electric field could no longer be prescribed \emph{a priori}, but would instead have to be determined self-consistently from the quasi-neutrality condition, following the approach commonly adopted in exospheric models of the solar wind \citep{Jockers1970,Lemaire1971,Maksimovic_al_2020,Zouganelis2004,Pierrard2023}.

The origin of this correction can already be understood within a two-component electron--proton plasma. If electrons and protons possess different temperature profiles, their hydrostatic equilibrium equations become

\begin{equation}
\frac{d}{dz}
\left(
n_e k_B\, T_e
\right)
=
-
m_e g\,n_e
-
eEn_e,
\end{equation}
\begin{equation}
\frac{d}{dz}
\left(
n_p k_B\, T_p
\right)
=
-
m_p g\,n_p
+
eEn_p.
\end{equation}

   Assuming quasi-neutrality,
\begin{equation}
n_e=n_p=n,
\end{equation}
    one obtains
\begin{equation}
k_B\,T_e\frac{dn}{dz}
+
k_B\,n\,\frac{dT_e}{dz}
=
-
m_e\,g\,n
-
e\,E\,n,
\end{equation}
\begin{equation}
k_B\,T_p\frac{dn}{dz}
+
k_Bn\frac{dT_p}{dz}
=
-
m_p\,g\,n
+
e\,E\,n.
\end{equation}
   Subtracting the two equations yields the generalized ambipolar electric field
\begin{equation}
E
=
\frac{1}{2\,e}\left[
(m_p-m_e)g
+
k_B
\left(
\displaystyle
\frac{dT_e}{dz}
-
\frac{dT_p}{dz}
\right)
+
\displaystyle
\frac{k_B}{n}
(T_e-T_p)
\frac{dn}{dz}\right].
\label{Eq:GeneralizedPR}
\end{equation}

Equation~\eqref{Eq:GeneralizedPR} naturally reduces to the classical Pannekoek--Rosseland field when the two species share the same temperature. It also shows explicitly that allowing different electron and proton temperatures introduces additional contributions associated with both the pressure gradients and the temperature gradients of the two species.

From a physical point of view, relaxing the assumption of identical stochastic heating is therefore expected to produce corrections to the classical Pannekoek--Rosseland electric field. The resulting ambipolar potential modifies the effective potential experienced by each particle species while preserving quasi-neutrality, thereby leading to quantitative changes in the corresponding density and temperature profiles.
Furthermore, the resulting temperature separation would emerge self-consistently from the interplay between species-dependent stochastic heating, gravitational velocity filtration, Coulomb collisions, and the ambipolar electric field.

This extension is also motivated by observations. Spectroscopic diagnostics of the solar corona indicate that electrons and ions are generally not heated in the same way, while indirect observational evidence suggests that proton and heavy-ion temperatures often exceed the electron temperature, particularly in low-density coronal regions such as coronal holes \citep{Kohl1997,Cranmer2009,Reale2010,Cranmer_2020,Zhu_2023}. These observations could therefore provide valuable constraints on the species-dependent stochastic heating distributions, $\gamma_{\alpha}$, and offer a means to assess whether the model is capable of reproducing the observed temperature separation between electrons and ions in the solar corona.

\subsection{Asymmetric boundary conditions and mass flows} \label{secAsymmetric}

As discussed in Sect.~\ref{subsec:basic}, the present model assumes symmetric boundary conditions at the two loop footpoints, resulting in velocity distribution functions that are even with respect to the parallel velocity and, consequently, in the absence of net particle and heat fluxes along the loop. Relaxing this assumption would naturally produce asymmetric velocity distribution functions, pressure differences between the two footpoints, and therefore siphon flows along the magnetic field.

The long-time-scale regime is expected to be particularly interesting. In this limit, each heating event lasts much longer than the particle crossing time, allowing the plasma within an individual magnetic loop to approach a quasi-steady state. If one footpoint is heated while the opposite one remains close to the chromospheric temperature, most of the loop is expected to become nearly isothermal at the temperature of the heated footpoint. The colder footpoint, however, continuously injects particles with much lower thermal energies. Because gravitational velocity filtration preferentially removes low-energy particles, the cold population remains concentrated near the corresponding footpoint, where its density decreases rapidly with height. As a consequence, a transition region naturally develops only in the vicinity of the colder footpoint, whereas the remainder of the loop remains approximately isothermal.

At the coarse-grained level, the situation is qualitatively different. The present surface coarse-graining averages over a large ensemble of unresolved magnetic flux tubes, each of which may have either the left or the right footpoint heated. Consequently, neighbouring flux tubes may exhibit opposite thermal configurations, with the hot and cold footpoints interchanged. Averaging over such an ensemble therefore produces two transition regions, one above each chromospheric footpoint, while preserving an approximately uniform coronal temperature over the central part of the loop. The existence of these two transition regions is therefore not a property of an individual flux tube but rather an emergent consequence of the stochastic heating process combined with the adopted surface coarse-graining.

The pressure asymmetry associated with different footpoint temperatures is also expected to drive siphon flows, whose velocity and mass flux depend on the degree of asymmetry between the two boundary conditions. At the kinetic level, these flows are accompanied by asymmetric velocity distribution functions, finite particle fluxes, and field-aligned heat fluxes.

In this situation, the classical Pannekoek--Rosseland electric field is no longer expected to provide a self-consistent description of quasi-neutrality. Instead, the ambipolar electric field must be determined from the momentum balance equations of the two plasma species. Assuming a one-dimensional geometry along the magnetic field, the proton and electron momentum equations can be written as
\begin{equation}
m_p\, n\, u_p\frac{du_p}{dz}
=
-\frac{dp_p}{dz}
-
m_p\,g\,n
+
e\,n\,E
+
R_p,
\end{equation}
\begin{equation}
m_e n\, u_e\frac{du_e}{dz}
=
-\frac{dp_e}{dz}
-
m_e g\,n
-
e\,n\,E
+
R_e,
\end{equation}
where \(u_\alpha\) denotes the bulk velocity of species \(\alpha\), while
\begin{equation}
R_\alpha
=
\sum_\beta
\int
m_\alpha\, v_\parallel\,
C_{\alpha\beta}(f_\alpha,f_\beta)
\,d^3v
\end{equation}
is the collisional momentum exchange, corresponding to the first velocity moment of the collision operator. Since momentum is conserved during Coulomb collisions,
\begin{equation}
R_p=-R_e\equiv R.
\end{equation}

In addition to quasi-neutrality,
\begin{equation}
n_e=n_p=n,
\end{equation}
we require the total electric current along the loop to vanish,
\begin{equation}
j
=
e\left(n_pu_p-n_eu_e\right)
=
0.
\end{equation}
Together with quasi-neutrality, this condition implies that electrons and protons have the same bulk velocity,
\begin{equation}
u_p=u_e\equiv u.
\end{equation}
Thus, although asymmetric heating can drive a finite mass flux along thevloop, the electron and proton particle fluxes remain equal and no net electric current is carried by the flow as in standard exospheric models for the solar wind \citep{Lemaire1971,Jockers1970,Maksimovic1997,Lamy2003,Zouganelis2004}.

Subtracting the electron momentum equation from the proton momentum
equation and using the above conditions, we obtain
\begin{equation}
\begin{aligned}
E
={}&
\frac{(m_p-m_e)g}{2e}
+
\frac{1}{2\,e\,n}
\left(
\frac{dp_p}{dz}
-
\frac{dp_e}{dz}
\right)
\\
&
+
\frac{m_p-m_e}{2e}\,
u\frac{du}{dz}
-
\frac{R}{e\,n}.
\end{aligned}
\label{Eq:GeneralizedPRflows}
\end{equation}

Equation~\eqref{Eq:GeneralizedPRflows} shows explicitly that the ambipolar electric field consists of four distinct contributions. The first term corresponds to the classical Pannekoek--Rosseland electric field, the second originates from the difference between the electron and proton pressure gradients, the third arises from the inertia associated with the common field-aligned bulk flow of electrons and protons, and the fourth represents the collisional momentum exchange between the two species.

Unlike the symmetric configurations considered in the present work, the velocity distribution functions associated with siphon flows are asymmetric with respect to the parallel velocity. Consequently, the first velocity moment of the collision operator is no longer zero, so that the friction term \(R\) contributes directly to the ambipolar electric field. Therefore, the electric field can no longer be prescribed \emph{a priori}, but must instead be determined self-consistently together with the electron and proton velocity distribution functions similarly to exospheric models \citep{Jockers1970,Lemaire1971,Maksimovic_al_2020,Lamy2003,Zouganelis2004,Pierrard2023}.

This more general formulation would also naturally allow different electron and proton temperature profiles to emerge self-consistently as a consequence of species-dependent stochastic heating, gravitational velocity filtration, Coulomb collisions, the ambipolar electric field, and the plasma flows themselves.

It is worth noting that persistent plasma flows along coronal loops have been inferred from Doppler shifts of coronal spectral lines \citep{Hara_2008}. Under suitable physical conditions, these field-aligned flows can be interpreted as siphon flows driven by pressure asymmetries between the two loop footpoints \citep{Reale2010}. Such flows have been extensively investigated using both hydrodynamic (HD) and magnetohydrodynamic (MHD) models, where asymmetric heating or pressure distributions naturally drive plasma motion along magnetic field lines \citep{Cargill1980,Noci1981,Orlando1995a,Orlando1995b,Mason_2023,Reep_2024}. Comprehensive reviews including observational evidence and theoretical modelling of plasma flows in coronal loops are provided by \citet{Reale2010} and \citet{Keppens2025}.

The extension proposed here would provide the kinetic counterpart of these HD and MHD descriptions. Rather than prescribing the pressure asymmetry phenomenologically, the present kinetic framework would allow it to emerge self-consistently from asymmetric stochastic heating at the two loop footpoints. Such a model would simultaneously predict the electron and proton velocity distribution functions, the ambipolar electric field, particle and heat fluxes, collisional momentum exchange, and the resulting siphon flows, thereby providing a fully kinetic description of field-aligned plasma transport in coronal loops.

Field-aligned plasma flows are also expected to influence the velocity-space anisotropy of the particle distribution functions. Unlike magnetic flux-tube expansion, which naturally produces an excess of parallel temperature (\(T_\parallel>T_\perp\)) through conservation of the magnetic moment, plasma flows modify the particle dynamics and therefore affect the resulting velocity-space anisotropy. By themselves, however, they do not necessarily produce an excess of perpendicular temperature. As is well known, additional physical mechanisms, such as resonant wave--particle interactions (e.g. ion-cyclotron heating), may also contribute to the formation of anisotropies with \(T_\perp>T_\parallel\) \citep{Cranmer_2000,Vocks2001,Vocks_2002a,Vocks_2002b,Hellinger2006,Cranmer2009,Vocks2016,Verscharen2019}. Such anisotropies are susceptible to ion-cyclotron and mirror instabilities, whereas the anisotropy generated by magnetic expansion favours the firehose instability. A kinetic model including asymmetric boundary conditions, magnetic expansion, plasma flows, Coulomb collisions, and stochastic heating would therefore make it possible to investigate how these competing mechanisms shape the particle distribution functions and determine which kinetic instabilities are most likely to develop in coronal loops.

\subsection{Distributed coronal heating and heat flux} \label{secistributed_heating}

Finally, we note that the implications of the present model differ in the short- and long-time-scale regimes, although both share the same fundamental limitation. In the long-time-scale regime, the coarse-grained temperature and density profiles are broadly consistent with coronal observations. At the level of the individual unresolved flux tubes, however, the plasma consists of a mixture of cold columns, remaining close to the chromospheric temperature \(T_0\sim10^4\,\mathrm{K}\), and hot columns reaching coronal temperatures of the order of \(10^6\,\mathrm{K}\). Consequently, the plasma between the heated structures does not become uniformly hot, in contrast with the nearly volume-filling coronal temperatures inferred from observations (within the limits of present instrument spatial resolution).

In the short-time-scale regime, each heated flux tube naturally develops a temperature inversion through gravitational filtering. Coulomb collisions preserve this inversion, although they significantly reduce the coronal density with respect to the collisionless case. Nevertheless, flux tubes that do not experience stochastic heating remain close to the chromospheric temperature. Thus, although the coarse-grained temperature profile again exhibits a transition region followed by a hot corona, the elementary plasma structures still consist of a mixture of hot and cold columns.

This limitation reflects the fact that the present work is specifically aimed at isolating the kinetic consequences of stochastic heating localized at the base of the transition region. It therefore neglects possible heating processes operating directly within the corona.

The above considerations suggest that additional heating mechanisms distributed throughout the corona are likely required to produce a more homogeneous thermal structure. Heating events occurring at coronal heights would also modify the kinetic boundary-value problem considered here. Particles energized in the corona would propagate along the magnetic field in both directions, producing both upward- and downward-moving populations. In particular, between the base of the transition region and the location of a coronal heating event, the downward-propagating hot particles generated in the corona would coexist with the upward-moving particles continuously injected from the lower boundary of the model. 

The coexistence of these two particle populations would naturally produce velocity distribution functions that are no longer symmetric with respect to the parallel velocity, giving rise to a finite field-aligned heat flux. Such heat transport is therefore expected to arise naturally once distributed coronal heating is included in the kinetic description.

Heat transport in weakly collisional plasmas has been extensively investigated over the past decades \citep{Lie-Svendsen_1999,Landi-Pantellini2001,Vocks2016,Dudk2017,Cranmer2021,Jeong_2022b,Coburn_2024}. Since the solar atmosphere generally operates in an intermediate regime that is neither fully collisional nor fully collisionless, the heat flux is expected to depart from the classical Spitzer--H\"arm expression, which is valid only under local thermodynamic equilibrium. Moreover, distributed coronal heating would also break the symmetry of the boundary conditions assumed in the present work, requiring a self-consistent determination of the ambipolar electric field together with the electron and proton distribution functions. A kinetic treatment including distributed coronal heating, asymmetric velocity distribution functions, ambipolar electric fields, and the resulting non-classical heat flux constitutes a natural extension of the present model.

\subsection{A multi-component plasma model} \label{secMulti-component}

Throughout the present work, the solar atmosphere has been modelled as a two-component plasma composed of electrons and protons. However, the transition region and the corona consist of a multi-component plasma containing several ion species with different masses, atomic numbers, ionization states, and abundances \citep{GolubPasachoff:book,Aschwandensolarcorona}. The presence of multiple ion species introduces an additional physical effect that is not included in the present treatment.

Assuming, for simplicity, that all species share the same temperature, the hydrostatic equilibrium equation for a generic species \(\alpha\) reads
\begin{equation}
k_B\, T
\frac{dn_\alpha}{dz}
=
-
m_\alpha\, g\,n_\alpha
+
e_\alpha\, E\,n_\alpha,
\end{equation}
where \(e_\alpha\) is the electric charge of species \(\alpha\).
The ambipolar electric field must then satisfy the quasi-neutrality condition
\begin{equation}
\sum_\alpha e_\alpha\, n_\alpha=0.
\end{equation}

Taking the derivative of the quasi-neutrality condition with respect to \(z\) and using the hydrostatic balance equations for all species yields

\begin{equation}
E(z)
=
g\,
\frac{
\displaystyle
\sum_\alpha e_\alpha\, m_\alpha\, n_\alpha(z)
}{
\displaystyle
\sum_\alpha\ e_\alpha^2\, n_\alpha(z)
}.
\label{eq:Emultispecies}
\end{equation}

Unlike the classical Pannekoek--Rosseland field, Eq.~\eqref{eq:Emultispecies} explicitly depends on the local densities, charges, and masses of all plasma species. Consequently, unless the relative abundances of the different species are fixed, the ambipolar electric field can no longer be prescribed \emph{a priori}, but must instead be determined self-consistently together with the density profiles of all species.

In practice, however, the elemental abundances of the solar atmosphere are relatively well constrained \citep{Feldman2003}, so that only hydrogen and helium are expected to provide significant contributions to Eq.~\eqref{eq:Emultispecies}. For a prescribed abundance ratio He/H, the remaining ingredient is the variation of the helium ionization state with height. This additional degree of freedom is expected to modify the ambipolar electric field and, consequently, the effective gravitational stratification of the plasma. Since helium possesses a smaller gravitational scale height than hydrogen, these effects may shift the transition region toward lower altitudes and increase the coronal density with respect to the present two-component model. Quantifying these effects requires a fully self-consistent treatment of the coupled ionization, composition, and ambipolar electric field, and is therefore left for future work.

\section{Summary and discussions}\label{secSummary}

In this work, we have extended the kinetic model of the solar atmosphere originally introduced in \citet{Barbieri2025c}. In that model, the collisionless coronal plasma is in steady contact with the chromosphere, represented as a thermal boundary. Motivated by the routine observation of small-scale, transient brightenings on the Sun (see Introduction and Sect.~\ref{sec1}), the upper chromosphere was modelled as a two-dimensional surface hosting localized heating events randomly distributed across it. 
So heated loops are present in a small fraction of the volume, while most of the unheated volume stays at the chromospheric temperature. 
In the previous formulation, the observed temperature inversion emerged solely as a localized spatial averaging (coarse-graining) resulting from the spatial distribution of these heating events.
In the present work, we have extended this description by incorporating also their temporal intermittency.  

In the collisionless case the boundary conditions depend on the relevant time-scale regime. In the one-dimensional model derived in \citet{Barbieri2025b}, it was shown that intermittent heating events can drive the plasma toward a non-thermal stationary configuration with temperature inversion only if their characteristic time scales are shorter than the electron crossing time, which is of the order of $15$~s for a loop length of approximately $0.1  \,R_{\odot}$.
The corresponding coarse-grained distribution function remains leptokurtic and is given by Eq.~\eqref{VDFphasespace}. In this work, by combining the temporal and spatial intermittency of stochastic heating, we derive a general expression for the coarse-grained velocity distribution function. The controlling parameter is \(A=A_t\,A_S\), where \(A_t\) denotes the fraction of time spent in the heating phase relative to the waiting time, and \(A_S\) is the spatial filling factor. When \(A\) is small, the temperature increases sharply with height between the chromosphere and the corona. In this regime, temperature inversion develops both within individual loops undergoing stochastic heating and after surface coarse-graining.

A second regime arises when the characteristic time scales of the heating events are comparable to or longer than the proton crossing time, which is of the order of $6$-$7\,\mathrm{min}$. In this long time-scale regime, by contrast, the system is composed of isothermal loops characterized by different temperatures. In this case, temperature inversion arises only at the surface coarse-graining level. The corresponding coarse-grained distribution function has the same formal expression as in the short time-scale regime, except that $A$ is replaced by the spatial filling factor $A_S \, n_T /n_0$ with $n_T$ the loop density at temperature $T$ and $n_0$ the top chromospheric density.
 
Although neighboring loops may experience different stationary states, the magnetic pressure largely dominates over the plasma pressure throughout the transition region and the corona ($\beta \ll 1$), ensuring lateral force balance and the global mechanical consistency of the system.

In the fast time-scale regime the loops undergoing stochastic heating are in a non-thermal stationary state, and Coulomb collisions can therefore modify the resulting macroscopic profiles.
In this paper we introduce the effect of collisions, we incorporate their cumulative action as a progressive attenuation of the suprathermal population with height. This reduced description appears to capture the dominant impact of collisional thermalization on macroscopic observables, such as density and temperature, without explicitly solving the full Fokker–Planck equation. Although approximate, this approach provides a simple and physically motivated framework to assess the role of collisions within the present model.
We find that temperature inversion persists and a hot corona is still produced. However, the coronal density is significantly reduced, falling below the lowest observed values in coronal holes, of the order of $10^7  \,\mathrm{cm}^{-3}$. This reduction is a consequence of strong dynamical friction at low heights, which limits the fraction of suprathermal particles able to reach coronal heights. We therefore conclude that, in this regime, the model qualitatively retains the temperature inversion but fails to quantitatively reproduce the observed coronal density.

In the long-timescale regime, the system behaves as a superposition of parallel loops in local thermal equilibrium. 
The inclusion of Coulomb collisions does not alter the global structure obtained in the collisionless case. 
Even when suprathermal distributions are imposed at the base, the very low Knudsen number ensures efficient thermalisation, so that the loops remain effectively isothermal at all heights and the coarse-grained temperature inversion is preserved. 

An order-of-magnitude estimate of radiative losses shows that the characteristic radiative length obtained from the ratio between the upward kinetic-energy flux of the hot population and the volumetric radiative losses is much larger than the typical loop length. Radiative cooling therefore does not modify the temperature and density profiles derived in this work at coronal heights.

The framework developed above can be applied not only to the solar corona, but also to other stars (see \cite{barbieri2024temperaturedensityprofilescorona} and references therein) and to astrophysical systems hosting hot coronal plasmas, such as active galactic nuclei (AGN) \citep{Haardt1991,Fabian2015,Wilkins2015,Palit2024,Zhao_2025}.

We finally emphasize that, although the present model possesses a number of limitations, as discussed in section \ref{secLimitations}, its primary objective is to isolate the respective roles of gravitational and collisional filtering in shaping the temperature and density profiles generated by spatially sparse and temporally intermittent stochastic heating at the base of the transition region.

\section*{Acknowledgements}
The authors wish to thank the anonymous reviewer for the valuable comments and suggestions that helped improve the clarity and quality of this work. L.B. wants to thank Sorbonne Université in the framework of the Initiative Physique des Infinis for financial support. L.B. thanks discussions with Susanna Parenti, Simone Landi, Lapo Casetti, Andrea Verdini, Emanuele Papini, Pierfrancesco Di Cintio, Arnaud Zaslavsky, Petr Hellinger, David Paipa-Leon, Etienne Berriot.

\bibliographystyle{elsarticle-harv}
\bibliography{manuscript}

@PREAMBLE{
 "\providecommand{\noopsort}[1]{}" 
 # "\providecommand{\singleletter}[1]{#1}%"
}

@STRING{aap       = "Astron.\ Astrophys." }

@STRING{apj       = "Astrophys.\ J." }

@STRING{apjl      = "Astrophys.\ J.\ Lett." }

@STRING{jgr       = "J.\ Geophys.\ Res." }

@STRING{ssr       = "Space Sci.\ Rev." }

@article{Pannekoek_1922,
	adsurl = {https://ui.adsabs.harvard.edu/abs/1922BAN.....1..107P},
	author = {{Pannekoek}, A.},
	journal = {Bull. Astr. Inst. Netherlandd},
	month = jul,
	pages = {107},
	title = {{Ionization in stellar atmospheres (Errata: 2 24)}},
	volume = {1},
	year = 1922}

@article{Rosseland_1924,
	adsurl = {https://ui.adsabs.harvard.edu/abs/1924MNRAS..84..720R},
	author = {{Rosseland}, S.},
	doi = {10.1093/mnras/84.9.720},
	journal = {MNRAS},
	month = jun,
	pages = {720-728},
	title = {{Electrical state of a star}},
	volume = {84},
	year = 1924
}

@article{Scudder1992a,
  doi = {10.1086/171858},
  year = {1992},
  month = oct,
  publisher = {American Astronomical Society},
  volume = {398},
  pages = {299},
  author = {Jack D.\ Scudder},
  title = {On the causes of temperature change in inhomogeneous low-density astrophysical plasmas},
  journal = {Astrophys.\  J.}
}

@article{Maksimovic_al_2020,
	author = {{Maksimovic}, M. and {Bale}, S.~D. and {Ber{\v{c}}i{\v{c}}}, L. and {Bonnell}, J.~W. and {Case}, A.~W. and {Dudok de Wit}, T. and {Goetz}, K. and {Halekas}, J.~S. and {Harvey}, P.~R. and {Issautier}, K. and {Kasper}, J.~C. and {Korreck}, K.~E. and {Jagarlamudi}, V. Krishna and {Lahmiti}, N. and {Larson}, D.~E. and {Lecacheux}, A. and {Livi}, R. and {MacDowall}, R.~J. and {Malaspina}, D.~M. and {Martinovi{\'c}}, M.~M. and {Meyer-Vernet}, N. and {Moncuquet}, M. and {Pulupa}, M. and {Salem}, C. and {Stevens}, M.~L. and {{\v{S}}tver{\'a}k}, {\v{S}}. and {Velli}, M. and {Whittlesey}, P.~L.},
	journal = {Astrophys.\ J.\ Suppl.\ Series},
	month = feb,
	number = {2},
	pages = {62},
	title = {{Anticorrelation between the Bulk Speed and the Electron Temperature in the Pristine Solar Wind: First Results from the Parker Solar Probe and Comparison with Helios}},
	volume = {246},
	year = 2020}

@article{Scudder1992b,
  doi = {10.1086/171859},
  year = {1992},
  month = oct,
  publisher = {American Astronomical Society},
  volume = {398},
  pages = {319},
  author = {Jack D. Scudder},
  title = {Why all stars should possess circumstellar temperature inversions},
  journal = {Astrophys.\  J.}
}

@book{nicholson1983introduction,
  title={Introduction to Plasma Theory},
  author={Nicholson, D.\ R.\ },
  isbn={9780471090458},
  year={1983},
  publisher={Wiley}
}

@article{Landi-Pantellini2001,
	author = {S.\ Landi and F.\ G.\ E.\ Pantellini},
	title = {On the temperature profile and heat flux in the solar corona:  
 Kinetic simulations},
	DOI= "10.1051/0004-6361:20010552",
	journal = {\aap},
	year = 2001,
	volume = 372,
	number = 2,
	pages = "686-701"
}

@book{belmont2013collisionless,
  title={Collisionless Plasmas in Astrophysics},
  author={Belmont, G.\ and Grappin, R.\ and Mottez, F.\ and Pantellini, F.\ and Pelletier, G.\  },
  isbn={9783527656240},
  year={2013},
  publisher={Wiley}
}

@BOOK{Aschwandensolarcorona,
       author = {{Aschwanden}, Markus J.},
        title = "{Physics of the Solar Corona. An Introduction with Problems and Solutions (2nd edition)}",
        publisher = {Publishing Ltd., Chichester, UK; Springer, New York, Berlin},
         year = 2005,
       adsurl = {https://ui.adsabs.harvard.edu/abs/2005psci.book.....A}
}

@ARTICLE{Klimchuk_2006,
       author = {{Klimchuk}, James A.},
        title = "{On Solving the Coronal Heating Problem}",
      journal = {Solar Physics},
         year = 2006,
        month = mar,
       volume = {234},
       number = {1},
        pages = {41-77},
          doi = {10.1007/s11207-006-0055-z},
       adsurl = {https://ui.adsabs.harvard.edu/abs/2006SoPh..234...41K}
}

@book{lazar2021kappa,
  title={Kappa Distributions: From Observational Evidences Via Controversial Predictions to a Consistent Theory of Nonequilibrium Plasmas},
  author={Lazar, M. and Fichtner, H.},
  isbn={9783030826239},
  series={Astrophysics and space science library},
  year={2021},
  publisher={Springer}
}

@ARTICLE{Heyvaerts_Priest_1983,
       author = {{Heyvaerts}, J. and {Priest}, E.~R.},
        title = "{Coronal heating by phase-mixed shear Alfven waves.}",
      journal = aap,
         year = 1983,
        month = jan,
       volume = {117},
        pages = {220-234},
       adsurl = {https://ui.adsabs.harvard.edu/abs/1983A&A...117..220H}
}

@ARTICLE{Ionson_1978,
       author = {{Ionson}, J.~A.},
        title = "{Resonant absorption of Alfv{\'e}nic surface waves and the heating of solar coronal loops.}",
      journal = apj,
         year = 1978,
        month = dec,
       volume = {226},
        pages = {650-673},
          doi = {10.1086/156648},
       adsurl = {https://ui.adsabs.harvard.edu/abs/1978ApJ...226..650I}
}

@article{Parker:1972wu,
        author = {E. N Parker},
        doi = {10.1086/151512},
        journal = {Astrophysical Journal},
        month = {Jun},
        pages = {499},
        title = {Topological Dissipation and the Small-Scale Fields in Turbulent Gases},
        url = {http://adsabs.harvard.edu/cgi-bin/nph-data_query?bibcode=1972ApJ...174..499P&link_type=ABSTRACT},
        volume = {174},
        year = {1972},
}

@article{Dmitruk:1997uf,
        author = {Pablo Dmitruk and Daniel O Gomez},
        doi = {10.1086/310760},
        journal = {Astrophysical Journal Letters v.484},
        month = {Jul},
        pages = {L83},
        title = {Turbulent Coronal Heating and the Distribution of Nanoflares},
        url = {http://adsabs.harvard.edu/cgi-bin/nph-data_query?bibcode=1997ApJ...484L..83D&link_type=ABSTRACT},
        volume = {484},
        year = {1997},
}

@ARTICLE{2013ApJ...773L...2R,
       author = {{Rappazzo}, A.~F. and {Parker}, E.~N.},
        title = "{Current Sheets Formation in Tangled Coronal Magnetic Fields}",
      journal = apjl,
         year = 2013,
        month = aug,
       volume = {773},
       number = {1},
          eid = {L2},
        pages = {L2},
          doi = {10.1088/2041-8205/773/1/L2},
       adsurl = {https://ui.adsabs.harvard.edu/abs/2013ApJ...773L...2R}
}

@article{Rappazzo:2008vl,
        author = {Franco Rappazzo and Marco Velli and Giorgio Einaudi and R. B Dahlburg},
        doi = {10.1086/528786},
        journal = {Astrophys.\  J.},
        month = {Apr},
        pages = {1348},
        title = {Nonlinear Dynamics of the Parker Scenario for Coronal Heating},
        url = {http://adsabs.harvard.edu/cgi-bin/nph-data_query?bibcode=2008ApJ...677.1348R&link_type=ABSTRACT},
        volume = {677},
        year = {2008},
}

@ARTICLE{2015RSPTA.37340265W,
       author = {{Wilmot-Smith}, A.~L.},
        title = "{An overview of flux braiding experiments}",
      journal = {Philosophical Transactions of the Royal Society of London Series A},
         year = 2015,
        month = apr,
       volume = {373},
       number = {2042},
        pages = {20140265-20140265},
          doi = {10.1098/rsta.2014.0265},
       adsurl = {https://ui.adsabs.harvard.edu/abs/2015RSPTA.37340265W}
}

@ARTICLE{2005ApJ...618.1020G,
       author = {{Gudiksen}, Boris Vilhelm and {Nordlund}, {\r{A}}ke},
        title = "{An Ab Initio Approach to the Solar Coronal Heating Problem}",
      journal = apj,
         year = 2005,
        month = jan,
       volume = {618},
       number = {2},
        pages = {1020-1030},
          doi = {10.1086/426063},
       adsurl = {https://ui.adsabs.harvard.edu/abs/2005ApJ...618.1020G}
}

@ARTICLE{Jockers1970,
       author = {{Jockers}, K.},
        title = "{Solar Wind Models Based on Exospheric Theory}",
      journal = aap,
         year = 1970,
        month = jun,
       volume = {6},
        pages = {219},
       adsurl = {https://ui.adsabs.harvard.edu/abs/1970A&A.....6..219J},
}

@ARTICLE{Lemaire1971,
       author = {{Lemaire}, J. and {Scherer}, M.},
        title = "{Kinetic models of the solar wind}",
      journal = jgr,
         year = 1971,
        month = jan,
       volume = {76},
       number = {31},
        pages = {7479},
          doi = {10.1029/JA076i031p07479},
       adsurl = {https://ui.adsabs.harvard.edu/abs/1971JGR....76.7479L},
}

@ARTICLE{Maksimovic1997,
       author = {{Maksimovic}, M. and {Pierrard}, V. and {Lemaire}, J.~F.},
        title = "{A kinetic model of the solar wind with Kappa distribution functions in the corona.}",
      journal = aap,
         year = 1997,
        month = aug,
       volume = {324},
        pages = {725-734},
       adsurl = {https://ui.adsabs.harvard.edu/abs/1997A&A...324..725M},
}

@ARTICLE{Lamy2003,
       author = {{Lamy}, H. and {Pierrard}, V. and {Maksimovic}, M. and {Lemaire}, J.~F.},
        title = "{A kinetic exospheric model of the solar wind with a nonmonotonic potential energy for the protons}",
      journal = {Journal of Geophysical Research (Space Physics)},
         year = 2003,
        month = jan,
       volume = {108},
       number = {A1},
          eid = {1047},
        pages = {1047},
          doi = {10.1029/2002JA009487},
       adsurl = {https://ui.adsabs.harvard.edu/abs/2003JGRA..108.1047L}
}

@ARTICLE{Zouganelis2004,
       author = {{Zouganelis}, I. and {Maksimovic}, M. and {Meyer-Vernet}, N. and {Lamy}, H. and {Issautier}, K.},
        title = "{A Transonic Collisionless Model of the Solar Wind}",
      journal = apj,
         year = 2004,
        month = may,
       volume = {606},
       number = {1},
        pages = {542-554},
          doi = {10.1086/382866},
       adsurl = {https://ui.adsabs.harvard.edu/abs/2004ApJ...606..542Z}
}

@article{Cauzzi:2009ta,
	author = {Gianna Cauzzi and Kevin Reardon and R. J Rutten and Alexandra Tritschler and H Uitenbroek},
	doi = {10.1051/0004-6361/200811595},
	journal = {\aap},
	month = {Aug},
	pages = {577},
	pmid = {2009A&A...503..577C},
	title = {The solar chromosphere at high resolution with IBIS. IV. Dual-line evidence of heating in chromospheric network},
	url = {http://adsabs.harvard.edu/cgi-bin/nph-data_query?bibcode=2009A%2526A...503..577C&link_type=ABSTRACT},
	volume = {503},
	year = {2009}
}

@book{GolubPasachoff:book,
  title = {The Solar Corona},
  author = {Golub, Leon and Pasachoff, Jay M.},
  edition = {second},
  year = {2009},
  publisher = {Cambridge University Press},
  address={Cambridge}
}

@article{Berghmans:2021wl,
        adsurl = {https://ui.adsabs.harvard.edu/abs/2021A&A...656L...4B},
        author = {{Berghmans}, D. and {Auch{\`e}re}, F. and {Long}, D.~M. and {Soubri{\'e}}, E. and {Mierla}, M. and {Zhukov}, A.~N. and {Sch{\"u}hle}, U. and {Antolin}, P. and {Harra}, L. and {Parenti}, S. and {Podladchikova}, O. and {Aznar Cuadrado}, R. and {Buchlin}, {\'E}. and {Dolla}, L. and {Verbeeck}, C. and {Gissot}, S. and {Teriaca}, L. and {Haberreiter}, M. and {Katsiyannis}, A.~C. and {Rodriguez}, L. and {Kraaikamp}, E. and {Smith}, P.~J. and {Stegen}, K. and {Rochus}, P. and {Halain}, J.~P. and {Jacques}, L. and {Thompson}, W.~T. and {Inhester}, B.},
        doi = {10.1051/0004-6361/202140380},
        journal = aap,
        month = dec,
        pages = {L4},
        title = {{Extreme-UV quiet Sun brightenings observed by the Solar Orbiter/EUI}},
        volume = {656},
        year = {2021}
}

@article{Dere:1989ux,
        adsurl = {https://ui.adsabs.harvard.edu/abs/1989SoPh..123...41D},
        author = {{Dere}, K.~P. and {Bartoe}, J. -D.~F. and {Brueckner}, G.~E.},
        doi = {10.1007/BF00150011},
        journal = {Solar Physics},
        month = mar,
        number = {1},
        pages = {41-68},
        title = {{Explosive Events in the Solar Transition Zone}},
        volume = {123},
        year = {1989}
}

@article{Peter:2014uz,
        adsurl = {https://ui.adsabs.harvard.edu/abs/2014Sci...346C.315P},
        author = {{Peter}, H. and {Tian}, H. and {Curdt}, W. and {Schmit}, D. and {Innes}, D. and {De Pontieu}, B. and {Lemen}, J. and {Title}, A. and {Boerner}, P. and {Hurlburt}, N. and {Tarbell}, T.~D. and {Wuelser}, J.~P. and {Mart{\'\i}nez-Sykora}, Juan and {Kleint}, L. and {Golub}, L. and {McKillop}, S. and {Reeves}, K.~K. and {Saar}, S. and {Testa}, P. and {Kankelborg}, C. and {Jaeggli}, S. and {Carlsson}, M. and {Hansteen}, V.},
        doi = {10.1126/science.1255726},
        eid = {1255726},
        journal = {Science},
        month = oct,
        number = {6207},
        pages = {1255726},
        title = {{Hot explosions in the cool atmosphere of the Sun}},
        volume = {346},
        year = 2014
}

@article{Teriaca:2004wy,
        adsurl = {https://ui.adsabs.harvard.edu/abs/2004A&A...427.1065T},
        author = {{Teriaca}, L. and {Banerjee}, D. and {Falchi}, A. and {Doyle}, J.~G. and {Madjarska}, M.~S.},
        doi = {10.1051/0004-6361:20040503},
        journal = aap,
        month = dec,
        pages = {1065-1074},
        title = {{Transition region small-scale dynamics as seen by SUMER on SOHO}},
        volume = {427},
        year = 2004
}

@article{Tiwari:2019us,
        adsurl = {https://ui.adsabs.harvard.edu/abs/2019ApJ...887...56T},
        author = {{Tiwari}, Sanjiv K. and {Panesar}, Navdeep K. and {Moore}, Ronald L. and {De Pontieu}, Bart and {Winebarger}, Amy R. and {Golub}, Leon and {Savage}, Sabrina L. and {Rachmeler}, Laurel A. and {Kobayashi}, Ken and {Testa}, Paola and {Warren}, Harry P. and {Brooks}, David H. and {Cirtain}, Jonathan W. and {McKenzie}, David E. and {Morton}, Richard J. and {Peter}, Hardi and {Walsh}, Robert W.},
        doi = {10.3847/1538-4357/ab54c1},
        eid = {56},
        journal = apj,
        month = dec,
        number = {1},
        pages = {56},
        title = {{Fine-scale Explosive Energy Release at Sites of Prospective Magnetic Flux Cancellation in the Core of the Solar Active Region Observed by Hi-C 2.1, IRIS, and SDO}},
        volume = {887},
        year = 2019
}

@article{Barbieri2024b, 
title={Temperature inversion in a confined plasma atmosphere: coarse-grained effect of temperature fluctuations at its base}, 
volume={90},
DOI={10.1017/S0022377824000849},
number={5},
journal={Journal of Plasma Physics},
author={Barbieri, Luca and Papini, Emanuele and Di Cintio, Pierfrancesco and Landi, Simone and Verdini, Andrea and Casetti, Lapo},
year={2024},
pages={905900511}
}

@ARTICLE{Lee:ApJ2020,
       author = {{Lee}, Kyoung-Sun and {Hara}, Hirohisa and {Watanabe}, Kyoko and {Joshi}, Anand D. and {Brooks}, David H. and {Imada}, Shinsuke and {Prasad}, Avijeet and {Dang}, Phillip and {Shimizu}, Toshifumi and {Savage}, Sabrina L. and {Moore}, Ronald and {Panesar}, Navdeep K. and {Reep}, Jeffrey W.},
        title = "{A Solar Magnetic-fan Flaring Arch Heated by Nonthermal Particles and Hot Plasma from an X-Ray Jet Eruption}",
      journal = {\apj},
         year = 2020,
        month = may,
       volume = {895},
       number = {1},
          eid = {42},
        pages = {42},
          doi = {10.3847/1538-4357/ab8bce}
}

@article{barbieri2023temperature,
	author = {Barbieri, Luca and Casetti, Lapo and Verdini, Andrea and Landi, Simone},
	title = {Temperature inversion in a gravitationally bound plasma: Case of 
        the solar corona},
	DOI= "10.1051/0004-6361/202348373",
	journal = aap,
	year = 2024,
	volume = 681,
	pages = "L5",
}

@ARTICLE{Rauoafi:ApJ2023,
       author = {{Raouafi}, Nour E. and {Stenborg}, G. and {Seaton}, D.~B. and {Wang}, H. and {Wang}, J. and {DeForest}, C.~E. and {Bale}, S.~D. and {Drake}, J.~F. and {Uritsky}, V.~M. and {Karpen}, J.~T. and {DeVore}, C.~R. and {Sterling}, A.~C. and {Horbury}, T.~S. and {Harra}, L.~K. and {Bourouaine}, S. and {Kasper}, J.~C. and {Kumar}, P. and {Phan}, T.~D. and {Velli}, M.},
        title = "{Magnetic Reconnection as the Driver of the Solar Wind}",
      journal = {\apj},
         year = 2023,
        month = mar,
       volume = {945},
       number = {1},
          eid = {28},
        pages = {28},
          doi = {10.3847/1538-4357/acaf6c}
}

@ARTICLE{Chamberlain1960,
       author = {{Chamberlain}, Joseph W.},
        title = "{Interplanetary Gas.II. Expansion of a Model Solar Corona.}",
      journal = {Astrophysical Journal},
         year = 1960,
        month = jan,
       volume = {131},
        pages = {47},
          doi = {10.1086/146805},
       adsurl = {https://ui.adsabs.harvard.edu/abs/1960ApJ...131...47C}
}

@article{Verscharen2019,
  title = {The multi-scale nature of the solar wind},
  volume = {16},
  ISSN = {1614-4961},
  url = {http://dx.doi.org/10.1007/s41116-019-0021-0},
  DOI = {10.1007/s41116-019-0021-0},
  number = {1},
  journal = {Living Reviews in Solar Physics},
  publisher = {Springer Science and Business Media LLC},
  author = {Verscharen,  Daniel and Klein,  Kristopher G. and Maruca,  Bennett A.},
  year = {2019},
  month = dec 
}

@article{Shoda2021,
	author = {{Shoda M.} and {Takasao S.}},
	title = {Corona and XUV emission modelling of the Sun and Sun-like stars},
	DOI= "10.1051/0004-6361/202141563",
	url= "https://doi.org/10.1051/0004-6361/202141563",
	journal = {\aap},
	year = 2021,
	volume = 656,
	pages = "A111",
}

@article{Airapetian_2021,
doi = {10.3847/1538-4357/ac081e},
url = {https://dx.doi.org/10.3847/1538-4357/ac081e},
year = {2021},
month = {aug},
publisher = {The American Astronomical Society},
volume = {916},
number = {2},
author = {Vladimir S. Airapetian and Meng Jin and Theresa Lüftinger and Sudeshna Boro Saikia and Oleg Kochukhov and Manuel Güdel and Bart Van Der Holst and W. Manchester IV},
title = {One Year in the Life of Young Suns: Data-constrained Corona-wind Model of κ1 Ceti},
journal = {The Astrophysical Journal},
}

@article{Shoda2024,
	author = {{Shoda M.} and {Namekata, Kosuke} and {Takasao, Shinsuke}},
	title = {Assessing the capability of a model-based stellar XUV estimation},
	DOI= "10.1051/0004-6361/202450129",
	url= "https://doi.org/10.1051/0004-6361/202450129",
	journal = {\aap},
	year = 2024,
	volume = 691,
	pages = "A152",
}

@article{barbieri2024temperaturedensityprofilescorona,
	author = {{Barbieri L.} and {Casetti, L.} and {Verdini A.} and {Landi S.}},
	title = {Temperature and density profiles in the corona of main-sequence stars induced by stochastic heating in the chromosphere},
	DOI= "10.1051/0004-6361/202452879",
	url= "https://doi.org/10.1051/0004-6361/202452879",
	journal = {\aap},
	year = 2025,
	volume = 694,
	pages = "A154",
}

@article{Mandrini_2000,
doi = {10.1086/308398},
url = {https://dx.doi.org/10.1086/308398},
year = {2000},
month = {feb},
publisher = {},
volume = {530},
number = {2},
pages = {999},
author = {Mandrini, C. H. and Démoulin, P. and Klimchuk, J. A.},
title = {Magnetic Field and Plasma Scaling Laws:
Their Implications for Coronal
Heating Models},
journal = {The Astrophysical Journal}, 
}

@article{Zhukov2021,
	author = {{Zhukov A. N.} and {Mierla, M.} and {Auchère, F.} and {Gissot, S.} and {Rodriguez, L.} and {Soubrié, E.} and {Thompson, W. T.}
	and {Inhester, B.} and {Nicula, B.} and {Antolin, P.} and {Parenti, S.} and {Buchlin, É.} and {Barczynski, K.} 
	and {Verbeeck, C.} and {Kraaikamp, E.} and {Smith, P. J.} and {Stegen, K.} and {Dolla, L.} and {Harra, L.} and {Long, D. M.} 
	and {Schühle, U.} and {Podladchikova, O.} and {Aznar Cuadrado, R.} and {Teriaca, L.} and {Haberreiter, M.} and {Katsiyannis, A. C.} 
	and {Rochus, P.} and {Halain, J.-P.} and {Jacques, L.} and {Berghmans, D.}},
	title = {Stereoscopy of extreme UV quiet Sun brightenings observed by Solar Orbiter/EUI},
	DOI= "10.1051/0004-6361/202141010",
	url= "https://doi.org/10.1051/0004-6361/202141010",
	journal = {\aap},
	year = 2021,
	volume = 656,
	pages = "A35",
}

@ARTICLE{Young2018-sp,
  title     = "Solar ultraviolet bursts",
  author    = "Young, Peter R and Tian, Hui and Peter, Hardi and Rutten, Robert
               J and Nelson, Chris J and Huang, Zhenghua and Schmieder,
               Brigitte and Vissers, Gregal J M and Toriumi, Shin and van der
               Voort, Luc H M Rouppe and Madjarska, Maria S and Danilovic,
               Sanja and Berlicki, Arkadiusz and Chitta, L P and Cheung, Mark C
               M and Madsen, Chad and Reardon, Kevin P and Katsukawa, Yukio and
               Heinzel, Petr",
  journal   = "Space Sci. Rev.",
  publisher = "Springer Science and Business Media LLC",
  volume    =  214,
  number    =  8,
  month     =  dec,
  year      =  2018,
  copyright = "https://creativecommons.org/licenses/by/4.0",
  language  = "en"
}

@article{narang2025,
	author = {{Narang N.} and {Verbeeck, Cis} and {Mierla, Marilena} and {Berghmans, David} and {Auchère, Frédéric} and {Shestov, Sergei} and {Delouille, Véronique} and {Chitta, Lakshmi Pradeep} and {Priest, Eric} and {Lim, Daye} and {Dolla, Laurent R.} and {Kraaikamp, Emil}},
	title = {Extreme-ultraviolet transient brightenings in the quiet-Sun corona - Closest perihelion observations with Solar Orbiter/EUI},
	DOI= "10.1051/0004-6361/202554650",
	url= "https://doi.org/10.1051/0004-6361/202554650",
	journal = aap,
	year = 2025,
	volume = 699,
	pages = "A138",
}

@article{Neslusan2001-rp,
  title     = {On the global electrostatic charge of stars},
  author    = {Neslu{\v s}an, L},
  journal   = {\aap},
  publisher = {EDP Sciences},
  volume    =  372,
  number    =  3,
  pages     = {913--915},
  month     =  jun,
  year      =  2001
}

@article{Harra2025,
  title = {The Dynamics of the Extreme Ultraviolet Quiet Sun and Coronal Holes in the Solar Orbiter Era},
  volume = {221},
  ISSN = {1572-9672},
  url = {http://dx.doi.org/10.1007/s11214-025-01177-3},
  DOI = {10.1007/s11214-025-01177-3},
  number = {4},
  journal = {Space Science Reviews},
  publisher = {Springer Science and Business Media LLC},
  author = {Harra,  Louise and Barczynski,  Krzysztof and Auchère,  Frédéric and Berghmans,  David and Chitta,  Lakshmi Pradeep and Parenti,  Susanna and Peter,  Hardi},
  year = {2025},
  month = jun 
}

@article{Dahlburg2012,
	author = {{Dahlburg R. B.} and {Einaudi, G.} and {Rappazzo, A. F.} and {Velli, M.}},
	title = {Turbulent coronal heating mechanisms:  coupling of dynamics and thermodynamics},
	DOI= "10.1051/0004-6361/201219752",
	url= "https://doi.org/10.1051/0004-6361/201219752",
	journal = {\aap},
	year = 2012,
	volume = 544,
	pages = "L20",
	month = "",
}

@article{Reale2010,
  title = {Coronal Loops: Observations and Modeling of Confined Plasma},
  volume = {7},
  ISSN = {1614-4961},
  url = {http://dx.doi.org/10.12942/lrsp-2010-5},
  DOI = {10.12942/lrsp-2010-5},
  journal = {Living Reviews in Solar Physics},
  publisher = {Springer Science and Business Media LLC},
  author = {Reale,  Fabio},
  year = {2010}
}

@article{Klimchuk2015,
  title = {Key aspects of coronal heating},
  volume = {373},
  ISSN = {1471-2962},
  url = {http://dx.doi.org/10.1098/rsta.2014.0256},
  DOI = {10.1098/rsta.2014.0256},
  number = {2042},
  journal = {Philosophical Transactions of the Royal Society A: Mathematical,  Physical and Engineering Sciences},
  publisher = {The Royal Society},
  author = {Klimchuk,  James A.},
  year = {2015},
  month = may,
  pages = {20140256}
}

@misc{Viall2021,
  title = {The Heating of the Solar Corona},
  ISBN = {9781119815600},
  ISSN = {2328-8779},
  url = {http://dx.doi.org/10.1002/9781119815600.ch2},
  DOI = {10.1002/9781119815600.ch2},
  journal = {Solar Physics and Solar Wind},
  publisher = {Wiley},
  author = {Viall,  Nicholeen and De Moortel,  Ineke and Downs,  Cooper and Klimchuk,  James A. and Parenti,  Susanna and Reale,  Fabio},
  year = {2021},
  month = apr,
  pages = {35–82}
}

@article{Barbieri_2025c,
title={Self-consistent generation of the ambipolar electric field in collisionless plasmas via multi-mode electrostatics},
volume={91},
DOI={10.1017/S0022377825100810},
number={5},
journal={Journal of Plasma Physics},
author={Barbieri, Luca},
year={2025},
pages={E135}
}

@article{Barbieri2025b,
title={Extended temporal coarse-graining in a stratified and confined plasma under thermal fluctuations}, volume={91},
DOI={10.1017/S0022377825100809},
number={5},
journal={Journal of Plasma Physics},
author={Barbieri, Luca and Landi, Simone and Casetti, Lapo and Verdini, Andrea},
year={2025},
pages={E134}
}

@ARTICLE{Barbieri2025c,
       author = {{Barbieri}, Luca and {D{\'e}moulin}, Pascal},
        title = "{Kinetic collisionless model of the solar transition region and corona with spatially intermittent heating}",
      journal = {\aap},
         year = 2025,
        month = dec,
       volume = {704},
          eid = {A84},
        pages = {A84},
          doi = {10.1051/0004-6361/202557356},
archivePrefix = {arXiv},
       eprint = {2510.05954},
 primaryClass = {astro-ph.SR},
       adsurl = {https://ui.adsabs.harvard.edu/abs/2025A&A...704A..84B}
}

@article{Lim2025,
  title = {Accessing the fine temporal scale of EUV brightenings and their quasi-periodic pulsations: 1-second cadence observations by Solar Orbiter/EUI},
  ISSN = {1432-0746},
  url = {http://dx.doi.org/10.1051/0004-6361/202557135},
  DOI = {10.1051/0004-6361/202557135},
  journal = aap,
  publisher = {EDP Sciences},
  author = {Lim,  Daye and Van Doorsselaere,  Tom and Narang,  Nancy and Hayes,  Laura A. and Kraaikamp,  Emil and Joshi,  Aadish and Loumou,  Konstantina and Verbeeck,  Cis and Berghmans,  David and Barczynski,  Krzysztof},
  year = {2025},
  month = oct 
}

@ARTICLE{Dolliou_2024,
       author = {{Dolliou}, A. and {Parenti}, S. and {Bocchialini}, K.},
        title = "{Spectroscopic evidence of cool plasma in quiet Sun small-scale brightenings detected by HRIEUV on board Solar Orbiter}",
      journal = {\aap},
         year = 2024,
        month = aug,
       volume = {688},
          eid = {A77},
        pages = {A77},
          doi = {10.1051/0004-6361/202450439},
archivePrefix = {arXiv},
       eprint = {2405.10790},
 primaryClass = {astro-ph.SR},
       adsurl = {https://ui.adsabs.harvard.edu/abs/2024A&A...688A..77D}
}

@ARTICLE{Milanovic_2025,
       author = {{Milanovi{\'c}}, N. and {Peter}, H. and {Chitta}, L.~P. and {Young}, P.~R.},
        title = "{Thermal structuring of the quiet solar corona}",
      journal = {\aap},
         year = 2025,
        month = aug,
       volume = {700},
          eid = {A247},
        pages = {A247},
          doi = {10.1051/0004-6361/202452955},
       adsurl = {https://ui.adsabs.harvard.edu/abs/2025A&A...700A.247M}
}

@article{Hau_2025,
doi = {10.3847/1538-4357/ada76f},
url = {https://dx.doi.org/10.3847/1538-4357/ada76f},
year = {2025},
month = {feb},
publisher = {The American Astronomical Society},
volume = {981},
number = {1},
pages = {18},
author = {Hau, L.-N. and Chang, C.-K. and Lazar, M. and Poedts, S.},
title = {Boltzmann–Poisson Theory of Nonthermal Self-gravitating Gases, Cold Dark Matter, and Solar Atmosphere},
journal = {The Astrophysical Journal}
}

@article{Huang2023,
	author = {{Huang Z.} and {Teriaca, L.} and {Aznar Cuadrado, R.} and {Chitta, L. P.} and {Mandal, S.} and {Peter, H.} 
	and {Schühle, U.} and {Solanki, S. K.} and {Auchère, F.} and {Berghmans, D.} and {Buchlin, É.} and {Carlsson, M.} and {Fludra, A.} 
	and {Fredvik, T.} and {Giunta, A.} and {Grundy, T.} and {Hassler, D.} and {Parenti, S.} and {Plaschke, F.}},
	title = {Imaging and spectroscopic observations of extreme-ultraviolet brightenings using EUI and SPICE on board Solar Orbiter⋆},
	DOI= "10.1051/0004-6361/202345988",
	url= "https://doi.org/10.1051/0004-6361/202345988",
	journal = {aap},
	year = 2023,
	volume = 673,
	pages = "A82",
}

@article{Usmanov_2025,
doi = {10.3847/1538-4357/ae019c},
url = {https://doi.org/10.3847/1538-4357/ae019c},
year = {2025},
month = {oct},
publisher = {The American Astronomical Society},
volume = {993},
number = {1},
pages = {87},
author = {Usmanov, Arcadi V. and Chhiber, Rohit and Matthaeus, William H. and Roy, Sohom and Goldstein, Melvyn L.},
title = {A Unified Three-dimensional Magnetohydrodynamic Model of the Solar Corona, Solar Wind, and Global Heliosphere with Turbulence Transport},
journal = {The Astrophysical Journal},
}

@article{Mullan_2025,
doi = {10.3847/2515-5172/adc454},
url = {https://doi.org/10.3847/2515-5172/adc454},
year = {2025},
month = {mar},
publisher = {The American Astronomical Society},
volume = {9},
number = {3},
pages = {66},
author = {Mullan, D. J.},
title = {The Corona in the Quiet Sun: A Global Substrate Driven by Velocity Filtration?},
journal = {Research Notes of the AAS},
}

@article{Yalim_2023,
doi = {10.3847/1538-4357/acb151},
url = {https://doi.org/10.3847/1538-4357/acb151},
year = {2023},
month = {feb},
publisher = {The American Astronomical Society},
volume = {944},
number = {2},
pages = {119},
author = {Yalim, M. S. and Zank, G. P. and Asgari-Targhi, M.},
title = {Coronal Loop Heating by Nearly Incompressible Magnetohydrodynamic and Reduced Magnetohydrodynamic Turbulence Models},
journal = {The Astrophysical Journal},
}

@article{Ayaz2025a,
    author = {Ayaz, Syed and Zank, Gary P and Khan, Imran A and Rivera, Yeimy J and Shalchi, Andreas and Zhao, L -L},
    title = {Solar coronal heating: role of kinetic and inertial Alfvén waves in heating and charged particle acceleration},
    journal = {Monthly Notices of the Royal Astronomical Society},
    volume = {540},
    number = {4},
    pages = {3583-3595},
    year = {2025},
    month = {06},
    issn = {0035-8711},
    doi = {10.1093/mnras/staf952},
    url = {https://doi.org/10.1093/mnras/staf952},
    eprint = {https://academic.oup.com/mnras/article-pdf/540/4/3583/63470389/staf952.pdf},
}

@article{Bose2024,
  title = {Chromospheric and coronal heating in an active region plage by dissipation of currents from braiding},
  volume = {8},
  ISSN = {2397-3366},
  url = {http://dx.doi.org/10.1038/s41550-024-02241-8},
  DOI = {10.1038/s41550-024-02241-8},
  number = {6},
  journal = {Nature Astronomy},
  publisher = {Springer Science and Business Media LLC},
  author = {Bose,  Souvik and De Pontieu,  Bart and Hansteen,  Viggo and Sainz Dalda,  Alberto and Savage,  Sabrina and Winebarger,  Amy},
  year = {2024},
  month = apr,
  pages = {697–705}
}

@article{Wang2025,
  title = {Multi-scale energy release events in the quiet Sun: a possible source for coronal heating},
  volume = {12},
  ISSN = {2296-987X},
  url = {http://dx.doi.org/10.3389/fspas.2025.1536035},
  DOI = {10.3389/fspas.2025.1536035},
  journal = {Frontiers in Astronomy and Space Sciences},
  publisher = {Frontiers Media SA},
  author = {Wang,  Rui and Jiao,  Yiming and Zhao,  Xiaowei and Huang,  Chong},
  year = {2025},
  month = mar 
}

@article{Chitta2022,
	author = {{Chitta L. P.} and {Peter, H.} and {Parenti, S.}
	and {Berghmans, D.} and {Auchère, F.} and {Solanki, S. K.}
	and {Aznar Cuadrado, R.} and {Schühle, U.} and {Teriaca, L.}
	and {Mandal, S.} and {Barczynski, K.} and {Buchlin, É.}
	and {Harra, L.} and {Kraaikamp, E.} and {Long, D. M.}
	and {Rodriguez, L.} and {Schwanitz, C.} and {Smith, P. J.}
	and {Verbeeck, C.} and {Zhukov, A. N.} and {Liu, W.} and {Cheung, M. C. M.}},
	title = {Solar coronal heating from small-scale magnetic braids⋆},
	DOI= "10.1051/0004-6361/202244170",
	url= "https://doi.org/10.1051/0004-6361/202244170",
	journal = aap,
	year = 2022,
	volume = 667,
	pages = "A166",
}

@INCOLLECTION{Arregui2024,
       author = {{Arregui}, I{\~n}igo and {Van Doorsselaere}, Tom},
        title = "{Coronal heating}",
    booktitle = {Magnetohydrodynamic Processes in Solar Plasmas},
         year = 2024,
       editor = {{Srivastava}, Abhishek Kumar and {Goossens}, Marcel and {Arregui}, Inigo},
        pages = {415-450},
          doi = {10.1016/B978-0-32-395664-2.00015-3},
       adsurl = {https://ui.adsabs.harvard.edu/abs/2024mpsp.book..415A}
}

@article{Mason2023,
  title = {Coronal Heating as Determined by the Solar Flare Frequency Distribution Obtained by Aggregating Case Studies},
  volume = {948},
  ISSN = {1538-4357},
  url = {http://dx.doi.org/10.3847/1538-4357/accc89},
  DOI = {10.3847/1538-4357/accc89},
  number = {2},
  journal = {The Astrophysical Journal},
  publisher = {American Astronomical Society},
  author = {Mason,  James Paul and Werth,  Alexandra and West,  Colin G. and Youngblood,  Allison and Woodraska,  Donald L. and et al.
  },
  year = {2023},
  month = may,
  pages = {71}
}

@article{Dolliou2023,
  title = {Temperature of quiet Sun small scale brightenings observed by EUI on board Solar Orbiter: Evidence for a cooler component},
  volume = {671},
  ISSN = {1432-0746},
  url = {http://dx.doi.org/10.1051/0004-6361/202244914},
  DOI = {10.1051/0004-6361/202244914},
  journal = aap,
  publisher = {EDP Sciences},
  author = {Dolliou,  A. and Parenti,  S. and Auchère,  F. and Bocchialini,  K. and Pelouze,  G. and Antolin,  P. and Berghmans,  D. and Harra,  L. and Long,  D. M. and Sch\"{u}hle,  U. and Kraaikamp,  E. and Stegen,  K. and Verbeeck,  C. and Gissot,  S. and Aznar Cuadrado,  R. and Buchlin,  E. and Mierla,  M. and Teriaca,  L. and Zhukov,  A. N.},
  year = {2023},
  month = mar,
  pages = {A64}
}

@article{Bhatnagar2025,
  title = {Magnetic topology of quiet-Sun Ellerman bombs and associated ultraviolet brightenings},
  volume = {693},
  ISSN = {1432-0746},
  url = {http://dx.doi.org/10.1051/0004-6361/202452822},
  DOI = {10.1051/0004-6361/202452822},
  journal = aap,
  publisher = {EDP Sciences},
  author = {Bhatnagar,  Aditi and Prasad,  Avijeet and Rouppe van der Voort,  Luc and Nóbrega-Siverio,  Daniel and Joshi,  Jayant},
  year = {2025},
  month = jan,
  pages = {A221}
}

@article{Dolliou2025,
  title = {The nature of small-scale extreme ultraviolet solar brightenings investigated as impulsive heating of short loops in 1D hydrodynamics simulations},
  volume = {699},
  ISSN = {1432-0746},
  url = {http://dx.doi.org/10.1051/0004-6361/202554451},
  DOI = {10.1051/0004-6361/202554451},
  journal = aap,
  publisher = {EDP Sciences},
  author = {Dolliou,  A. and Klimchuk,  J. A. and Parenti,  S. and Bocchialini,  K.},
  year = {2025},
  month = jun,
  pages = {A61}
}

@article{Lim2025a,
	author = {{Lim D.} and {Van Doorsselaere, Tom} 
	and {Berghmans, David} and {Hayes, Laura A.} 
	and {Verbeeck, Cis} and {Narang, Nancy} 
	and {Dominique, Marie} and {Inglis, Andrew R.}},
	title = {Quasi-periodic pulsations in extreme-ultraviolet brightenings},
	DOI= "10.1051/0004-6361/202554587",
	url= "https://doi.org/10.1051/0004-6361/202554587",
	journal = aap,
	year = 2025,
	volume = 698,
	pages = "A65",
}

@ARTICLE{Solanki_1991,
       author = {{Solanki}, S.~K. and {Steiner}, O. and {Uitenbroeck}, H.},
        title = "{Two-dimensional models of the solar chromosphere. I - The CA II K line as a diagnostic: 1.5-D radiative transfer}",
      journal = {\aap},
         year = 1991,
        month = oct,
       volume = {250},
       number = {1},
        pages = {220-234},
       adsurl = {https://ui.adsabs.harvard.edu/abs/1991A&A...250..220S}
}

@article{Johnston_2025,
doi = {10.3847/1538-4357/ae08a2},
url = {https://doi.org/10.3847/1538-4357/ae08a2},
year = {2025},
month = {nov},
publisher = {The American Astronomical Society},
volume = {994},
number = {2},
pages = {139},
author = {Johnston, Craig D. and Daldorff,
Lars K. S. and Klimchuk, James A. and Mondal,
Shanwlee Sow and Barnes, Will T. and Leake, James E. and Reid,
Jack and Parker, Jacob D.},
title = {Self-Consistent Heating of the Magnetically Closed Solar Corona:
Generation of Nanoflares, Thermodynamic Response of the Plasma and
Observational Signatures},
journal = {The Astrophysical Journal},
}

@article{Mondal_2025,
doi = {10.3847/1538-4357/ae0cac},
url = {https://doi.org/10.3847/1538-4357/ae0cac},
year = {2025},
month = {nov},
publisher = {The American Astronomical Society},
volume = {994},
number = {1},
pages = {71},
author = {Sow Mondal, Shanwlee and Daldorff, Lars K. S.
and Klimchuk, James A. and Johnston, Craig. D.},
title = {Characterizing Nanoflare Energy and Frequency through Field Line Analysis},
journal = {The Astrophysical Journal},
}

@book{Meyer-Vernet_2007, place={Cambridge}, series={Cambridge Atmospheric and Space Science Series}, title={Basics of the Solar Wind}, publisher={Cambridge University Press}, author={Meyer-Vernet, Nicole}, year={2007}, collection={Cambridge Atmospheric and Space Science Series}}

@misc{parenti2025,
      title={Small EUV Brightenings in the Quiet Solar Atmosphere: New Insights from the Solar Orbiter Mission}, 
      author={Susanna Parenti},
      year={2025},
      eprint={2512.07286},
      archivePrefix={arXiv},
      primaryClass={astro-ph.SR},
      url={https://arxiv.org/abs/2512.07286}, 
}

@misc{Banik2026,
  doi = {10.48550/ARXIV.2601.03344},
  url = {https://arxiv.org/abs/2601.03344},
  author = {Banik,  Uddipan and Bhattacharjee,  Amitava},
  title = {Non-thermal particle acceleration in multi-species kinetic plasmas: universal power-law distribution functions and temperature inversion in the solar corona},
  publisher = {arXiv},
  year = {2026},
  copyright = {Creative Commons Attribution 4.0 International}
}

@book{LandiDeglInnocenti2019,
  title = {Elementi di Meccanica dei Fluidi,  Termodinamica e Fisica Statistica},
  ISBN = {9788847039919},
  url = {http://dx.doi.org/10.1007/978-88-470-3991-9},
  DOI = {10.1007/978-88-470-3991-9},
  publisher = {Springer Milan},
  author = {Landi Degl’Innocenti,  Egidio},
  year = {2019}
}

@book{cowan1998statistical,
  title={Statistical Data Analysis},
  author={Cowan, G.},
  isbn={9780198501558},
  lccn={98014554},
  series={Oxford science publications},
  url={https://books.google.it/books?id=ff8ZyW0nlJAC},
  year={1998},
  publisher={Clarendon Press}
}

@article{Palit2024,
  title = {X-ray view of dissipative warm corona in active galactic nuclei},
  volume = {690},
  ISSN = {1432-0746},
  url = {http://dx.doi.org/10.1051/0004-6361/202450111},
  DOI = {10.1051/0004-6361/202450111},
  journal = aap,
  publisher = {EDP Sciences},
  author = {Palit,  B. and Różańska,  A. and Petrucci,  P. O. and Gronkiewicz,  D. and Barnier,  S. and Bianchi,  S. and Ballantyne,  D. R. and Gianolli,  V. E. and Middei,  R. and Belmont,  R. and Ursini,  F.},
  year = {2024},
  month = oct,
  pages = {A308}
}

@article{BECKsuper,
title = {Superstatistics},
journal = {Physica A: Statistical Mechanics and its Applications},
volume = {322},
pages = {267-275},
year = {2003},
issn = {0378-4371},
doi = {https://doi.org/10.1016/S0378-4371(03)00019-0},
url = {https://www.sciencedirect.com/science/article/pii/S0378437103000190},
author = {C. Beck and E.G.D. Cohen},
}

@article{Bian2010,
	author = {{Bian, N. H.} and {Kontar, E. P.} and {Brown, J. C.}},
	title = {Parallel electric field generation by Alfv\'en wave turbulence},
	DOI= "10.1051/0004-6361/201014048",
	url= "https://doi.org/10.1051/0004-6361/201014048",
	journal = {A\&A},
	year = 2010,
	volume = 519,
	pages = "A114",
	month = "",
}

@article{Chandran_2010,
doi = {10.1088/0004-637X/720/1/503},
url = {https://doi.org/10.1088/0004-637X/720/1/503},
year = {2010},
month = {aug},
publisher = {The American Astronomical Society},
volume = {720},
number = {1},
pages = {503},
author = {Chandran, Benjamin D. G. and Li, Bo and Rogers, Barrett N. and Quataert, Eliot and Germaschewski, Kai},
title = {PERPENDICULAR ION HEATING BY LOW-FREQUENCY ALFVÉN-WAVE TURBULENCE IN THE SOLAR WIND},
journal = {The Astrophysical Journal},
}

@article{Bian2011,
	author = {{Bian, N. H.} and {Kontar, E. P.}},
	title = {Parallel electric field amplification by phase mixing of Alfven
          waves},
	DOI= "10.1051/0004-6361/201015385",
	url= "https://doi.org/10.1051/0004-6361/201015385",
	journal = {A\&A},
	year = 2011,
	volume = 527,
	pages = "A130",
	month = "",
}

@book{pollard1966celestial,
  author    = {Harry Pollard},
  title     = {Mathematical Introduction to Celestial Mechanics},
  series    = {Prentice-Hall Mathematics Series},
  publisher = {Prentice-Hall},
  address   = {Englewood Cliffs, NJ},
  year      = {1966},
  isbn      = {9780135613325}
}

@article{DelZanna2021,
  author = {Del Zanna, G. and Dere, K. P. and Young, P. R. and Landi, E. and Sutherland, R. S.},
  title = {CHIANTI—An atomic database for spectroscopic diagnostics of astrophysical plasmas},
  journal = {The Astrophysical Journal},
  volume = {909},
  pages = {38},
  year = {2021}
}

@article{Fabian2015,
author = {Fabian, A. C. et al.},
title = {Properties of AGN coronae in the NuSTAR era},
journal = {MNRAS},
volume = {451},
pages = {4375--4383},
year = {2015}
}

@article{Davis2019,
  author = {Davis, S. and Avaria, G. and Bora, B. and Jain, J. and Moreno, J. and Pavez, C. and Soto, L.},
  title = {Single-particle velocity distributions of collisionless, steady-state plasmas must follow superstatistics},
  journal = {Physical Review E},
  volume = {100},
  number = {2},
  pages = {023205},
  year = {2019},
  doi = {10.1103/PhysRevE.100.023205}
}

@article{Davis2020,
  author = {Davis, S.},
  title = {On the possible distributions of temperature in nonequilibrium steady states},
  journal = {Journal of Physics A: Mathematical and Theoretical},
  volume = {53},
  number = {4},
  pages = {045004},
  year = {2020},
  doi = {10.1088/1751-8121/ab5c1b}
}

@article{Tamburrini2025,
  author = {Tamburrini, Abiam and Davis, Sergio and Moya, Pablo S.},
  title = {Unifying Kappa Distribution Models for Non-Equilibrium Space Plasmas: A Superstatistical Approach Based on Moments},
  journal = {arXiv},
  eprint = {2509.07927},
  archivePrefix = {arXiv},
  primaryClass = {physics.plasm-ph},
  year = {2025},
  doi = {10.48550/arXiv.2509.07927}
}

@article{Chavanis2006,
  author  = {Chavanis, Pierre-Henri},
  title   = {Coarse-grained distributions and superstatistics},
  journal = {Physica A: Statistical Mechanics and its Applications},
  volume  = {359},
  pages   = {177--212},
  year    = {2006},
  doi     = {10.1016/j.physa.2005.06.043}
}

@article{Wilkins2015,
    author = {Wilkins, D. R. and Gallo, L. C.},
    title = {The Comptonization of accretion disc X-ray emission: consequences for X-ray reflection and the geometry of AGN coronae},
    journal = {Monthly Notices of the Royal Astronomical Society},
    volume = {448},
    number = {1},
    pages = {703-712},
    year = {2015},
    month = {02},
    issn = {0035-8711},
    doi = {10.1093/mnras/stu2524},
    url = {https://doi.org/10.1093/mnras/stu2524},
    eprint = {https://academic.oup.com/mnras/article-pdf/448/1/703/9376612/stu2524.pdf},
}

@ARTICLE{Haardt1991,
       author = {{Haardt}, F. and {Maraschi}, L.},
        title = "{A Two-Phase Model for the X-Ray Emission from Seyfert Galaxies}",
      journal = {\apjl},
         year = 1991,
        month = oct,
       volume = {380},
        pages = {L51},
          doi = {10.1086/186171},
       adsurl = {https://ui.adsabs.harvard.edu/abs/1991ApJ...380L..51H}
}

@article{Zhao_2025,
doi = {10.3847/1538-4357/ada35e},
url = {https://doi.org/10.3847/1538-4357/ada35e},
year = {2025},
month = {jan},
publisher = {The American Astronomical Society},
volume = {979},
number = {2},
pages = {201},
author = {Zhao, Yu and Li, Yan-Rong and Wang, Jian-Min},
title = {A New Interpretation for the Hot Corona in Active Galactic Nuclei},
journal = {The Astrophysical Journal},
}

@article{Che_2021,
doi = {10.3847/1538-4357/ac1fe7},
url = {https://doi.org/10.3847/1538-4357/ac1fe7},
year = {2021},
month = {nov},
publisher = {The American Astronomical Society},
volume = {921},
number = {2},
pages = {135},
author = {Che, H. and Zank, G. P. and Benz, A. O.},
title = {Ion Acceleration and the Development of a Power-law Energy Spectrum in Magnetic Reconnection},
journal = {The Astrophysical Journal},
}

@article{Hellinger2006,
author = {Hellinger, Petr and Trávníček, Pavel and Kasper, Justin C. and Lazarus, Alan J.},
title = {Solar wind proton temperature anisotropy: Linear theory and WIND/SWE observations},
journal = {Geophysical Research Letters},
volume = {33},
number = {9},
pages = {},
doi = {https://doi.org/10.1029/2006GL025925},
year = {2006}
}

@article{Gary2001,
  title = {Plasma Beta above a Solar Active Region: Rethinking the Paradigm},
  volume = {203},
  ISSN = {1573-093X},
  url = {http://dx.doi.org/10.1023/A:1012722021820},
  DOI = {10.1023/a:1012722021820},
  number = {1},
  journal = {Solar Physics},
  publisher = {Springer Science and Business Media LLC},
  author = {Gary,  G. Allen},
  year = {2001},
  month = oct,
  pages = {71–86}
}

@article{Lie-Svendsen_1999,
doi = {10.1086/307936},
url = {https://doi.org/10.1086/307936},
year = {1999},
month = {nov},
publisher = {},
volume = {525},
number = {2},
pages = {1056},
author = {Lie-Svendsen, {\O}ystein and Holzer, Thomas E. and Leer, Egil},
title = {Electron Heat Conduction in the Solar Transition Region: Validity of the Classical Description},
journal = {The Astrophysical Journal},
}

@article{Vocks2016,
  title = {Suprathermal electron distributions in the solar transition region},
  volume = {596},
  ISSN = {1432-0746},
  url = {http://dx.doi.org/10.1051/0004-6361/201629209},
  DOI = {10.1051/0004-6361/201629209},
  journal = {Astronomy \& Astrophysics},
  publisher = {EDP Sciences},
  author = {Vocks,  C. and Dzifčáková,  E. and Mann,  G.},
  year = {2016},
  month = Nov,
  pages = {A41}
}

@article{Hara_2008,
doi = {10.1086/588252},
url = {https://doi.org/10.1086/588252},
year = {2008},
month = {mar},
publisher = {},
volume = {678},
number = {1},
pages = {L67},
author = {Hara, Hirohisa and Watanabe, Tetsuya and Harra, Louise K. and Culhane, J. Leonard and Young, Peter R. and Mariska, John T. and Doschek, George A.},
title = {Coronal Plasma Motions near Footpoints of Active Region Loops Revealed from Spectroscopic Observations with Hinode EIS},
journal = {The Astrophysical Journal},
}

@article{Noci1981,
  title = {Siphon flows in the solar corona},
  volume = {69},
  ISSN = {1573-093X},
  url = {http://dx.doi.org/10.1007/BF00151256},
  DOI = {10.1007/bf00151256},
  number = {1},
  journal = {Solar Physics},
  publisher = {Springer Science and Business Media LLC},
  author = {Noci,  G.},
  year = {1981},
  month = Jan,
  pages = {63–76}
}

@article{Cargill1980,
  title = {Siphon flows in coronal loops: I. Adiabatic flow},
  volume = {65},
  ISSN = {1573-093X},
  url = {http://dx.doi.org/10.1007/BF00152793},
  DOI = {10.1007/bf00152793},
  number = {2},
  journal = {Solar Physics},
  publisher = {Springer Science and Business Media LLC},
  author = {Cargill,  P. J. and Priest,  E. R.},
  year = {1980},
  month = Mar,
  pages = {251–269}
}

@ARTICLE{Orlando1995a,
       author = {{Orlando}, S. and {Peres}, G. and {Serio}, S.},
        title = "{Models of stationary siphon flows in stratified, thermally conducting coronal loops. I. Regular solutions.}",
      journal = {\aap},
         year = 1995,
        month = feb,
       volume = {294},
        pages = {861-873},
       adsurl = {https://ui.adsabs.harvard.edu/abs/1995A&A...294..861O}
}

@ARTICLE{Orlando1995b,
       author = {{Orlando}, S. and {Peres}, G. and {Serio}, S.},
        title = "{Models of stationary siphon flows in stratified, thermally conducting coronal loops. II. Shocked solutions.}",
      journal = {\aap},
         year = 1995,
        month = aug,
       volume = {300},
        pages = {549},
       adsurl = {https://ui.adsabs.harvard.edu/abs/1995A&A...300..549O}
}

@article{Mason_2023,
doi = {10.3847/1538-4357/acac85},
url = {https://doi.org/10.3847/1538-4357/acac85},
year = {2023},
month = {jan},
publisher = {The American Astronomical Society},
volume = {943},
number = {2},
pages = {84},
author = {Mason, Emily I and Antiochos, Spiro K and Bradshaw, Stephen},
title = {The Equilibrium of Coronal Loops Near Separatrices},
journal = {The Astrophysical Journal},
}

@article{Reep_2024,
doi = {10.3847/1538-4357/ad3c3c},
url = {https://doi.org/10.3847/1538-4357/ad3c3c},
year = {2024},
month = {may},
publisher = {The American Astronomical Society},
volume = {967},
number = {1},
pages = {53},
author = {Reep, Jeffrey W. and Scott, Roger B. and Chhabra, Sherry and Unverferth, John and Knizhnik, Kalman J.},
title = {Mass Flows in Expanding Coronal Loops},
journal = {The Astrophysical Journal},
}

@article{Keppens2025,
  title = {Modeling multiphase plasma in the corona: prominences and rain},
  volume = {22},
  ISSN = {1614-4961},
  url = {http://dx.doi.org/10.1007/s41116-025-00043-2},
  DOI = {10.1007/s41116-025-00043-2},
  number = {1},
  journal = {Living Reviews in Solar Physics},
  publisher = {Springer Science and Business Media LLC},
  author = {Keppens,  Rony and Zhou,  Yuhao and Xia,  Chun},
  year = {2025},
  month = Dec 
}

@article{Verscharen2026,
  title = {Transport of Electrons in Tangled Magnetic Fields},
  volume = {222},
  ISSN = {1572-9672},
  url = {http://dx.doi.org/10.1007/s11214-026-01303-9},
  DOI = {10.1007/s11214-026-01303-9},
  number = {4},
  journal = {Space Science Reviews},
  publisher = {Springer Science and Business Media LLC},
  author = {Verscharen,  Daniel and Jeffrey,  Natasha and Artemyev,  Anton and Coburn,  Jesse T. and Kunz,  Matthew W. and Pezzi,  Oreste and Riquelme,  Mario and Svenningsson,  Ida and Wilson III,  Lynn B.},
  year = {2026},
  month = May 
}

@article{Kohl1997,
  title = {First Results from the Soho Ultraviolet Coronagraph Spectrometer},
  volume = {175},
  ISSN = {1573-093X},
  url = {http://dx.doi.org/10.1023/A:1004903206467},
  DOI = {10.1023/a:1004903206467},
  number = {2},
  journal = {Solar Physics},
  publisher = {Springer Science and Business Media LLC},
  author = {Kohl,  J. L. and Noci,  G. and Antonucci,  E. and Tondello,  G. and Huber,  M. C. E. and Gardner,  L. D. and Nicolosi,  P. and Strachan,  L. and Fineschi,  S. and Raymond,  J. C. and Romoli,  M. and Spadaro,  D. and Panasyuk,  A. and Siegmund,  O. H. W. and Benna,  C. and Ciaravella,  A. and Cranmer,  S. R. and Giordano,  S. and Karovska,  M. and Martin,  R. and Michels,  J. and Modigliani,  A. and Naletto,  G. and Pernechele,  C. and Poletto,  G. and Smith,  P. L.},
  year = {1997},
  month = Oct,
  pages = {613–644}
}

@article{Cranmer2009,
  title = {Coronal Holes},
  volume = {6},
  ISSN = {1614-4961},
  url = {http://dx.doi.org/10.12942/lrsp-2009-3},
  DOI = {10.12942/lrsp-2009-3},
  journal = {Living Reviews in Solar Physics},
  publisher = {Springer Science and Business Media LLC},
  author = {Cranmer,  Steven R.},
  year = {2009}
}

@article{Cranmer_2020,
doi = {10.3847/1538-4357/abab04},
url = {https://doi.org/10.3847/1538-4357/abab04},
year = {2020},
month = {sep},
publisher = {The American Astronomical Society},
volume = {900},
number = {2},
pages = {105},
author = {Cranmer, Steven R.},
title = {Heating Rates for Protons and Electrons in Polar Coronal Holes: Empirical Constraints from the Ultraviolet Coronagraph Spectrometer},
journal = {The Astrophysical Journal},
}

@article{Zhu_2023,
doi = {10.3847/1538-4357/acc187},
url = {https://doi.org/10.3847/1538-4357/acc187},
year = {2023},
month = {may},
publisher = {The American Astronomical Society},
volume = {948},
number = {2},
pages = {90},
author = {Zhu, Yingjie and Szente, Judit and Landi, Enrico},
title = {Estimating Ion Temperatures at the Polar Coronal Hole Boundary},
journal = {The Astrophysical Journal},
}

@article{Gabriel1976,
    author = {Gabriel, A. H.},
    title = {A Discussion on the physics of the solar atmosphere - A magnetic model of the solar transition region},
    journal = {Philosophical Transactions of the Royal Society of London, Series A: Mathematical and Physical Sciences},
    volume = {281},
    number = {1304},
    pages = {339-352},
    year = {1976},
    month = {05},
    issn = {0080-4614},
    doi = {10.1098/rsta.1976.0031},
    url = {https://doi.org/10.1098/rsta.1976.0031},
    eprint = {https://royalsocietypublishing.org/rsta/article-pdf/281/1304/339/269658/rsta.1976.0031.pdf},
}

@article{Dowdy1986,
  title = {On the magnetic structure of the quiet transition region},
  volume = {105},
  ISSN = {1573-093X},
  url = {http://dx.doi.org/10.1007/BF00156374},
  DOI = {10.1007/bf00156374},
  number = {1},
  journal = {Solar Physics},
  publisher = {Springer Science and Business Media LLC},
  author = {Dowdy,  James F. and Rabin,  Douglas and Moore,  Ronald L.},
  year = {1986},
  month = May,
  pages = {35–45}
}

@article{Peter2001,
	author = {{Peter, H.}},
	title = {On the nature of the transition region from the chromosphere to the
    corona of the Sun},
	DOI= "10.1051/0004-6361:20010697",
	url= "https://doi.org/10.1051/0004-6361:20010697",
	journal = {A\&A},
	year = 2001,
	volume = 374,
	number = 3,
	pages = "1108-1120",
}

@article{Wiegelmann2014,
  title = {The magnetic field in the solar atmosphere},
  volume = {22},
  ISSN = {1432-0754},
  url = {http://dx.doi.org/10.1007/s00159-014-0078-7},
  DOI = {10.1007/s00159-014-0078-7},
  number = {1},
  journal = {The Astronomy and Astrophysics Review},
  publisher = {Springer Science and Business Media LLC},
  author = {Wiegelmann,  Thomas and Thalmann,  Julia K. and Solanki,  Sami K.},
  year = {2014},
  month = Oct 
}

@article{Judge_2021,
doi = {10.3847/1538-4357/abf8ad},
url = {https://doi.org/10.3847/1538-4357/abf8ad},
year = {2021},
month = {jun},
publisher = {The American Astronomical Society},
volume = {914},
number = {1},
pages = {70},
author = {Judge, Philip},
title = {Magnetic Connections across the Chromosphere–Corona Transition Region},
journal = {The Astrophysical Journal},
}

@article{Vocks2001,
author = {Vocks, C. and Marsch, E.},
title = {A semi-kinetic model of wave-ion interaction in the solar corona},
journal = {Geophysical Research Letters},
volume = {28},
number = {10},
pages = {1917-1920},
doi = {https://doi.org/10.1029/2000GL012764},
url = {https://agupubs.onlinelibrary.wiley.com/doi/abs/10.1029/2000GL012764},
eprint = {https://agupubs.onlinelibrary.wiley.com/doi/pdf/10.1029/2000GL012764},
year = {2001}
}

@article{Vocks_2002a,
doi = {10.1086/338884},
url = {https://doi.org/10.1086/338884},
year = {2002},
month = {apr},
publisher = {},
volume = {568},
number = {2},
pages = {1017},
author = {Vocks, C.},
title = {A Kinetic Model for Ions in the Solar Corona Including Wave-Particle Interactions and Coulomb Collisions},
journal = {The Astrophysical Journal},
}

@article{Vocks_2002b,
doi = {10.1086/338885},
url = {https://doi.org/10.1086/338885},
year = {2002},
month = {apr},
publisher = {},
volume = {568},
number = {2},
pages = {1030},
author = {Vocks, C. and Marsch, E.},
title = {Kinetic Results for Ions in the Solar Corona with Wave-Particle Interactions and Coulomb Collisions},
journal = {The Astrophysical Journal},
}

@INPROCEEDINGS{Vocks2021,
       author = {{Vocks}, Christian},
        title = "{Kinetic Models of Wave-Electron Interaction in the Solar Corona and Wind}",
    booktitle = {Kappa Distributions; From Observational Evidences via Controversial Predictions to a Consistent Theory of Nonequilibrium Plasmas},
         year = 2021,
       editor = {{Lazar}, Marian and {Fichtner}, Horst},
       series = {Astrophysics and Space Science Library},
       volume = {464},
        month = jan,
        pages = {125-143},
          doi = {10.1007/978-3-030-82623-9_7},
       adsurl = {https://ui.adsabs.harvard.edu/abs/2021ASSL..464..125V}
}

@article{Marsch2006,
  title = {Kinetic Physics of the Solar Corona and Solar Wind},
  volume = {3},
  ISSN = {1614-4961},
  url = {http://dx.doi.org/10.12942/lrsp-2006-1},
  DOI = {10.12942/lrsp-2006-1},
  journal = {Living Reviews in Solar Physics},
  publisher = {Springer Science and Business Media LLC},
  author = {Marsch,  Eckart},
  year = {2006}
}

@Article{Marsch2003,
AUTHOR = {Marsch, E. and Vocks, C. and Tu, C.-Y.},
TITLE = {On ion-cyclotron-resonance heating of the corona and solar wind},
JOURNAL = {Nonlinear Processes in Geophysics},
VOLUME = {10},
YEAR = {2003},
NUMBER = {1/2},
PAGES = {101--112},
URL = {https://npg.copernicus.org/articles/10/101/2003/},
DOI = {10.5194/npg-10-101-2003}
}

@article{Dudk2017,
  title = {Nonequilibrium Processes in the Solar Corona,  Transition Region,  Flares,  and Solar Wind (Invited Review)},
  volume = {292},
  ISSN = {1573-093X},
  url = {http://dx.doi.org/10.1007/s11207-017-1125-0},
  DOI = {10.1007/s11207-017-1125-0},
  number = {8},
  journal = {Solar Physics},
  publisher = {Springer Science and Business Media LLC},
  author = {Dudík,  Jaroslav and Dzifčáková,  Elena and Meyer-Vernet,  Nicole and Del Zanna,  Giulio and Young,  Peter R. and Giunta,  Alessandra and Sylwester,  Barbara and Sylwester,  Janusz and Oka,  Mitsuo and Mason,  Helen E. and Vocks,  Christian and Matteini,  Lorenzo and Krucker,  S\"{a}m and Williams,  David R. and Mackovjak,  Simon},
  year = {2017},
  month = {July}, 
}

@article{Cranmer2021,
author = {Cranmer, Steven R. and Schiff, Avery J.},
title = {Electron Heat Flux in the Solar Wind: Generalized Approaches to Fluid Transport With a Variety of Skewed Velocity Distributions},
journal = {Journal of Geophysical Research: Space Physics},
volume = {126},
number = {10},
pages = {e2021JA029666},
doi = {https://doi.org/10.1029/2021JA029666},
url = {https://agupubs.onlinelibrary.wiley.com/doi/abs/10.1029/2021JA029666},
eprint = {https://agupubs.onlinelibrary.wiley.com/doi/pdf/10.1029/2021JA029666},
note = {e2021JA029666 2021JA029666},
year = {2021}
}

@article{Ayaz2025b,
  title = {A study of particle acceleration,  heating,  power deposition,  and the damping length of kinetic Alfvén waves in non-Maxwellian coronal plasma},
  volume = {694},
  ISSN = {1432-0746},
  url = {http://dx.doi.org/10.1051/0004-6361/202452376},
  DOI = {10.1051/0004-6361/202452376},
  journal = {Astronomy \& Astrophysics},
  publisher = {EDP Sciences},
  author = {Ayaz,  S. and Zank,  G. P. and Khan,  I. A. and Li,  G. and Rivera,  Y. J.},
  year = {2025},
  month = Feb,
  pages = {A23}
}

@article{Pierrard2014,
author = {Pierrard, V. and Pieters, M.},
title = {Coronal heating and solar wind acceleration for electrons, protons, and minor ions obtained from kinetic models based on kappa distributions},
journal = {Journal of Geophysical Research: Space Physics},
volume = {119},
number = {12},
pages = {9441-9455},
doi = {https://doi.org/10.1002/2014JA020678},
url = {https://agupubs.onlinelibrary.wiley.com/doi/abs/10.1002/2014JA020678},
eprint = {https://agupubs.onlinelibrary.wiley.com/doi/pdf/10.1002/2014JA020678},
year = {2014}
}

@Article{Pierrard2023,
AUTHOR = {Pierrard, Viviane and Péters de Bonhome, Maximilien and Halekas, Jasper and Audoor, Charline and Whittlesey, Phyllis and Livi, Roberto},
TITLE = {Exospheric Solar Wind Model Based on Regularized Kappa Distributions for the Electrons Constrained by Parker Solar Probe Observations},
JOURNAL = {Plasma},
VOLUME = {6},
YEAR = {2023},
NUMBER = {3},
PAGES = {518--540},
URL = {https://www.mdpi.com/2571-6182/6/3/36},
ISSN = {2571-6182},
DOI = {10.3390/plasma6030036}
}

@article{Cranmer_2000,
doi = {10.1086/308620},
url = {https://doi.org/10.1086/308620},
year = {2000},
month = {apr},
publisher = {},
volume = {532},
number = {2},
pages = {1197},
author = {Cranmer, Steven R.},
title = {Ion Cyclotron Wave Dissipation in the Solar Corona: The Summed
Effect of More than 2000 Ion
Species},
journal = {The Astrophysical Journal},
}

@book{huang1987statistical,
  title={Statistical mechanics},
  author={Huang, K.},
  isbn={9780471815181},
  lccn={86032466},
  url={https://books.google.it/books?id=M8PvAAAAMAAJ},
  year={1987},
  publisher={Wiley}
}

@article{Dolliou2026,
  title = {Small-scale impulsive extreme-UV emission enhancements along network loops},
  volume = {709},
  ISSN = {1432-0746},
  url = {http://dx.doi.org/10.1051/0004-6361/202558068},
  DOI = {10.1051/0004-6361/202558068},
  journal = {Astronomy \& Astrophysics},
  publisher = {EDP Sciences},
  author = {Dolliou,  A. and Peter,  H. and Mandal,  S. and Chitta,  L. P. and Teriaca,  L. and Chen,  Y. and Calchetti,  D.},
  year = {2026},
  month = May,
  pages = {A71}
}

@article{Pierrard2011,
  title = {Evolution of the Electron Distribution Function in the Whistler Wave Turbulence of the Solar Wind},
  volume = {269},
  ISSN = {1573-093X},
  url = {http://dx.doi.org/10.1007/s11207-010-9700-7},
  DOI = {10.1007/s11207-010-9700-7},
  number = {2},
  journal = {Solar Physics},
  publisher = {Springer Science and Business Media LLC},
  author = {Pierrard,  V. and Lazar,  M. and Schlickeiser,  R.},
  year = {2011},
  month = jan,
  pages = {421–438}
}

@article{PtersdeBonhome2025,
  title = {A kinetic model of solar wind acceleration driven by ambipolar electric potential and velocity-space diffusion},
  volume = {697},
  ISSN = {1432-0746},
  url = {http://dx.doi.org/10.1051/0004-6361/202554250},
  DOI = {10.1051/0004-6361/202554250},
  journal = {Astronomy \& Astrophysics},
  publisher = {EDP Sciences},
  author = {Péters de Bonhome,  M. and Pierrard,  V. and Bacchini,  F.},
  year = {2025},
  month = may,
  pages = {A104}
}

@article{Vinogradov2026,
author = {Vinogradov, A. and Lazar, M. and Zouganelis, I. and Pierrard, V. and Poedts, S.},
title = {Kinetic-Based Macro-Modeling of the Solar Wind at Large Heliocentric Distances: Kappa Electrons at the Exobase},
journal = {Journal of Geophysical Research: Space Physics},
volume = {131},
number = {2},
pages = {e2025JA034770},
doi = {https://doi.org/10.1029/2025JA034770},
url = {https://agupubs.onlinelibrary.wiley.com/doi/abs/10.1029/2025JA034770},
eprint = {https://agupubs.onlinelibrary.wiley.com/doi/pdf/10.1029/2025JA034770},
note = {e2025JA034770 2025JA034770},
year = {2026}
}

@ARTICLE{Feldman2003,
       author = {{Feldman}, U. and {Widing}, K.~G.},
        title = "{Elemental Abundances in the Solar Upper Atmosphere Derived by Spectroscopic Means}",
      journal = {\ssr},
         year = 2003,
        month = jul,
       volume = {107},
       number = {3},
        pages = {665-720},
          doi = {10.1023/A:1026103726147},
       adsurl = {https://ui.adsabs.harvard.edu/abs/2003SSRv..107..665F}
}

@article{Ayaz2026a,
author = {Ayaz, S. and Zank, Gary P. and Khan, Imran A. and Shalchi, Andreas and Zhao, L.-L. and Rivera, Yeimy J.},
title = {Kinetic Modeling of Inertial Alfvén Waves in the Solar Corona: Implications for Heating and Particle Acceleration},
journal = {Journal of Geophysical Research: Space Physics},
volume = {131},
number = {1},
pages = {e2025JA034989},
doi = {https://doi.org/10.1029/2025JA034989},
url = {https://agupubs.onlinelibrary.wiley.com/doi/abs/10.1029/2025JA034989},
eprint = {https://agupubs.onlinelibrary.wiley.com/doi/pdf/10.1029/2025JA034989},
note = {e2025JA034989 2025JA034989},
year = {2026}
}

@article{Ayaz2026b,
doi = {10.3847/1538-4357/ae77f7},
url = {https://doi.org/10.3847/1538-4357/ae77f7},
year = {2026},
month = {jul},
publisher = {The American Astronomical Society},
volume = {1005},
number = {2},
pages = {133},
author = {Ayaz, Syed and Zank, Gary P. and Zhao, L.-L. and Rivera, Yeimy J. and Khan, Imran A.},
title = {Parker Solar Probe Observations of Dust Mass and Charge Densities and Their Impact on Kinetic Alfvén Wave Dynamics in Solar Coronal Heating},
journal = {The Astrophysical Journal},
}

@article{Ayaz_2024,
doi = {10.3847/1538-4357/ad5bdc},
url = {https://doi.org/10.3847/1538-4357/ad5bdc},
year = {2024},
month = {jul},
publisher = {The American Astronomical Society},
volume = {970},
number = {2},
pages = {140},
author = {Ayaz, Syed and Li, Gang and Khan, Imran A.},
title = {Solar Coronal Heating by Kinetic Alfvén Waves},
journal = {The Astrophysical Journal},
}

@ARTICLE{Verscharen2022,
    
AUTHOR={Verscharen, Daniel  and Chandran, B. D. G.  and Boella, E.  and Halekas, J.  and Innocenti, M. E.  and Jagarlamudi, V. K.  and Micera, A.  and Pierrard, V.  and Štverák, S. and Vasko, I. Y.  and Velli, M.  and Whittlesey, P. L.}, 
           
TITLE={Electron-Driven Instabilities in the Solar Wind},
          
JOURNAL={Frontiers in Astronomy and Space Sciences},
          
VOLUME={Volume 9 - 2022},
  
YEAR={2022},
  
URL={https://www.frontiersin.org/journals/astronomy-and-space-sciences/articles/10.3389/fspas.2022.951628},
  
DOI={10.3389/fspas.2022.951628},
  
ISSN={2296-987X},
}

@article{Jeong_2022,
doi = {10.3847/2041-8213/ac4dff},
url = {https://doi.org/10.3847/2041-8213/ac4dff},
year = {2022},
month = {feb},
publisher = {The American Astronomical Society},
volume = {926},
number = {2},
pages = {L26},
author = {Jeong, Seong-Yeop and Abraham, Joel B. and Verscharen, Daniel and Berčič, Laura and Stansby, 
David and Nicolaou, Georgios and Owen, Christopher J. and Wicks, Robert T. and Fazakerley, Andrew N. and Agudelo Rueda,
Jeffersson A. and Bakrania, Mayur},
title = {The Stability of the Electron Strahl against the Oblique Fast-magnetosonic/Whistler Instability in the Inner Heliosphere},
journal = {The Astrophysical Journal Letters},
}

@article{Opie_2022,
doi = {10.3847/1538-4357/ac982f},
url = {https://doi.org/10.3847/1538-4357/ac982f},
year = {2022},
month = {dec},
publisher = {The American Astronomical Society},
volume = {941},
number = {2},
pages = {176},
author = {Opie, Simon and Verscharen, Daniel and Chen, Christopher H. K. and Owen, Christopher J. and Isenberg, Philip A.},
title = {Conditions for Proton Temperature Anisotropy to Drive Instabilities in the Solar Wind},
journal = {The Astrophysical Journal},
}

@article{Stverak2008,
author = {Štverák, Stepan and Travnicek, Pavel and Maksimovic, Milan and Marsch, Eckart and Fazakerley, Andrew N. and Scime, Earl E.},
title = {Electron temperature anisotropy constraints in the solar wind},
journal = {Journal of Geophysical Research: Space Physics},
volume = {113},
number = {A3},
pages = {},
doi = {https://doi.org/10.1029/2007JA012733},
url = {https://agupubs.onlinelibrary.wiley.com/doi/abs/10.1029/2007JA012733},
eprint = {https://agupubs.onlinelibrary.wiley.com/doi/pdf/10.1029/2007JA012733},
year = {2008}
}

@article{Coburn_2024,
doi = {10.3847/1538-4357/ad1329},
url = {https://doi.org/10.3847/1538-4357/ad1329},
year = {2024},
month = {mar},
publisher = {The American Astronomical Society},
volume = {964},
number = {1},
pages = {100},
author = {Coburn, Jesse T. and Verscharen, Daniel and Owen, Christopher J. and Maksimovic, Milan and Horbury, Timothy S.
and Chen, Christopher H. K. and Guo, Fan and Fu, Xiangrong and Liu, Jingting and Abraham, Joel B. and Nicolaou, Georgios and
Innocenti, Maria Elena and Micera, Alfredo and Jagarlamudi, Vamsee Krishna},
title = {The Regulation of the Solar Wind Electron Heat Flux by Wave–Particle Interactions},
journal = {The Astrophysical Journal},
}

@article{Jeong_2022b,
doi = {10.3847/1538-4357/ac4805},
url = {https://doi.org/10.3847/1538-4357/ac4805},
year = {2022},
month = {mar},
publisher = {The American Astronomical Society},
volume = {927},
number = {2},
pages = {162},
author = {Jeong, Seong-Yeop and Verscharen, Daniel and Vocks, Christian and Abraham, Joel B. and Owen, Christopher J. and Wicks, Robert T.
 and Fazakerley, Andrew N. and Stansby, David and Berčič, Laura and Nicolaou, Georgios and Agudelo Rueda, Jeffersson A. and Bakrania, Mayur},
title = {The Kinetic Expansion of Solar-wind Electrons: Transport Theory and Predictions for the Very Inner Heliosphere},
journal = {The Astrophysical Journal},
}

\appendix

\section{Derivation of Eq.~\eqref{lossprobability}}
\label{appendix_lossprobability}

The expression in Eq.~\eqref{lossprobability} can be motivated by a standard kinetic-theory argument \citep{LandiDeglInnocenti2019}. We consider a suprathermal test particle generated by a heating event at temperature $T$ propagating through a background plasma of number density $n_c(z)$. Within our coarse-grained description, let $\sigma(z)$ denote an effective cross section associated with the thermalization of suprathermal particles into the central Maxwellian core.

Over an infinitesimal displacement $dz$, the test particle sweeps a volume
\begin{equation}
dV = \sigma(z)  \,dz    \,,
\end{equation}
which contains a random number $N$ of statistically independent targets. The random variable $N$ therefore follows a Poisson distribution with mean
\begin{equation}
\mu(z) = \langle N \rangle = n_c(z)  \,\sigma(z)  \,dz  \,.
\end{equation}
where $n_c(z)$ is local density of the plasma given by Eq. \eqref{densitycollisions}.

The probability that a collision (i.e., a thermalization event) occurs within the interval $dz$ is given by the probability that at least one target is present in the swept volume,
\begin{equation}
P_{\rm loss}(z,dz) = P(N \ge 1)  \,.
\end{equation}

A Poisson-distributed random variable with mean $\mu(z)$ is described by \citep{cowan1998statistical}
\begin{equation}
P(N) = \frac{\mu^N(z)}{N!} \, e^{-\mu(z)}  \,.
\end{equation}
Hence, the probability of at least one encounter is
\begin{equation}
P(N \ge 1) = 1 - P(N=0) = 1 - e^{-\mu(z)} \,.
\end{equation}

In the limit of an infinitesimal path length $dz \to 0$, the mean number of encounters satisfies $\mu(z) \ll 1$, and the exponential can be expanded to first order,
\begin{equation}
e^{-\mu(z)} \simeq 1 - \mu(z)  \,.
\end{equation}
Consequently, the collision probability reduces to
\begin{equation}
P_{\rm loss}(z,dz) \simeq \mu(z) = n_c(z) \,\sigma(z) \,dz  \,.
\end{equation}

Using the standard kinetic-theory definition of the mean free path \citep{nicholson1983introduction},
\begin{equation}
\lambda(z)
\equiv
\frac{1}{n_c(z)  \,\sigma(z)} \,,   
\label{Hotmeanfreepathdef}
\end{equation}
the local thermalization probability can finally be written in compact form as
\begin{equation}
P_{\rm loss}(z,dz) = \frac{dz}{\lambda(z)}  \,.
\label{lossprobability_kinetic}
\end{equation}

Equation~\eqref{lossprobability_kinetic} is the analogue of the standard kinetic-theory expression for the mean free path \citep{LandiDeglInnocenti2019} and provides a microscopic justification for the Markovian loss model adopted in Eq.~\eqref{lossprobability}. 

Now an explicit expression for $\lambda$ is required. For Coulomb interactions, the mean free path of suprathermal particles with temperature $T$ can be estimated as \citep{nicholson1983introduction}
\begin{equation}\label{lambdaH_appendix}
\lambda(T,z)=\frac{(3k_B\, T)^2}{4\pi e^4 n_c(z)\,\ln\Lambda} \, .
\end{equation}
where the Coulomb logarithm is defined as
\begin{equation}
\ln{\Lambda} = \ln\!\left(\frac{b_{\mathrm{max}}}{b_{\mathrm{min}}}\right)  \,.
\end{equation}

The minimum impact parameter $b_{\mathrm{min}}$ is obtained by equating the average kinetic energy of the suprathermal particles to the electrostatic interaction energy,
\begin{equation}
\frac{3}{2}k_B\, T = \frac{e^2}{b_{\mathrm{min}}}   \,,
\end{equation}
which gives
\begin{equation}
b_{\mathrm{min}} = \frac{2e^2}{3k_B\, T}  \,.
\end{equation}

The maximum impact parameter $b_{\mathrm{max}}$ is determined by plasma screening and can be approximated by the local Debye length,
\begin{equation}
b_{\mathrm{max}} = \lambda_D(z)
=
\sqrt{\frac{k_B\, T_c(z)}{4\pi\, e^2\, n_c(z)}}   \,,
\end{equation}
where $T_c(z)$ is the local temperature of the plasma given by Eq.~\eqref{temperaturecollisions}.

The Coulomb logarithm then becomes
\begin{equation}
\ln{\Lambda}
=
\ln\!\left(
\frac{3\,k_B\, T}{2\,e^2}
\sqrt{\frac{k_B\, T_c(z)}{4\pi\, e^2\, n_c(z)}}
\right)  \,.
\end{equation}

\section{Numerical scheme for Eqs. \eqref{C_equation}, \eqref{A_diffusive} and \eqref{TM_temperaturecore}}
\label{AppendixNumerics}

This Appendix describes the numerical implementation used to solve the coupled system formed by
Eqs. \eqref{C_equation},~\eqref{A_diffusive} and \eqref{TM_temperaturecore}. Temperature integrals are discretized using a trapezoidal rule on a fixed temperature grid, while the spatial integration in $z$ is performed with a stiff, variable-step implicit solver (MATLAB \texttt{ode15s}). This choice is motivated by the strongly stiff relaxation term $dw/dz=-w/\lambda$, since the mean free paths $\lambda(T,z)$ may become much smaller than the typical spatial step required to resolve macroscopic profiles.

\subsection{Temperature discretization (trapezoidal rule)}

The temperature interval $[T_0,T_{\max}]$ is discretized using $N_T$ nodes
\begin{equation}
T_j = T_0 + (j-1)\delta T,
\qquad j=1,\dots,N_T ,
\end{equation}
where
\begin{equation}
\delta T = \frac{T_{\max}-T_0}{N_T-1}.
\end{equation}

The trapezoidal weights are
\begin{equation}
\omega_1 = \omega_{N_T} = \frac{\delta T}{2},
\qquad
\omega_j = \delta T \quad (2 \le j \le N_T-1).
\end{equation}

Any temperature integral of the form
\begin{equation}
\int_{T_0}^{T_{\max}} \gamma(T)\,F(T)\,dT
\end{equation}
is approximated as
\begin{equation}
\int_{T_0}^{T_{\max}} \gamma(T)\,F(T)\,dT
\approx
\sum_{j=1}^{N_T} \omega_j\,\gamma(T_j)\,F(T_j).
\end{equation}

We denote
\begin{equation}
w_j(z)\equiv w(T_j,z).
\end{equation}

\subsection{Macroscopic quantities}

Using the trapezoidal rule, the halo contributions to the density and pressure are computed as
\begin{equation}
\begin{aligned}
S_0(z)
&=
\int_{T_0}^{T_{\max}}
\gamma(T)\,w(T,z)\,
e^{-m\,g\,z/(k_B\,T)}\,dT
\\
&\approx
\sum_{j=1}^{N_T}
\omega_j\,\gamma(T_j)\,
w_j(z)\,
e^{-m\,g\,z/(k_B\,T_j)} ,
\end{aligned}
\end{equation}
\begin{equation}
\begin{aligned}
S_1(z)
&=
\int_{T_0}^{T_{\max}}
T\,\gamma(T)\,w(T,z)\,
e^{-m\,g\,z/(k_B\,T)}\,dT
\\
&\approx
\sum_{j=1}^{N_T}
\omega_j\,T_j\,\gamma(T_j)\,
w_j(z)\,
e^{-m\,g\,z/(k_B\,T_j)} .
\end{aligned}
\end{equation}

The density is
\begin{equation}\label{trapezoidaldensity}
\begin{aligned}
n_c(z)
=
n_0
\Big[
A_t\,S_0(z)
+
C(z)\,
e^{-m\,g\,z/(k_B\,T_M(z))}
\Big] .
\end{aligned}
\end{equation}

The pressure is
\begin{equation}\label{trapezoidapressure}
\begin{aligned}
p_c(z)
=
n_0\, k_B
\Big[
A_t\,S_1(z)
+
C(z)\,T_M(z)\,
e^{-m\,g\,z/(k_B\,T_M(z))}
\Big] .
\end{aligned}
\end{equation}

The local temperature follows from
\begin{equation}\label{trapezoidaltemperature}
T_c(z)=\frac{p_c(z)}{k_B\,n_c(z)}.
\end{equation}

\subsection{Mean free path}

For each temperature node $T_j$ we compute
\begin{equation}
\ln\Lambda_j(z)
=
\ln\!\left(
\frac{3\,k_B\, T_j}{2\,e^2}
\sqrt{
\frac{k_B\, T_c(z)}
{4\pi\, e^2\, n_c(z)}
}
\right),
\end{equation}
\begin{equation}
\lambda_j(z)
=
\frac{(3\,k_B\, T_j)^2}
{4\pi\, e^4\,n_c(z)\,\ln\Lambda_j(z)} .
\end{equation}

The evolution equation for each $w_j$ is
\begin{equation}
\frac{dw_j}{dz}
=
-\frac{w_j}{\lambda_j}.
\end{equation}

\subsection{Evaluation of the $C(z)$ and $T_M(z)$ equations}

The integral appearing in Eq.~\eqref{C_equation} is discretized as
\begin{equation}
\begin{aligned}
I_C(z)
&=
\int_{T_0}^{T_{\max}}
\gamma(T)\,
\frac{w(T,z)}{\lambda(T,z)}
e^{-m\,g\,z/(k_B\,T)}\,dT
\\
&\approx
\sum_{j=1}^{N_T}
\omega_j\,\gamma(T_j)\,
\frac{w_j(z)}{\lambda_j(z)}
e^{-m\,g\,z/(k_B\,T_j)} .
\end{aligned}
\end{equation}

The two integrals appearing in Eq.~\eqref{TM_temperaturecore} are discretized as
\begin{equation}
\begin{aligned}
I_1(z)
&=
\int_{T_0}^{T_{\max}}
T\,\gamma(T)\,
\frac{w(T,z)}{\lambda(T,z)}
e^{-m\,g\,z/(k_B\,T)}\,dT
\\
&\approx
\sum_{j=1}^{N_T}
\omega_j\,T_j\,\gamma(T_j)\,
\frac{w_j(z)}{\lambda_j(z)}
e^{-m\,g\,z/(k_B\,T_j)} ,
\end{aligned}
\end{equation}
\begin{equation}
\begin{aligned}
I_0(z)
&=
\int_{T_0}^{T_{\max}}
\gamma(T)\,
\frac{w(T,z)}{\lambda(T,z)}
e^{-m\,g\,z/(k_B\,T)}\,dT
\\
&\approx
\sum_{j=1}^{N_T}
\omega_j\,\gamma(T_j)\,
\frac{w_j(z)}{\lambda_j(z)}
e^{-m\,g\,z/(k_B\,T_j)} .
\end{aligned}
\end{equation}

\subsection{Spatial integration}

Let
\begin{equation}
\mathbf{y}(z)
=
\left(
w_1(z),\dots,w_{N_T}(z),C(z),T_M(z)
\right)^{\mathsf{T}} .
\end{equation}

The system can be written in compact form as
\begin{equation}
\frac{d\mathbf{y}}{dz}
=
\mathbf{F}(z,\mathbf{y}) .
\end{equation}

The first $N_T$ components are
\begin{equation}
F_j(z,\mathbf{y})
=
-\frac{w_j(z)}{\lambda_j(z)},
\qquad
j=1,\dots,N_T .
\end{equation}

The $(N_T+1)$-th component corresponds to the evolution of the core coefficient $C(z)$:
\begin{equation}
\frac{dC}{dz}
=
F_{N_T+1}(z,\mathbf{y}),
\end{equation}
with
\begin{equation}
F_{N_T+1}(z,\mathbf{y})
=
A_t\,
e^{\frac{m\,g\,z}{k_B\, T_M(z)}}\,
I_C(z).
\end{equation}

The last component corresponds to the evolution of the core temperature:
\begin{equation}
\frac{dT_M}{dz}
=
F_{N_T+2}(z,\mathbf{y}),
\end{equation}
with
\begin{equation}
F_{N_T+2}(z,\mathbf{y})
=
\frac{
A_t\left[I_1(z)-T_M(z)\,I_0(z)\right]
}{
C(z)\,
e^{-m\,g\,z/(k_B\,T_M(z))}
\left(
1+\frac{m\,g\,z}{k_B\, T_M(z)}
\right)
}.
\end{equation}

The system is integrated on $z\in[0,1]$ using MATLAB \texttt{ode15s}. At each integration step the macroscopic quantities $n_c(z)$, $p_c(z)$, and $T_c(z)$, together with the Coulomb logarithms $\ln\Lambda_j(z)$ and the mean free paths $\lambda_j(z)$, are recomputed consistently from the current values of $(w_j,C,T_M)$.

The initial conditions are
\begin{equation}
w_j(0)=1 \quad \forall j,
\qquad
C(0)=1-A_t,
\qquad
T_M(0)=T_0 .
\end{equation}

The numerical scheme was validated through standard convergence and consistency tests. In particular, increasing the number of temperature grid points \(N_T\) does not produce appreciable changes in the density and temperature profiles. Moreover, the solver correctly reproduces the collisionless limit, confirming the consistency of the implementation with the analytical behaviour expected in the limit of vanishing collisional effects.

\end{document}